\documentclass[longauth]{aa}

\usepackage{graphicx}
\usepackage{natbib}
\usepackage{scalerel}
\usepackage{makecell}

\usepackage[table]{xcolor}

\usepackage{soul}
\usepackage{comment}

\def\ba#1\ea{\begin{align}#1\end{align}}

\newcommand\kk{{\vec k}}

\newcommand\mr{\mathrm}
\newcommand\bs{\boldsymbol}

\newcommand\omegac{\omega_\mathrm{cdm}}
\newcommand\omegab{\omega_\mathrm{b}}

\newcommand\invMpc{h/\mr{Mpc}}
\newcommand\Mpc{\mr{Mpc}/h}
\newcommand\Gpc{\mr{Gpc}/h}

\newcommand\kmax{k_\mr{max}}

\newcommand\dif{{\rm d}}

\newcommand{\covmos}{\texttt{COVMOS}}
\newcommand{\demcov}{\texttt{DEMNUni-Cov}}

\newcommand{\halofit}{\texttt{Halofit}}

\newcommand{\lcal}{ {\mathcal L} }

\newcommand{\gcal}{ {\mathcal G} }
\newcommand{\scal}{ {\mathcal S} }

\newcommand{\dirac}{\delta^{\rm D}}

\newcommand{\rmin}{r_{\rm min}}
\newcommand{\rmax}{r_{\rm max}}

\newcommand{\acos}{{\rm acos}}

\defcitealias{KP1}{B-2025}

\usepackage{txfonts}
\usepackage[pdfencoding=auto,psdextra]{hyperref}
\hypersetup{
    colorlinks=true,
    linkcolor=blue,
    filecolor=magenta,      
    urlcolor=blue,
    citecolor=blue
}
\makeatletter
\renewcommand*\aa@pageof{, page \thepage{} of 25 }
\makeatother

\usepackage[utf8]{inputenc}

\usepackage[switch, modulo]{lineno}

\renewcommand{\linenumbers}[0]{}

\usepackage{euclid}

\begin{document}

\newcommand{\orcid}[1]{} %% if already defined in aa.cls: comment, or use renewcommand			   
\author{Euclid Collaboration: S.~Gouyou~Beauchamps\thanks{\email{gouyou@ice.csic.es}}\inst{\ref{aff1},\ref{aff2}}
\and J.~Bel\inst{\ref{aff3}}
\and P.~Baratta\orcid{0000-0001-5533-8437}\inst{\ref{aff4},\ref{aff5}}
\and C.~Carbone\orcid{0000-0003-0125-3563}\inst{\ref{aff6}}
\and B.~Altieri\orcid{0000-0003-3936-0284}\inst{\ref{aff7}}
\and S.~Andreon\orcid{0000-0002-2041-8784}\inst{\ref{aff8}}
\and N.~Auricchio\orcid{0000-0003-4444-8651}\inst{\ref{aff9}}
\and C.~Baccigalupi\orcid{0000-0002-8211-1630}\inst{\ref{aff10},\ref{aff11},\ref{aff12},\ref{aff13}}
\and M.~Baldi\orcid{0000-0003-4145-1943}\inst{\ref{aff14},\ref{aff9},\ref{aff15}}
\and S.~Bardelli\orcid{0000-0002-8900-0298}\inst{\ref{aff9}}
\and P.~Battaglia\orcid{0000-0002-7337-5909}\inst{\ref{aff9}}
\and F.~Bernardeau\orcid{0009-0007-3015-2581}\inst{\ref{aff16},\ref{aff17}}
\and A.~Biviano\orcid{0000-0002-0857-0732}\inst{\ref{aff11},\ref{aff10}}
\and E.~Branchini\orcid{0000-0002-0808-6908}\inst{\ref{aff18},\ref{aff19},\ref{aff8}}
\and M.~Brescia\orcid{0000-0001-9506-5680}\inst{\ref{aff20},\ref{aff21}}
\and S.~Camera\orcid{0000-0003-3399-3574}\inst{\ref{aff22},\ref{aff23},\ref{aff24}}
\and G.~Ca\~nas-Herrera\orcid{0000-0003-2796-2149}\inst{\ref{aff25},\ref{aff26}}
\and V.~Capobianco\orcid{0000-0002-3309-7692}\inst{\ref{aff24}}
\and V.~F.~Cardone\inst{\ref{aff27},\ref{aff28}}
\and J.~Carretero\orcid{0000-0002-3130-0204}\inst{\ref{aff29},\ref{aff30}}
\and S.~Casas\orcid{0000-0002-4751-5138}\inst{\ref{aff31},\ref{aff32}}
\and M.~Castellano\orcid{0000-0001-9875-8263}\inst{\ref{aff27}}
\and G.~Castignani\orcid{0000-0001-6831-0687}\inst{\ref{aff9}}
\and S.~Cavuoti\orcid{0000-0002-3787-4196}\inst{\ref{aff21},\ref{aff33}}
\and K.~C.~Chambers\orcid{0000-0001-6965-7789}\inst{\ref{aff34}}
\and C.~Colodro-Conde\inst{\ref{aff35}}
\and G.~Congedo\orcid{0000-0003-2508-0046}\inst{\ref{aff25}}
\and L.~Conversi\orcid{0000-0002-6710-8476}\inst{\ref{aff36},\ref{aff7}}
\and Y.~Copin\orcid{0000-0002-5317-7518}\inst{\ref{aff37}}
\and F.~Courbin\orcid{0000-0003-0758-6510}\inst{\ref{aff38},\ref{aff39},\ref{aff40}}
\and H.~M.~Courtois\orcid{0000-0003-0509-1776}\inst{\ref{aff41}}
\and M.~Crocce\orcid{0000-0002-9745-6228}\inst{\ref{aff2},\ref{aff1}}
\and A.~Da~Silva\orcid{0000-0002-6385-1609}\inst{\ref{aff42},\ref{aff43}}
\and H.~Degaudenzi\orcid{0000-0002-5887-6799}\inst{\ref{aff44}}
\and S.~de~la~Torre\inst{\ref{aff45}}
\and G.~De~Lucia\orcid{0000-0002-6220-9104}\inst{\ref{aff11}}
\and H.~Dole\orcid{0000-0002-9767-3839}\inst{\ref{aff46}}
\and F.~Dubath\orcid{0000-0002-6533-2810}\inst{\ref{aff44}}
\and X.~Dupac\inst{\ref{aff7}}
\and S.~Dusini\orcid{0000-0002-1128-0664}\inst{\ref{aff47}}
\and S.~Escoffier\orcid{0000-0002-2847-7498}\inst{\ref{aff4}}
\and M.~Farina\orcid{0000-0002-3089-7846}\inst{\ref{aff48}}
\and R.~Farinelli\inst{\ref{aff9}}
\and S.~Farrens\orcid{0000-0002-9594-9387}\inst{\ref{aff49}}
\and S.~Ferriol\inst{\ref{aff37}}
\and F.~Finelli\orcid{0000-0002-6694-3269}\inst{\ref{aff9},\ref{aff50}}
\and P.~Fosalba\orcid{0000-0002-1510-5214}\inst{\ref{aff1},\ref{aff2}}
\and S.~Fotopoulou\orcid{0000-0002-9686-254X}\inst{\ref{aff51}}
\and N.~Fourmanoit\orcid{0009-0005-6816-6925}\inst{\ref{aff4}}
\and M.~Frailis\orcid{0000-0002-7400-2135}\inst{\ref{aff11}}
\and E.~Franceschi\orcid{0000-0002-0585-6591}\inst{\ref{aff9}}
\and M.~Fumana\orcid{0000-0001-6787-5950}\inst{\ref{aff6}}
\and S.~Galeotta\orcid{0000-0002-3748-5115}\inst{\ref{aff11}}
\and K.~George\orcid{0000-0002-1734-8455}\inst{\ref{aff52}}
\and W.~Gillard\orcid{0000-0003-4744-9748}\inst{\ref{aff4}}
\and B.~Gillis\orcid{0000-0002-4478-1270}\inst{\ref{aff25}}
\and C.~Giocoli\orcid{0000-0002-9590-7961}\inst{\ref{aff9},\ref{aff15}}
\and J.~Gracia-Carpio\orcid{0000-0003-4689-3134}\inst{\ref{aff53}}
\and A.~Grazian\orcid{0000-0002-5688-0663}\inst{\ref{aff54}}
\and F.~Grupp\inst{\ref{aff53},\ref{aff55}}
\and S.~V.~H.~Haugan\orcid{0000-0001-9648-7260}\inst{\ref{aff56}}
\and W.~Holmes\inst{\ref{aff57}}
\and A.~Hornstrup\orcid{0000-0002-3363-0936}\inst{\ref{aff58},\ref{aff59}}
\and K.~Jahnke\orcid{0000-0003-3804-2137}\inst{\ref{aff60}}
\and B.~Joachimi\orcid{0000-0001-7494-1303}\inst{\ref{aff61}}
\and S.~Kermiche\orcid{0000-0002-0302-5735}\inst{\ref{aff4}}
\and A.~Kiessling\orcid{0000-0002-2590-1273}\inst{\ref{aff57}}
\and B.~Kubik\orcid{0009-0006-5823-4880}\inst{\ref{aff37}}
\and M.~Kunz\orcid{0000-0002-3052-7394}\inst{\ref{aff62}}
\and H.~Kurki-Suonio\orcid{0000-0002-4618-3063}\inst{\ref{aff63},\ref{aff64}}
\and A.~M.~C.~Le~Brun\orcid{0000-0002-0936-4594}\inst{\ref{aff65}}
\and S.~Ligori\orcid{0000-0003-4172-4606}\inst{\ref{aff24}}
\and P.~B.~Lilje\orcid{0000-0003-4324-7794}\inst{\ref{aff56}}
\and V.~Lindholm\orcid{0000-0003-2317-5471}\inst{\ref{aff63},\ref{aff64}}
\and I.~Lloro\orcid{0000-0001-5966-1434}\inst{\ref{aff66}}
\and G.~Mainetti\orcid{0000-0003-2384-2377}\inst{\ref{aff67}}
\and E.~Maiorano\orcid{0000-0003-2593-4355}\inst{\ref{aff9}}
\and O.~Mansutti\orcid{0000-0001-5758-4658}\inst{\ref{aff11}}
\and S.~Marcin\inst{\ref{aff68}}
\and O.~Marggraf\orcid{0000-0001-7242-3852}\inst{\ref{aff69}}
\and K.~Markovic\orcid{0000-0001-6764-073X}\inst{\ref{aff57}}
\and M.~Martinelli\orcid{0000-0002-6943-7732}\inst{\ref{aff27},\ref{aff28}}
\and N.~Martinet\orcid{0000-0003-2786-7790}\inst{\ref{aff45}}
\and F.~Marulli\orcid{0000-0002-8850-0303}\inst{\ref{aff70},\ref{aff9},\ref{aff15}}
\and R.~J.~Massey\orcid{0000-0002-6085-3780}\inst{\ref{aff71}}
\and E.~Medinaceli\orcid{0000-0002-4040-7783}\inst{\ref{aff9}}
\and S.~Mei\orcid{0000-0002-2849-559X}\inst{\ref{aff72},\ref{aff73}}
\and M.~Meneghetti\orcid{0000-0003-1225-7084}\inst{\ref{aff9},\ref{aff15}}
\and E.~Merlin\orcid{0000-0001-6870-8900}\inst{\ref{aff27}}
\and G.~Meylan\inst{\ref{aff74}}
\and A.~Mora\orcid{0000-0002-1922-8529}\inst{\ref{aff75}}
\and M.~Moresco\orcid{0000-0002-7616-7136}\inst{\ref{aff70},\ref{aff9}}
\and L.~Moscardini\orcid{0000-0002-3473-6716}\inst{\ref{aff70},\ref{aff9},\ref{aff15}}
\and R.~Nakajima\orcid{0009-0009-1213-7040}\inst{\ref{aff69}}
\and C.~Neissner\orcid{0000-0001-8524-4968}\inst{\ref{aff76},\ref{aff30}}
\and S.-M.~Niemi\orcid{0009-0005-0247-0086}\inst{\ref{aff77}}
\and C.~Padilla\orcid{0000-0001-7951-0166}\inst{\ref{aff76}}
\and S.~Paltani\orcid{0000-0002-8108-9179}\inst{\ref{aff44}}
\and F.~Pasian\orcid{0000-0002-4869-3227}\inst{\ref{aff11}}
\and K.~Pedersen\inst{\ref{aff78}}
\and W.~J.~Percival\orcid{0000-0002-0644-5727}\inst{\ref{aff79},\ref{aff80},\ref{aff81}}
\and V.~Pettorino\orcid{0000-0002-4203-9320}\inst{\ref{aff77}}
\and S.~Pires\orcid{0000-0002-0249-2104}\inst{\ref{aff49}}
\and G.~Polenta\orcid{0000-0003-4067-9196}\inst{\ref{aff82}}
\and M.~Poncet\inst{\ref{aff83}}
\and L.~A.~Popa\inst{\ref{aff84}}
\and F.~Raison\orcid{0000-0002-7819-6918}\inst{\ref{aff53}}
\and A.~Renzi\orcid{0000-0001-9856-1970}\inst{\ref{aff85},\ref{aff47},\ref{aff9}}
\and J.~Rhodes\orcid{0000-0002-4485-8549}\inst{\ref{aff57}}
\and G.~Riccio\inst{\ref{aff21}}
\and E.~Romelli\orcid{0000-0003-3069-9222}\inst{\ref{aff11}}
\and M.~Roncarelli\orcid{0000-0001-9587-7822}\inst{\ref{aff9}}
\and C.~Rosset\orcid{0000-0003-0286-2192}\inst{\ref{aff72}}
\and R.~Saglia\orcid{0000-0003-0378-7032}\inst{\ref{aff55},\ref{aff53}}
\and Z.~Sakr\orcid{0000-0002-4823-3757}\inst{\ref{aff86},\ref{aff87},\ref{aff88}}
\and A.~G.~S\'anchez\orcid{0000-0003-1198-831X}\inst{\ref{aff53}}
\and D.~Sapone\orcid{0000-0001-7089-4503}\inst{\ref{aff89}}
\and B.~Sartoris\orcid{0000-0003-1337-5269}\inst{\ref{aff55},\ref{aff11}}
\and P.~Schneider\orcid{0000-0001-8561-2679}\inst{\ref{aff69}}
\and A.~Secroun\orcid{0000-0003-0505-3710}\inst{\ref{aff4}}
\and G.~Seidel\orcid{0000-0003-2907-353X}\inst{\ref{aff60}}
\and E.~Sihvola\orcid{0000-0003-1804-7715}\inst{\ref{aff90}}
\and P.~Simon\inst{\ref{aff69}}
\and C.~Sirignano\orcid{0000-0002-0995-7146}\inst{\ref{aff85},\ref{aff47}}
\and G.~Sirri\orcid{0000-0003-2626-2853}\inst{\ref{aff15}}
\and P.~Tallada-Cresp\'{i}\orcid{0000-0002-1336-8328}\inst{\ref{aff29},\ref{aff30}}
\and A.~N.~Taylor\inst{\ref{aff25}}
\and I.~Tereno\orcid{0000-0002-4537-6218}\inst{\ref{aff42},\ref{aff91}}
\and N.~Tessore\orcid{0000-0002-9696-7931}\inst{\ref{aff92}}
\and S.~Toft\orcid{0000-0003-3631-7176}\inst{\ref{aff93},\ref{aff94}}
\and R.~Toledo-Moreo\orcid{0000-0002-2997-4859}\inst{\ref{aff95}}
\and F.~Torradeflot\orcid{0000-0003-1160-1517}\inst{\ref{aff30},\ref{aff29}}
\and I.~Tutusaus\orcid{0000-0002-3199-0399}\inst{\ref{aff2},\ref{aff1},\ref{aff87}}
\and J.~Valiviita\orcid{0000-0001-6225-3693}\inst{\ref{aff63},\ref{aff64}}
\and T.~Vassallo\orcid{0000-0001-6512-6358}\inst{\ref{aff11},\ref{aff52}}
\and G.~Verdoes~Kleijn\orcid{0000-0001-5803-2580}\inst{\ref{aff96}}
\and A.~Veropalumbo\orcid{0000-0003-2387-1194}\inst{\ref{aff8},\ref{aff19},\ref{aff18}}
\and Y.~Wang\orcid{0000-0002-4749-2984}\inst{\ref{aff97}}
\and J.~Weller\orcid{0000-0002-8282-2010}\inst{\ref{aff55},\ref{aff53}}
\and G.~Zamorani\orcid{0000-0002-2318-301X}\inst{\ref{aff9}}
\and E.~Zucca\orcid{0000-0002-5845-8132}\inst{\ref{aff9}}
\and M.~Ballardini\orcid{0000-0003-4481-3559}\inst{\ref{aff98},\ref{aff99},\ref{aff9}}
\and A.~Boucaud\orcid{0000-0001-7387-2633}\inst{\ref{aff72}}
\and E.~Bozzo\orcid{0000-0002-8201-1525}\inst{\ref{aff44}}
\and C.~Burigana\orcid{0000-0002-3005-5796}\inst{\ref{aff100},\ref{aff50}}
\and R.~Cabanac\orcid{0000-0001-6679-2600}\inst{\ref{aff87}}
\and M.~Calabrese\orcid{0000-0002-2637-2422}\inst{\ref{aff101},\ref{aff6}}
\and A.~Cappi\inst{\ref{aff102},\ref{aff9}}
\and T.~Castro\orcid{0000-0002-6292-3228}\inst{\ref{aff11},\ref{aff12},\ref{aff10},\ref{aff103}}
\and J.~A.~Escartin~Vigo\inst{\ref{aff53}}
\and L.~Gabarra\orcid{0000-0002-8486-8856}\inst{\ref{aff104}}
\and J.~Garc\'ia-Bellido\orcid{0000-0002-9370-8360}\inst{\ref{aff86}}
\and J.~Macias-Perez\orcid{0000-0002-5385-2763}\inst{\ref{aff105}}
\and R.~Maoli\orcid{0000-0002-6065-3025}\inst{\ref{aff106},\ref{aff27}}
\and N.~Mauri\orcid{0000-0001-8196-1548}\inst{\ref{aff107},\ref{aff15}}
\and R.~B.~Metcalf\orcid{0000-0003-3167-2574}\inst{\ref{aff70},\ref{aff9}}
\and P.~Monaco\orcid{0000-0003-2083-7564}\inst{\ref{aff108},\ref{aff11},\ref{aff12},\ref{aff10}}
\and A.~A.~Nucita\inst{\ref{aff109},\ref{aff110},\ref{aff111}}
\and A.~Pezzotta\orcid{0000-0003-0726-2268}\inst{\ref{aff8}}
\and M.~P\"ontinen\orcid{0000-0001-5442-2530}\inst{\ref{aff63}}
\and I.~Risso\orcid{0000-0003-2525-7761}\inst{\ref{aff8},\ref{aff19}}
\and V.~Scottez\orcid{0009-0008-3864-940X}\inst{\ref{aff112},\ref{aff113}}
\and M.~Sereno\orcid{0000-0003-0302-0325}\inst{\ref{aff9},\ref{aff15}}
\and M.~Tenti\orcid{0000-0002-4254-5901}\inst{\ref{aff15}}
\and M.~Tucci\inst{\ref{aff44}}
\and M.~Viel\orcid{0000-0002-2642-5707}\inst{\ref{aff10},\ref{aff11},\ref{aff13},\ref{aff12},\ref{aff103}}
\and M.~Wiesmann\orcid{0009-0000-8199-5860}\inst{\ref{aff56}}
\and Y.~Akrami\orcid{0000-0002-2407-7956}\inst{\ref{aff86},\ref{aff114}}
\and I.~T.~Andika\orcid{0000-0001-6102-9526}\inst{\ref{aff52}}
\and S.~Anselmi\orcid{0000-0002-3579-9583}\inst{\ref{aff47},\ref{aff85},\ref{aff115}}
\and M.~Archidiacono\orcid{0000-0003-4952-9012}\inst{\ref{aff116},\ref{aff117}}
\and F.~Atrio-Barandela\orcid{0000-0002-2130-2513}\inst{\ref{aff118}}
\and L.~Bazzanini\orcid{0000-0003-0727-0137}\inst{\ref{aff98},\ref{aff9}}
\and D.~Bertacca\orcid{0000-0002-2490-7139}\inst{\ref{aff85},\ref{aff54},\ref{aff47}}
\and M.~Bethermin\orcid{0000-0002-3915-2015}\inst{\ref{aff119}}
\and F.~Beutler\orcid{0000-0003-0467-5438}\inst{\ref{aff25}}
\and A.~Blanchard\orcid{0000-0001-8555-9003}\inst{\ref{aff87}}
\and L.~Blot\orcid{0000-0002-9622-7167}\inst{\ref{aff120},\ref{aff65}}
\and M.~Bonici\orcid{0000-0002-8430-126X}\inst{\ref{aff79},\ref{aff6}}
\and M.~L.~Brown\orcid{0000-0002-0370-8077}\inst{\ref{aff121}}
\and S.~Bruton\orcid{0000-0002-6503-5218}\inst{\ref{aff122}}
\and A.~Calabro\orcid{0000-0003-2536-1614}\inst{\ref{aff27}}
\and B.~Camacho~Quevedo\orcid{0000-0002-8789-4232}\inst{\ref{aff10},\ref{aff13},\ref{aff11}}
\and F.~Caro\inst{\ref{aff27}}
\and C.~S.~Carvalho\inst{\ref{aff91}}
\and F.~Cogato\orcid{0000-0003-4632-6113}\inst{\ref{aff70},\ref{aff9}}
\and A.~R.~Cooray\orcid{0000-0002-3892-0190}\inst{\ref{aff123}}
\and S.~Davini\orcid{0000-0003-3269-1718}\inst{\ref{aff19}}
\and F.~De~Paolis\orcid{0000-0001-6460-7563}\inst{\ref{aff109},\ref{aff110},\ref{aff111}}
\and G.~Desprez\orcid{0000-0001-8325-1742}\inst{\ref{aff96}}
\and A.~D\'iaz-S\'anchez\orcid{0000-0003-0748-4768}\inst{\ref{aff124}}
\and S.~Di~Domizio\orcid{0000-0003-2863-5895}\inst{\ref{aff18},\ref{aff19}}
\and J.~M.~Diego\orcid{0000-0001-9065-3926}\inst{\ref{aff125}}
\and V.~Duret\orcid{0009-0009-0383-4960}\inst{\ref{aff4}}
\and M.~Y.~Elkhashab\orcid{0000-0001-9306-2603}\inst{\ref{aff108},\ref{aff11},\ref{aff12},\ref{aff10}}
\and A.~Enia\orcid{0000-0002-0200-2857}\inst{\ref{aff9}}
\and Y.~Fang\orcid{0000-0002-0334-6950}\inst{\ref{aff55}}
\and A.~G.~Ferrari\orcid{0009-0005-5266-4110}\inst{\ref{aff15}}
\and A.~Finoguenov\orcid{0000-0002-4606-5403}\inst{\ref{aff63}}
\and A.~Franco\orcid{0000-0002-4761-366X}\inst{\ref{aff110},\ref{aff109},\ref{aff111}}
\and K.~Ganga\orcid{0000-0001-8159-8208}\inst{\ref{aff72}}
\and T.~Gasparetto\orcid{0000-0002-7913-4866}\inst{\ref{aff27}}
\and E.~Gaztanaga\orcid{0000-0001-9632-0815}\inst{\ref{aff2},\ref{aff1},\ref{aff126}}
\and F.~Giacomini\orcid{0000-0002-3129-2814}\inst{\ref{aff15}}
\and F.~Gianotti\orcid{0000-0003-4666-119X}\inst{\ref{aff9}}
\and E.~J.~Gonzalez\orcid{0000-0002-0226-9893}\inst{\ref{aff76},\ref{aff30},\ref{aff127}}
\and G.~Gozaliasl\orcid{0000-0002-0236-919X}\inst{\ref{aff128},\ref{aff63}}
\and A.~Gruppuso\orcid{0000-0001-9272-5292}\inst{\ref{aff9},\ref{aff15}}
\and M.~Guidi\orcid{0000-0001-9408-1101}\inst{\ref{aff14},\ref{aff9}}
\and C.~M.~Gutierrez\orcid{0000-0001-7854-783X}\inst{\ref{aff35},\ref{aff129}}
\and A.~Hall\orcid{0000-0002-3139-8651}\inst{\ref{aff25}}
\and H.~Hildebrandt\orcid{0000-0002-9814-3338}\inst{\ref{aff130}}
\and J.~Hjorth\orcid{0000-0002-4571-2306}\inst{\ref{aff78}}
\and J.~J.~E.~Kajava\orcid{0000-0002-3010-8333}\inst{\ref{aff131},\ref{aff132},\ref{aff133}}
\and Y.~Kang\orcid{0009-0000-8588-7250}\inst{\ref{aff44}}
\and V.~Kansal\orcid{0000-0002-4008-6078}\inst{\ref{aff134},\ref{aff135}}
\and D.~Karagiannis\orcid{0000-0002-4927-0816}\inst{\ref{aff98},\ref{aff136}}
\and K.~Kiiveri\inst{\ref{aff90}}
\and J.~Kim\orcid{0000-0003-2776-2761}\inst{\ref{aff104}}
\and C.~C.~Kirkpatrick\inst{\ref{aff90}}
\and S.~Kruk\orcid{0000-0001-8010-8879}\inst{\ref{aff7}}
\and F.~Lacasa\orcid{0000-0002-7268-3440}\inst{\ref{aff137},\ref{aff46}}
\and M.~Lattanzi\orcid{0000-0003-1059-2532}\inst{\ref{aff99}}
\and J.~Le~Graet\orcid{0000-0001-6523-7971}\inst{\ref{aff4}}
\and L.~Legrand\orcid{0000-0003-0610-5252}\inst{\ref{aff138},\ref{aff139}}
\and M.~Lembo\orcid{0000-0002-5271-5070}\inst{\ref{aff17}}
\and F.~Lepori\orcid{0009-0000-5061-7138}\inst{\ref{aff140}}
\and G.~Leroy\orcid{0009-0004-2523-4425}\inst{\ref{aff141},\ref{aff71}}
\and G.~F.~Lesci\orcid{0000-0002-4607-2830}\inst{\ref{aff70},\ref{aff9}}
\and J.~Lesgourgues\orcid{0000-0001-7627-353X}\inst{\ref{aff31}}
\and T.~I.~Liaudat\orcid{0000-0002-9104-314X}\inst{\ref{aff142}}
\and S.~J.~Liu\orcid{0000-0001-7680-2139}\inst{\ref{aff48}}
\and M.~Magliocchetti\orcid{0000-0001-9158-4838}\inst{\ref{aff48}}
\and F.~Mannucci\orcid{0000-0002-4803-2381}\inst{\ref{aff143}}
\and C.~J.~A.~P.~Martins\orcid{0000-0002-4886-9261}\inst{\ref{aff144},\ref{aff145}}
\and L.~Maurin\orcid{0000-0002-8406-0857}\inst{\ref{aff46}}
\and M.~Miluzio\inst{\ref{aff7},\ref{aff146}}
\and C.~Moretti\orcid{0000-0003-3314-8936}\inst{\ref{aff11},\ref{aff10},\ref{aff12}}
\and G.~Morgante\inst{\ref{aff9}}
\and C.~Murray\inst{\ref{aff72}}
\and S.~Nadathur\orcid{0000-0001-9070-3102}\inst{\ref{aff126}}
\and K.~Naidoo\orcid{0000-0002-9182-1802}\inst{\ref{aff126},\ref{aff60}}
\and A.~Navarro-Alsina\orcid{0000-0002-3173-2592}\inst{\ref{aff69}}
\and S.~Nesseris\orcid{0000-0002-0567-0324}\inst{\ref{aff86}}
\and L.~Pagano\orcid{0000-0003-1820-5998}\inst{\ref{aff98},\ref{aff99}}
\and D.~Paoletti\orcid{0000-0003-4761-6147}\inst{\ref{aff9},\ref{aff50}}
\and F.~Passalacqua\orcid{0000-0002-8606-4093}\inst{\ref{aff85},\ref{aff47}}
\and K.~Paterson\orcid{0000-0001-8340-3486}\inst{\ref{aff60}}
\and L.~Patrizii\inst{\ref{aff15}}
\and C.~Pattison\orcid{0000-0003-3272-2617}\inst{\ref{aff126}}
\and R.~Paviot\orcid{0009-0002-8108-3460}\inst{\ref{aff49}}
\and A.~Pisani\orcid{0000-0002-6146-4437}\inst{\ref{aff4}}
\and D.~Potter\orcid{0000-0002-0757-5195}\inst{\ref{aff147}}
\and G.~W.~Pratt\inst{\ref{aff49}}
\and S.~Quai\orcid{0000-0002-0449-8163}\inst{\ref{aff70},\ref{aff9}}
\and M.~Radovich\orcid{0000-0002-3585-866X}\inst{\ref{aff54}}
\and W.~Roster\orcid{0000-0002-9149-6528}\inst{\ref{aff53}}
\and S.~Sacquegna\orcid{0000-0002-8433-6630}\inst{\ref{aff148}}
\and M.~Sahl\'en\orcid{0000-0003-0973-4804}\inst{\ref{aff149}}
\and D.~B.~Sanders\orcid{0000-0002-1233-9998}\inst{\ref{aff34}}
\and A.~Schneider\orcid{0000-0001-7055-8104}\inst{\ref{aff147}}
\and D.~Sciotti\orcid{0009-0008-4519-2620}\inst{\ref{aff27},\ref{aff28}}
\and E.~Sellentin\inst{\ref{aff150},\ref{aff26}}
\and L.~C.~Smith\orcid{0000-0002-3259-2771}\inst{\ref{aff151}}
\and K.~Tanidis\orcid{0000-0001-9843-5130}\inst{\ref{aff152}}
\and C.~Tao\orcid{0000-0001-7961-8177}\inst{\ref{aff4}}
\and F.~Tarsitano\orcid{0000-0002-5919-0238}\inst{\ref{aff153},\ref{aff44}}
\and G.~Testera\inst{\ref{aff19}}
\and R.~Teyssier\orcid{0000-0001-7689-0933}\inst{\ref{aff154}}
\and S.~Tosi\orcid{0000-0002-7275-9193}\inst{\ref{aff18},\ref{aff8},\ref{aff19}}
\and A.~Troja\orcid{0000-0003-0239-4595}\inst{\ref{aff85},\ref{aff47}}
\and A.~Venhola\orcid{0000-0001-6071-4564}\inst{\ref{aff155}}
\and D.~Vergani\orcid{0000-0003-0898-2216}\inst{\ref{aff9}}
\and F.~Vernizzi\orcid{0000-0003-3426-2802}\inst{\ref{aff16}}
\and G.~Verza\orcid{0000-0002-1886-8348}\inst{\ref{aff156},\ref{aff157}}
\and P.~Vielzeuf\orcid{0000-0003-2035-9339}\inst{\ref{aff4}}
\and S.~Vinciguerra\orcid{0009-0005-4018-3184}\inst{\ref{aff45}}
\and N.~A.~Walton\orcid{0000-0003-3983-8778}\inst{\ref{aff151}}
\and A.~H.~Wright\orcid{0000-0001-7363-7932}\inst{\ref{aff130}}}
										   
%%%% please do not edit the affiliation list -- contact ECEB Bureau for changes
\institute{Institut d'Estudis Espacials de Catalunya (IEEC),  Edifici RDIT, Campus UPC, 08860 Castelldefels, Barcelona, Spain\label{aff1}
\and
Institute of Space Sciences (ICE, CSIC), Campus UAB, Carrer de Can Magrans, s/n, 08193 Barcelona, Spain\label{aff2}
\and
Aix-Marseille Universit\'e, Universit\'e de Toulon, CNRS, CPT, Marseille, France\label{aff3}
\and
Aix-Marseille Universit\'e, CNRS/IN2P3, CPPM, Marseille, France\label{aff4}
\and
Aix Marseille Univ, INSERM, MMG, Marseille, France\label{aff5}
\and
INAF-IASF Milano, Via Alfonso Corti 12, 20133 Milano, Italy\label{aff6}
\and
ESAC/ESA, Camino Bajo del Castillo, s/n., Urb. Villafranca del Castillo, 28692 Villanueva de la Ca\~nada, Madrid, Spain\label{aff7}
\and
INAF-Osservatorio Astronomico di Brera, Via Brera 28, 20122 Milano, Italy\label{aff8}
\and
INAF-Osservatorio di Astrofisica e Scienza dello Spazio di Bologna, Via Piero Gobetti 93/3, 40129 Bologna, Italy\label{aff9}
\and
IFPU, Institute for Fundamental Physics of the Universe, via Beirut 2, 34151 Trieste, Italy\label{aff10}
\and
INAF-Osservatorio Astronomico di Trieste, Via G. B. Tiepolo 11, 34143 Trieste, Italy\label{aff11}
\and
INFN, Sezione di Trieste, Via Valerio 2, 34127 Trieste TS, Italy\label{aff12}
\and
SISSA, International School for Advanced Studies, Via Bonomea 265, 34136 Trieste TS, Italy\label{aff13}
\and
Dipartimento di Fisica e Astronomia, Universit\`a di Bologna, Via Gobetti 93/2, 40129 Bologna, Italy\label{aff14}
\and
INFN-Sezione di Bologna, Viale Berti Pichat 6/2, 40127 Bologna, Italy\label{aff15}
\and
Institut de Physique Th\'eorique, CEA, CNRS, Universit\'e Paris-Saclay 91191 Gif-sur-Yvette Cedex, France\label{aff16}
\and
Institut d'Astrophysique de Paris, UMR 7095, CNRS, and Sorbonne Universit\'e, 98 bis boulevard Arago, 75014 Paris, France\label{aff17}
\and
Dipartimento di Fisica, Universit\`a di Genova, Via Dodecaneso 33, 16146, Genova, Italy\label{aff18}
\and
INFN-Sezione di Genova, Via Dodecaneso 33, 16146, Genova, Italy\label{aff19}
\and
Department of Physics "E. Pancini", University Federico II, Via Cinthia 6, 80126, Napoli, Italy\label{aff20}
\and
INAF-Osservatorio Astronomico di Capodimonte, Via Moiariello 16, 80131 Napoli, Italy\label{aff21}
\and
Dipartimento di Fisica, Universit\`a degli Studi di Torino, Via P. Giuria 1, 10125 Torino, Italy\label{aff22}
\and
INFN-Sezione di Torino, Via P. Giuria 1, 10125 Torino, Italy\label{aff23}
\and
INAF-Osservatorio Astrofisico di Torino, Via Osservatorio 20, 10025 Pino Torinese (TO), Italy\label{aff24}
\and
Institute for Astronomy, University of Edinburgh, Royal Observatory, Blackford Hill, Edinburgh EH9 3HJ, UK\label{aff25}
\and
Leiden Observatory, Leiden University, Einsteinweg 55, 2333 CC Leiden, The Netherlands\label{aff26}
\and
INAF-Osservatorio Astronomico di Roma, Via Frascati 33, 00078 Monteporzio Catone, Italy\label{aff27}
\and
INFN-Sezione di Roma, Piazzale Aldo Moro, 2 - c/o Dipartimento di Fisica, Edificio G. Marconi, 00185 Roma, Italy\label{aff28}
\and
Centro de Investigaciones Energ\'eticas, Medioambientales y Tecnol\'ogicas (CIEMAT), Avenida Complutense 40, 28040 Madrid, Spain\label{aff29}
\and
Port d'Informaci\'{o} Cient\'{i}fica, Campus UAB, C. Albareda s/n, 08193 Bellaterra (Barcelona), Spain\label{aff30}
\and
Institute for Theoretical Particle Physics and Cosmology (TTK), RWTH Aachen University, 52056 Aachen, Germany\label{aff31}
\and
Deutsches Zentrum f\"ur Luft- und Raumfahrt e. V. (DLR), Linder H\"ohe, 51147 K\"oln, Germany\label{aff32}
\and
INFN section of Naples, Via Cinthia 6, 80126, Napoli, Italy\label{aff33}
\and
Institute for Astronomy, University of Hawaii, 2680 Woodlawn Drive, Honolulu, HI 96822, USA\label{aff34}
\and
Instituto de Astrof\'{\i}sica de Canarias, E-38205 La Laguna, Tenerife, Spain\label{aff35}
\and
European Space Agency/ESRIN, Largo Galileo Galilei 1, 00044 Frascati, Roma, Italy\label{aff36}
\and
Universit\'e Claude Bernard Lyon 1, CNRS/IN2P3, IP2I Lyon, UMR 5822, Villeurbanne, F-69100, France\label{aff37}
\and
Institut de Ci\`{e}ncies del Cosmos (ICCUB), Universitat de Barcelona (IEEC-UB), Mart\'{i} i Franqu\`{e}s 1, 08028 Barcelona, Spain\label{aff38}
\and
Instituci\'o Catalana de Recerca i Estudis Avan\c{c}ats (ICREA), Passeig de Llu\'{\i}s Companys 23, 08010 Barcelona, Spain\label{aff39}
\and
Institut de Ciencies de l'Espai (IEEC-CSIC), Campus UAB, Carrer de Can Magrans, s/n Cerdanyola del Vall\'es, 08193 Barcelona, Spain\label{aff40}
\and
UCB Lyon 1, CNRS/IN2P3, IUF, IP2I Lyon, 4 rue Enrico Fermi, 69622 Villeurbanne, France\label{aff41}
\and
Departamento de F\'isica, Faculdade de Ci\^encias, Universidade de Lisboa, Edif\'icio C8, Campo Grande, PT1749-016 Lisboa, Portugal\label{aff42}
\and
Instituto de Astrof\'isica e Ci\^encias do Espa\c{c}o, Faculdade de Ci\^encias, Universidade de Lisboa, Campo Grande, 1749-016 Lisboa, Portugal\label{aff43}
\and
Department of Astronomy, University of Geneva, ch. d'Ecogia 16, 1290 Versoix, Switzerland\label{aff44}
\and
Aix-Marseille Universit\'e, CNRS, CNES, LAM, Marseille, France\label{aff45}
\and
Universit\'e Paris-Saclay, CNRS, Institut d'astrophysique spatiale, 91405, Orsay, France\label{aff46}
\and
INFN-Padova, Via Marzolo 8, 35131 Padova, Italy\label{aff47}
\and
INAF-Istituto di Astrofisica e Planetologia Spaziali, via del Fosso del Cavaliere, 100, 00100 Roma, Italy\label{aff48}
\and
Universit\'e Paris-Saclay, Universit\'e Paris Cit\'e, CEA, CNRS, AIM, 91191, Gif-sur-Yvette, France\label{aff49}
\and
INFN-Bologna, Via Irnerio 46, 40126 Bologna, Italy\label{aff50}
\and
School of Physics, HH Wills Physics Laboratory, University of Bristol, Tyndall Avenue, Bristol, BS8 1TL, UK\label{aff51}
\and
University Observatory, LMU Faculty of Physics, Scheinerstr.~1, 81679 Munich, Germany\label{aff52}
\and
Max \Planck Institute for Extraterrestrial Physics, Giessenbachstr. 1, 85748 Garching, Germany\label{aff53}
\and
INAF-Osservatorio Astronomico di Padova, Via dell'Osservatorio 5, 35122 Padova, Italy\label{aff54}
\and
Universit\"ats-Sternwarte M\"unchen, Fakult\"at f\"ur Physik, Ludwig-Maximilians-Universit\"at M\"unchen, Scheinerstr.~1, 81679 M\"unchen, Germany\label{aff55}
\and
Institute of Theoretical Astrophysics, University of Oslo, P.O. Box 1029 Blindern, 0315 Oslo, Norway\label{aff56}
\and
Jet Propulsion Laboratory, California Institute of Technology, 4800 Oak Grove Drive, Pasadena, CA, 91109, USA\label{aff57}
\and
Technical University of Denmark, Elektrovej 327, 2800 Kgs. Lyngby, Denmark\label{aff58}
\and
Cosmic Dawn Center (DAWN), Denmark\label{aff59}
\and
Max-\Planck-Institut f\"ur Astronomie, K\"onigstuhl 17, 69117 Heidelberg, Germany\label{aff60}
\and
Department of Physics and Astronomy, University College London, Gower Street, London WC1E 6BT, UK\label{aff61}
\and
Universit\'e de Gen\`eve, D\'epartement de Physique Th\'eorique and Centre for Astroparticle Physics, 24 quai Ernest-Ansermet, CH-1211 Gen\`eve 4, Switzerland\label{aff62}
\and
Department of Physics, P.O. Box 64, University of Helsinki, 00014 Helsinki, Finland\label{aff63}
\and
Helsinki Institute of Physics, Gustaf H{\"a}llstr{\"o}min katu 2, University of Helsinki, 00014 Helsinki, Finland\label{aff64}
\and
Laboratoire d'etude de l'Univers et des phenomenes eXtremes, Observatoire de Paris, Universit\'e PSL, Sorbonne Universit\'e, CNRS, 92190 Meudon, France\label{aff65}
\and
SKAO, Jodrell Bank, Lower Withington, Macclesfield SK11 9FT, UK\label{aff66}
\and
Centre de Calcul de l'IN2P3/CNRS, 21 avenue Pierre de Coubertin 69627 Villeurbanne Cedex, France\label{aff67}
\and
University of Applied Sciences and Arts of Northwestern Switzerland, School of Computer Science, 5210 Windisch, Switzerland\label{aff68}
\and
Universit\"at Bonn, Argelander-Institut f\"ur Astronomie, Auf dem H\"ugel 71, 53121 Bonn, Germany\label{aff69}
\and
Dipartimento di Fisica e Astronomia "Augusto Righi" - Alma Mater Studiorum Universit\`a di Bologna, via Piero Gobetti 93/2, 40129 Bologna, Italy\label{aff70}
\and
Department of Physics, Institute for Computational Cosmology, Durham University, South Road, Durham, DH1 3LE, UK\label{aff71}
\and
Universit\'e Paris Cit\'e, CNRS, Astroparticule et Cosmologie, 75013 Paris, France\label{aff72}
\and
CNRS-UCB International Research Laboratory, Centre Pierre Bin\'etruy, IRL2007, CPB-IN2P3, Berkeley, USA\label{aff73}
\and
Institute of Physics, Laboratory of Astrophysics, Ecole Polytechnique F\'ed\'erale de Lausanne (EPFL), Observatoire de Sauverny, 1290 Versoix, Switzerland\label{aff74}
\and
Telespazio UK S.L. for European Space Agency (ESA), Camino bajo del Castillo, s/n, Urbanizacion Villafranca del Castillo, Villanueva de la Ca\~nada, 28692 Madrid, Spain\label{aff75}
\and
Institut de F\'{i}sica d'Altes Energies (IFAE), The Barcelona Institute of Science and Technology, Campus UAB, 08193 Bellaterra (Barcelona), Spain\label{aff76}
\and
European Space Agency/ESTEC, Keplerlaan 1, 2201 AZ Noordwijk, The Netherlands\label{aff77}
\and
DARK, Niels Bohr Institute, University of Copenhagen, Jagtvej 155, 2200 Copenhagen, Denmark\label{aff78}
\and
Waterloo Centre for Astrophysics, University of Waterloo, Waterloo, Ontario N2L 3G1, Canada\label{aff79}
\and
Department of Physics and Astronomy, University of Waterloo, Waterloo, Ontario N2L 3G1, Canada\label{aff80}
\and
Perimeter Institute for Theoretical Physics, Waterloo, Ontario N2L 2Y5, Canada\label{aff81}
\and
Space Science Data Center, Italian Space Agency, via del Politecnico snc, 00133 Roma, Italy\label{aff82}
\and
Centre National d'Etudes Spatiales -- Centre spatial de Toulouse, 18 avenue Edouard Belin, 31401 Toulouse Cedex 9, France\label{aff83}
\and
Institute of Space Science, Str. Atomistilor, nr. 409 M\u{a}gurele, Ilfov, 077125, Romania\label{aff84}
\and
Dipartimento di Fisica e Astronomia "G. Galilei", Universit\`a di Padova, Via Marzolo 8, 35131 Padova, Italy\label{aff85}
\and
Instituto de F\'isica Te\'orica UAM-CSIC, Campus de Cantoblanco, 28049 Madrid, Spain\label{aff86}
\and
Institut de Recherche en Astrophysique et Plan\'etologie (IRAP), Universit\'e de Toulouse, CNRS, UPS, CNES, 14 Av. Edouard Belin, 31400 Toulouse, France\label{aff87}
\and
Universit\'e St Joseph; Faculty of Sciences, Beirut, Lebanon\label{aff88}
\and
Departamento de F\'isica, FCFM, Universidad de Chile, Blanco Encalada 2008, Santiago, Chile\label{aff89}
\and
Department of Physics and Helsinki Institute of Physics, Gustaf H\"allstr\"omin katu 2, University of Helsinki, 00014 Helsinki, Finland\label{aff90}
\and
Instituto de Astrof\'isica e Ci\^encias do Espa\c{c}o, Faculdade de Ci\^encias, Universidade de Lisboa, Tapada da Ajuda, 1349-018 Lisboa, Portugal\label{aff91}
\and
Mullard Space Science Laboratory, University College London, Holmbury St Mary, Dorking, Surrey RH5 6NT, UK\label{aff92}
\and
Cosmic Dawn Center (DAWN)\label{aff93}
\and
Niels Bohr Institute, University of Copenhagen, Jagtvej 128, 2200 Copenhagen, Denmark\label{aff94}
\and
Universidad Polit\'ecnica de Cartagena, Departamento de Electr\'onica y Tecnolog\'ia de Computadoras,  Plaza del Hospital 1, 30202 Cartagena, Spain\label{aff95}
\and
Kapteyn Astronomical Institute, University of Groningen, PO Box 800, 9700 AV Groningen, The Netherlands\label{aff96}
\and
Caltech/IPAC, 1200 E. California Blvd., Pasadena, CA 91125, USA\label{aff97}
\and
Dipartimento di Fisica e Scienze della Terra, Universit\`a degli Studi di Ferrara, Via Giuseppe Saragat 1, 44122 Ferrara, Italy\label{aff98}
\and
Istituto Nazionale di Fisica Nucleare, Sezione di Ferrara, Via Giuseppe Saragat 1, 44122 Ferrara, Italy\label{aff99}
\and
INAF, Istituto di Radioastronomia, Via Piero Gobetti 101, 40129 Bologna, Italy\label{aff100}
\and
Astronomical Observatory of the Autonomous Region of the Aosta Valley (OAVdA), Loc. Lignan 39, I-11020, Nus (Aosta Valley), Italy\label{aff101}
\and
Universit\'e C\^{o}te d'Azur, Observatoire de la C\^{o}te d'Azur, CNRS, Laboratoire Lagrange, Bd de l'Observatoire, CS 34229, 06304 Nice cedex 4, France\label{aff102}
\and
ICSC - Centro Nazionale di Ricerca in High Performance Computing, Big Data e Quantum Computing, Via Magnanelli 2, Bologna, Italy\label{aff103}
\and
Department of Physics, Oxford University, Keble Road, Oxford OX1 3RH, UK\label{aff104}
\and
Univ. Grenoble Alpes, CNRS, Grenoble INP, LPSC-IN2P3, 53, Avenue des Martyrs, 38000, Grenoble, France\label{aff105}
\and
Dipartimento di Fisica, Sapienza Universit\`a di Roma, Piazzale Aldo Moro 2, 00185 Roma, Italy\label{aff106}
\and
Dipartimento di Fisica e Astronomia "Augusto Righi" - Alma Mater Studiorum Universit\`a di Bologna, Viale Berti Pichat 6/2, 40127 Bologna, Italy\label{aff107}
\and
Dipartimento di Fisica - Sezione di Astronomia, Universit\`a di Trieste, Via Tiepolo 11, 34131 Trieste, Italy\label{aff108}
\and
Department of Mathematics and Physics E. De Giorgi, University of Salento, Via per Arnesano, CP-I93, 73100, Lecce, Italy\label{aff109}
\and
INFN, Sezione di Lecce, Via per Arnesano, CP-193, 73100, Lecce, Italy\label{aff110}
\and
INAF-Sezione di Lecce, c/o Dipartimento Matematica e Fisica, Via per Arnesano, 73100, Lecce, Italy\label{aff111}
\and
Institut d'Astrophysique de Paris, 98bis Boulevard Arago, 75014, Paris, France\label{aff112}
\and
ICL, Junia, Universit\'e Catholique de Lille, LITL, 59000 Lille, France\label{aff113}
\and
CERCA/ISO, Department of Physics, Case Western Reserve University, 10900 Euclid Avenue, Cleveland, OH 44106, USA\label{aff114}
\and
Laboratoire Univers et Th\'eorie, Observatoire de Paris, Universit\'e PSL, Universit\'e Paris Cit\'e, CNRS, 92190 Meudon, France\label{aff115}
\and
Dipartimento di Fisica "Aldo Pontremoli", Universit\`a degli Studi di Milano, Via Celoria 16, 20133 Milano, Italy\label{aff116}
\and
INFN-Sezione di Milano, Via Celoria 16, 20133 Milano, Italy\label{aff117}
\and
Departamento de F{\'\i}sica Fundamental. Universidad de Salamanca. Plaza de la Merced s/n. 37008 Salamanca, Spain\label{aff118}
\and
Universit\'e de Strasbourg, CNRS, Observatoire astronomique de Strasbourg, UMR 7550, 67000 Strasbourg, France\label{aff119}
\and
Center for Data-Driven Discovery, Kavli IPMU (WPI), UTIAS, The University of Tokyo, Kashiwa, Chiba 277-8583, Japan\label{aff120}
\and
Jodrell Bank Centre for Astrophysics, Department of Physics and Astronomy, University of Manchester, Oxford Road, Manchester M13 9PL, UK\label{aff121}
\and
California Institute of Technology, 1200 E California Blvd, Pasadena, CA 91125, USA\label{aff122}
\and
Department of Physics \& Astronomy, University of California Irvine, Irvine CA 92697, USA\label{aff123}
\and
Departamento F\'isica Aplicada, Universidad Polit\'ecnica de Cartagena, Campus Muralla del Mar, 30202 Cartagena, Murcia, Spain\label{aff124}
\and
Instituto de F\'isica de Cantabria, Edificio Juan Jord\'a, Avenida de los Castros, 39005 Santander, Spain\label{aff125}
\and
Institute of Cosmology and Gravitation, University of Portsmouth, Portsmouth PO1 3FX, UK\label{aff126}
\and
Instituto de Astronomia Teorica y Experimental (IATE-CONICET), Laprida 854, X5000BGR, C\'ordoba, Argentina\label{aff127}
\and
Department of Computer Science, Aalto University, PO Box 15400, Espoo, FI-00 076, Finland\label{aff128}
\and
Universidad de La Laguna, Dpto. Astrof\'\i sica, E-38206 La Laguna, Tenerife, Spain\label{aff129}
\and
Ruhr University Bochum, Faculty of Physics and Astronomy, Astronomical Institute (AIRUB), German Centre for Cosmological Lensing (GCCL), 44780 Bochum, Germany\label{aff130}
\and
Department of Physics and Astronomy, Vesilinnantie 5, University of Turku, 20014 Turku, Finland\label{aff131}
\and
Finnish Centre for Astronomy with ESO (FINCA), Quantum, Vesilinnantie 5, University of Turku, 20014 Turku, Finland\label{aff132}
\and
Serco for European Space Agency (ESA), Camino bajo del Castillo, s/n, Urbanizacion Villafranca del Castillo, Villanueva de la Ca\~nada, 28692 Madrid, Spain\label{aff133}
\and
ARC Centre of Excellence for Dark Matter Particle Physics, Melbourne, Australia\label{aff134}
\and
Centre for Astrophysics \& Supercomputing, Swinburne University of Technology,  Hawthorn, Victoria 3122, Australia\label{aff135}
\and
Department of Physics and Astronomy, University of the Western Cape, Bellville, Cape Town, 7535, South Africa\label{aff136}
\and
Universit\'e Libre de Bruxelles (ULB), Service de Physique Th\'eorique CP225, Boulevard du Triophe, 1050 Bruxelles, Belgium\label{aff137}
\and
DAMTP, Centre for Mathematical Sciences, Wilberforce Road, Cambridge CB3 0WA, UK\label{aff138}
\and
Kavli Institute for Cosmology Cambridge, Madingley Road, Cambridge, CB3 0HA, UK\label{aff139}
\and
Departement of Theoretical Physics, University of Geneva, Switzerland\label{aff140}
\and
Department of Physics, Centre for Extragalactic Astronomy, Durham University, South Road, Durham, DH1 3LE, UK\label{aff141}
\and
IRFU, CEA, Universit\'e Paris-Saclay 91191 Gif-sur-Yvette Cedex, France\label{aff142}
\and
INAF-Osservatorio Astrofisico di Arcetri, Largo E. Fermi 5, 50125, Firenze, Italy\label{aff143}
\and
Centro de Astrof\'{\i}sica da Universidade do Porto, Rua das Estrelas, 4150-762 Porto, Portugal\label{aff144}
\and
Instituto de Astrof\'isica e Ci\^encias do Espa\c{c}o, Universidade do Porto, CAUP, Rua das Estrelas, PT4150-762 Porto, Portugal\label{aff145}
\and
HE Space for European Space Agency (ESA), Camino bajo del Castillo, s/n, Urbanizacion Villafranca del Castillo, Villanueva de la Ca\~nada, 28692 Madrid, Spain\label{aff146}
\and
Department of Astrophysics, University of Zurich, Winterthurerstrasse 190, 8057 Zurich, Switzerland\label{aff147}
\and
INAF - Osservatorio Astronomico d'Abruzzo, Via Maggini, 64100, Teramo, Italy\label{aff148}
\and
Theoretical astrophysics, Department of Physics and Astronomy, Uppsala University, Box 516, 751 37 Uppsala, Sweden\label{aff149}
\and
Mathematical Institute, University of Leiden, Einsteinweg 55, 2333 CA Leiden, The Netherlands\label{aff150}
\and
Institute of Astronomy, University of Cambridge, Madingley Road, Cambridge CB3 0HA, UK\label{aff151}
\and
Center for Astrophysics and Cosmology, University of Nova Gorica, Nova Gorica, Slovenia\label{aff152}
\and
Institute for Particle Physics and Astrophysics, Dept. of Physics, ETH Zurich, Wolfgang-Pauli-Strasse 27, 8093 Zurich, Switzerland\label{aff153}
\and
Department of Astrophysical Sciences, Peyton Hall, Princeton University, Princeton, NJ 08544, USA\label{aff154}
\and
Space physics and astronomy research unit, University of Oulu, Pentti Kaiteran katu 1, FI-90014 Oulu, Finland\label{aff155}
\and
International Centre for Theoretical Physics (ICTP), Strada Costiera 11, 34151 Trieste, Italy\label{aff156}
\and
Center for Computational Astrophysics, Flatiron Institute, 162 5th Avenue, 10010, New York, NY, USA\label{aff157}}  

\title{\Euclid\/ preparation}
\subtitle{CII. Non-Gaussianity of two-point statistics likelihood: Parameter inference with a non-Gaussian likelihood in Fourier and configuration space}   

%\institute{$^{1}$ Institute of Space Sciences (ICE, CSIC), Campus UAB, Carrer de Can Magrans, s/n, 08193 Barcelona, Spain\\
%$^{2}$Institut d’Estudis Espacials de Catalunya (IEEC), Carrer Gran Capitá 2-4, 08034 Barcelona, Spain\\
%$^{3}$  Aix Marseille Univ, Universit\'e de Toulon, CNRS, CPT, Marseille, France\\
%$^{4}$ Aix Marseille Universit\'e, CNRS/IN2P3, CPPM, IPhU, Marseille, France\\
%$^{5}$ Istituto di Astrofisica Spaziale e Fisica cosmica Milano, Via A. %Corti 12, I-20133 Milano, Italy \\
%$^{6}$  INFN Sezione di Milano, Via G. Celoria 16, I-20133, Milano, Italy\\}

% 
% Put your abstract here:
%
\abstract{
%We assess the impact of non-Gaussian features in the distribution of two-point statistics on cosmological inferencing in the context of \Euclid. Using the \covmos~method, we generated $10~000$ realisations of the nonlinear matter field across four survey geometries and measured both the power spectrum and two-point correlation function. We validated the finding reported in our 2025 companion paper, confirming that the skewness of these statistics increases as the survey volume decreases.

The extraction of cosmological information from two-point statistics critically relies on the assumed form of their likelihood. Although a Gaussian likelihood is generally adopted, the first paper of this series showed that the distribution of power-spectrum estimates exhibits non-Gaussian features on both large and small scales ($k<0.5\, \invMpc$), with their amplitudes depending on the survey volume, masking, and shot noise. In this work, we account for this skewness in parameter inference by modelling the likelihood through an Edgeworth expansion. This procedure involves the complete skewness tensor, composed of one-point, two-point, and three-point correlators. To simplify the calculations of this expansion, we performed a change in the basis, which reduced the precision matrix to the identity. In this basis, the off-diagonal elements of the skewness tensor are consistent with zero, while the amplitude of its diagonal match the level expected for a Gaussian underlying field. We performed a parameter inference with this likelihood model and found that including only the diagonal part of the skewness is sufficient; whereas incorporating the full skewness tensor injects noise without improving accuracy. Despite the estimated excess skewness in the original basis, the cosmological constraints remain effectively unchanged when adopting a Gaussian likelihood or considering the more complete Edgeworth expansion, with variations in the figure of merit of cosmological parameters between the two cases below $5\%$. This result remains unchanged against variations of the survey volume and geometry, scale-cut, as well as the two-point statistics (i.e. power spectrum or correlation function). Using $10\, 000$ cloned \Euclid large mocks based on realistic galaxy catalogues with characteristics approximating future \Euclid data, we found no detectable excess skewness on intermediate scales, due to the level of shot noise expected for the \Euclid spectroscopic sample. We conclude that the Gaussian likelihood assumption is robust for \Euclid two-point statistics analyses in both Fourier and the configuration space.}

%
% Provide up to five key words:
%
    \keywords{Non-Gaussian likelihood -- Euclid -- Large-scale structures -- two-point statistics}
%    from the list in
%     https://www.aanda.org/for-authors/latex-issues/information-files#pop}
%
% Add short versions of title and author list for page headings
%
   
\titlerunning{Non-Gaussianity of the two-point statistics likelihood}
  \authorrunning{Euclid Collaboration: S.~Gouyou~Beauchamps et al.}
   
   \maketitle

%
%-------------------------------------------------------------------
%
%
%   Start the main text of your paper here
%

\section{Introduction}

The primary summary statistic, used by modern cosmological analyses to measure cosmological parameters, is the two-point statistics of the specific observed fields. Such statistics have been (and will continue to be) used to analyse the temperature and polarisation of the cosmic microwave background (CMB) observed by \Planck~\citep{planck_20}, the South Pole Telescope~\citep[SPT,][]{SPT_22}, or the Atacama Cosmology Telescope~\citep[ACT,][]{ACT_spectra_DR6}, as well as studies of the clustering of galaxies in three dimensions (3D), for example, with Dark Energy Spectroscopic Instrument~\citep[DESI,][]{aghamousa_16}, or galaxy weak lensing the {\it Vera C. Rubin} Observatory Legacy Survey of Space and Time~\citep[LSST,][]{ivezic2018lsst} is soon scheduled to observe. The primary probes of \Euclid, 3D clustering and a combination of weak lensing and photometric clustering (i.e. the so-called $3\times2$-point analysis), are also based on two-point statistics~\citep{EuclidSkyOverview}.

The extraction of cosmological information from two-point statistics is generally performed through a likelihood-based analysis, where a key assumption has to be made on the form of the likelihood function. The most common assumption is to assume the distribution of two-point statistics to be Gaussian. However, it is known that this distribution presents deviations from Gaussianity, especially at large scales, where the number of modes or pairs is too low to approximate the distribution to be Gaussian, {although they are actually $\chi^2$-distributed~\citep{hamimeche_08, takahashi_09}. We studied this distribution in detail, in our companion paper~\citep[][hereafter B-2025]{EP-Bel}, where we focussed on the matter power spectrum. We observed, as expected, deviations from a Gaussian distribution at large scales, but also at intermediate scales~\citep[already observed in][]{blot_15}. We could link the latter to the higher order $N$-point statistics of the density field itself. In addition, we explored a variety of setups to understand the dominant effects are on the non-Gaussianity of the power spectrum distributions. We found that deviations from a Gaussian distribution at both large and intermediate scales were greatly enhanced when a survey window function was applied. In the present work, we explore whether if these deviations from Gaussianity could have an impact on cosmological constraints.

Several works in the literature have studied the limits of the Gaussian likelihood assumption. The first to do so were carried out in the context of CMB analyses, finding that incorrectly assuming a Gaussian likelihood on large scales could cause significant biases in the estimation of cosmological parameters~\citep{chu_05, smith_06, rocha_10}. Thus, in the past two decades, CMB analyses have used different types of non-Gaussian likelihood on both large and small scales~\citep{planck_20_like}. On the side of weak lensing and galaxy clustering, some works have shown that a Gaussian likelihood assumption is sufficient~\citep{hartlap_2009, wilking_2012, simon_2015, Taylor_19, DES_cov, upham_21, hall_22, Oehl_2025} or that it only has a mild impact on parameter inference~\citep{hahn_18}. 

In B-2025, we showed that the amount of non-Gaussianity in the power spectrum distribution depends on many factors, such as the survey volume and the density of the galaxy sample; thus, it will vary from one survey to another. In this work, we want to gauge whether the Gaussian likelihood assumption can significantly bias cosmological constraints, depending on the level of non-Gaussianity. We focus on the volume and shape of the survey, in particular, considering different survey window functions with variations in terms of size and geometry. We then estimate the level of non-Gaussianity expected in the distribution of two-point statistics for a realistic \Euclid spectroscopic sample. This allows us to draw a conclusion on whether the Gaussian likelihood assumption is sufficient for the analysis of the two-point statistics of the \Euclid spectroscopic data.

To test the impact of the Gaussian likelihood assumption, we compare the result of a parameter inference using a Gaussian or a non-Gaussian likelihood in a realistic case where the data have a non-Gaussian distribution. To accurately model the non-Gaussian distribution of the data, we developed a likelihood model based on an Edgeworth expansion as presented in~\citet{Amendola96}. It provides a flexible way to incorporate non-Gaussian features to the likelihood without making any assumptions on the underlying field. Following B-2025, we now focus on the skewness of the distribution, as we have seen it is the dominant non-Gaussian feature. To simplify the computation of the correction terms, we rely on the properties of the Edgeworth expansion, which allow us to perform the likelihood analysis in a basis where the data are uncorrelated. This is equivalent to a principal component analysis, as in~\citet{lin_20}.

As in B-2025, we rely extensively  on the \covmos~method~\citep{baratta_19, baratta_22} for the fast production of a large set of mock catalogues of the nonlinear dark matter field. Here, we have produced $10\,000$ of these simulations and cut different volumes from their periodic boxes to mimick a survey window function. We also use \covmos~to clone more realistic galaxy catalogues specifically tailored to simulate the \Euclid spectroscopic sample, the \Euclid large mocks~\citep[ELM,][]{EP-Monaco1}, based on \texttt{PINOCCHIO}~\citep{monaco_13}. We are confident in our use of \covmos~simulations for this type of analysis, based on our findings in B-2025, where we demonstrated that \covmos~is able to accurately model the skewness of the power spectrum distribution, up to $k=1.3\, \invMpc$.

In B-2025, our study was focussed on Fourier space, whereas in the present work, we also extend our analysis to the configuration space. Since both the power spectrum and the two-point correlation function (2PCF) will be used for \Euclid data analyses, we want to gauge the validity of the Gaussian likelihood approximation for both statistics. In addition, this allowed us to carry out a consistency check between the cosmological constraints obtained from the two statistics.

In Sect.~\ref{sect:simu}, we present the \texttt{COVMOS} method and the different samples of simulations we produced for this study, as well as the two-point statistics estimators. In Sect.~\ref{sect:measurements}, we present the modelling of the two-point statistics affected by a survey window function and the skewness estimated from the simulations in each geometry. In Sect.~\ref{sect:nglik}, we present the non-Gaussian likelihood model that we used for parameter inference, which is based on an Edgeworth expansion together with the effect of a proper change in the basis on the skewness of the power spectrum and the 2PCF. The result of the parameter inference analysis is presented in Sect.~\ref{sect:param}. In Sect.~\ref{sect:pinoc} we study the skewness of the power spectrum for a more realistic case, with \Euclid-like simulations containing galaxies rather than dark matter only. Finally, we present our conclusions in Sect.~\ref{sect:conclu}.

\section{Simulations and estimators}
\label{sect:simu}
In this section, we present the \covmos\ set of realisations that we generated together with the various geometries we have defined. In addition, we describe both the Fourier space and configuration space two-point statistics estimators we employed.

\subsection{\covmos~simulations with varying geometries}

\covmos\footnote{  \href{https://github.com/PhilippeBaratta/COVMOS}{github.com/PhilippeBaratta/COVMOS}}~is a public code first developed to generate fast approximate simulations to accurately model the covariance of two-point statistics~\citep{baratta_19, baratta_22, gouyou_25}. We have also shown in B-2025 that \covmos~is well suited to reproduce the distribution of dark-matter two-point statistics in Fourier space and especially its skewness on a large range of scales ($k \leq 1.3\, \invMpc$). As a result, in the present work, we chose to produce \covmos\ realisations to generate a synthetic two-point statistics dataset to accurately estimate its covariance matrix and its skewness.

Similarly to commonly used log-normal mock catalogue methods~\citep{Xavier_16, Alonso_14, Agrawal_17}, \covmos~takes a target power spectrum as its input. But the main difference with log-normal methods is that it also takes as input a target density one-point probability density function (PDF), which is arbitrary. 

As in B-2025, the target one-point PDF was estimated from the 50 $\Lambda$CDM \texttt{DEMNUni-Cov} simulations~\citep{carbone_16, parimbelli_21, baratta_22, EP-Ingoglia, gouyou_25}. The cosmological parameters of these simulations were taken from \Planck 2013~\citep{Planck_13}, namely, $(\Omega_\mr{m}, \Omega_\mr{b}, h, n_\mr{s}, A_\mr{s}) = (0.32, 0.05, 0.67, 0.96, 2.1265\times 10^{-9})$. However, in the present work, we required the target power spectrum to comprise the non-linear matter power spectrum predicted with \texttt{Halofit}~\citep{smith_02, takahashi_12} in the corresponding cosmology. This choice ensures that the model employed when fitting the two-point statistics accurately describes our synthetic dataset. More details are given in Sect.~\ref{sect:model}.

Given these target parameters, we generated a sample of $10\, 000$ dark matter catalogues in real space, with periodic boundary conditions within a comoving volume of $L_{\rm box}^3=1\, (\mathrm{Gpc}/h)^3$.  Since we went on to use them to study cosmological parameter inference, we decided to generate two sets at two different redshifts $z=0$ and $z=0.5$. This allowsed us to maximise the skewness of the distribution of the power spectrum estimator (see B-2025). Each catalogue contains $2\times10^7$ particles, giving a number density of $0.02\, (\mathrm{Mpc}/h)^{-3}$. Setting this process in real space allowed us to maximise the skewness (see B-2025) of the two-point statistics distribution and makes the likelihood modelling easier.

In B-2025, we found that the skewness was greatly enhanced when applying a survey window function; namely, when going from a periodic box to a more realistic case of a non-periodic volume. Thus, it is interesting to study how different levels of skewness in the two-point statistics estimators propagate to the parameter inference by employing different geometries.

To do so, we simply cut different volumes from the \covmos~periodic boxes. We considered two spheres of radius $R=0.5\ \Gpc$ and $R=0.2\ \Gpc$, respectively, and a cone with a half-aperture angle $\theta=15\degree$ and radial extension $(R_\mathrm{min},\ R_\mathrm{max}) = (1.6,\ 1.9)\ \Gpc$ (i.e. with the observer outside the periodic box). These are named S500, S200, and CONE, respectively, and their corresponding volumes are reported in Table~\ref{tab:geometries}. The S500 geometry corresponds to a volume of about one-third of the expected first spectroscopic bin of the first \Euclid data release (DR1). The periodic box is referred to as BOX and used as the reference.

\begin{table*}
\centering
\caption{Summary of all the geometries considered in this work, available for the two redshifts: $z=0$ and $z=0.5$.}
\begin{tabular}{l|c|c}
  Name & Geometry (lengths in $\mathrm{Gpc}/h$) & Volume in $[\mathrm{Gpc}/h]^3$ \\
  \hline
  \hline

  BOX & Box with $L=1$ & $1$ \\
  \hline
  S500 & Sphere with $R=0.5$ & $0.5$ \\
  \hline
  S200 & Sphere with $R=0.2$ & $0.03$ \\
  \hline
  CONE & Cone with $(\theta,\ R_\mathrm{min},\ R_\mathrm{max}) = (15\degree,\ 1.6,\ 1.9)$ & $0.2$ \\
\end{tabular}
\label{tab:geometries}
\end{table*}

We note that we sampled particles lying in these different geometries from the comoving outputs at fixed redshift. Thus, there is no redshift evolution in the spheres or in the cone. In addition, the two comoving outputs at redshifts $z=0$ and $z=0.5$ were generated independently, so there is no correlation between them.

\subsection{Estimators}\label{sect:2pt_estim}

In this section, we provide the details of how we estimated the two-point statistics both in Fourier and configuration space from our sample of \covmos\ realisations. While the case of the periodic box is straightforward, in the three other cases (S500, S200, and CONE) it is necessary to take into consideration the corresponding window functions. We note that since we decided to perform the analysis without redshift space distortions (RSD), we only considered the monopole of the power spectrum and 2PCF.

\subsubsection{Fourier space}\label{sect:2pt_estim_pk}

Throughout this work, we adopted the following convention for the spatial Cartesian Fourier transform (FT) $f_{\vec k}$ of a function, $f(\vec x)$,
\be
f_{\vec k} := \dfrac{1}{(2\pi)^3} \int f(\vec x) \, {\rm e}^{-\mathrm{i}\vec k \cdot \vec x}\ \dif ^3 x\; .
\ee
Since FTs are computed using the fast Fourier transform (FFT) algorithm, the Fourier space is assumed to be discrete and regularly sampled on a 3D grid, which corresponds to the fundamental frequency, $k_{\rm F}:= 2\pi/L$, associated with the bounding box of size $L=1000\, \invMpc$. By defining the density contrast field as $\delta(\vec x) := n(\vec x)/\bar n - 1$, with $n(\vec x)$ the number density at spatial position $\vec x$ and $\bar n$ the mean number density of the Universe, the estimator of the monopole power spectrum can be written as the averaged power in spherical Fourier shells of width $k_{\rm F}$,
\be\label{eq:pk_estim}
\hat P(k) := \frac{ \tilde k_{\rm F}^3}{N_k}\sum_{i=1}^{N_k} |\delta_{\vec k_i}|^2 \; ,
\ee
where $N_k$ is the number of independent modes in each $k$-shell, $\delta_{\vec{k}_i}$  is the FT of the density contrast, and $\tilde k_{\rm F}^3 := (2\pi)^3/V$ is the effective fundamental mode associated with the volume of the survey. Note that we do not consider the weights introduced by \citet{fkp_1994} in this work because the mean number density does not vary across the volume.

Without any loss of generality, we can consider the survey window function $W(\vec{x})$ which is $1$ within the survey and $0$ outside. As a result, the observed density field $n^{\rm o}(\vec x)$ is related to the true density field $n(\vec{x})$ via
\be\label{eq:rho_obs}
 n^{\mathrm{o}} (\vec{x}) = n(\vec{x})\, W(\vec{x}) \; .
\ee
By defining the observed density contrast field, 
\begin{equation}
\delta^{\mathrm{o}}(\vec{x}) :=  n^{\mathrm{o}}(\vec{x})/\bar n_\mathrm{V}-1\;,
\label{eq:d_obs}
\end{equation}
with the mean density, $\bar n_\mathrm{V}, = V^{-1}\int n^{\rm o}(\vec x)\;\dif^3 x$, and taking its FT, we get
\be\label{eq:conv_delta}
\delta^{\mathrm{o}}_{\vec k}= \tilde{\delta}_{\kk} + W_{\vec k} \; ,
\ee
where $\tilde{\delta}_{\kk}$ is the convolution of the true density contrast in Fourier space with the FT of the window function, $W_{\kk}$,
\be\label{eq:obs_field}
\Tilde{\delta}_{\kk} := \int \delta_{\kk'} W_{\kk-\kk'}\; \mr{d}^3 k' \; .
\ee

In Fourier space, there are two different effects introduced by the survey window function: (i) the observed density contrast is biased by $W_{\kk}$ (see Eq.~\ref{eq:conv_delta}) and (ii) the density contrast field is convolved with the survey window function (see Eq.~\ref{eq:obs_field}). From the observed contrast density field, it is possible to remove the bias if $W_{\kk}$ is known. However, it is not trivial to deconvolve the survey window function. It is common practice~\citep{peacock_1991} 
to define the convolved power spectrum $\tilde P_{\kk} := \tilde k_{\rm F}^3\,\langle |\tilde\delta_{\kk}|^2 \rangle$, which we can express as 
\be\label{eq:pk_conv}
\Tilde{P}_{\kk} = \int P_{\kk'} |W_{\kk-\kk'}|^2\ \mr{d}^3k' \; ,
\ee
where $P_{\kk}$ is the true three-dimensional (3D) underlying power spectrum. In that case, if we want to compare this estimate to a model, it is needed to forward model the effect of the convolution to the theory power spectrum.

From the above equation, it is clear that we need to know the FT of the survey window functions we consider. The case of the spherical window function centred at the origin is particularly interesting because the expression of its FT is trivial  
\be\label{eq:wsphere_fourier}
W^\mr{S}_{\kk} = 4\pi R^3 j_1(kR)/(kR) \; ,
\ee
where $R$ is the radius of the sphere and $j_1$ the spherical Bessel function of the order of $1$. For the cone geometry, the formal expression of $W_{\vec k}$ takes the form of an infinite series. As a result, in the same way as B-2025, it is more efficient to evaluate it from a random uniform distribution which has 20 times the density of the \covmos~catalogues. We also checked that it is equivalent and more efficient to directly assign this window function to a real space 3D grid and take its FFT.

For the estimation of the power spectra, we used the \texttt{NBodyKit}\footnote{\href{https://nbodykit.readthedocs.io/en/latest/}{nbodykit.readthedocs.io}} software~\citep{nbodykit}. The 3D grid has a sampling parameter of $N_\mr{s} = 512$, we use a piecewise cubic spline (PCS) mass assignment scheme and we apply the interlacing method~\citep{sefusatti_15} to reduce the effect of aliasing on small scales. 

\subsubsection{Configuration space}\label{sect:2pt_estim_xi}

The estimation of the multipoles of the 2PCF is critical for this work. The most commonly used estimator is the Landy \& Szalay~\citep{landy_szalay_93}, defined as

\begin{equation}
\hat \xi (r,\mu) =  \frac{N_{\rm R}(N_{\rm R}-1)}{N_{\rm D}(N_{\rm D}-1)}\frac{{\rm DD} (r, \mu)}{{\rm RR}(r,\mu)} - 2 \frac{(N_{\rm R}-1)}{N_{\rm D}}\frac{{\rm DR}(r, \mu)}{{\rm RR}(r,\mu)} + 1 \; ,
\label{LS}
\end{equation}
where we need to specify a binning in $r$ (i.e. the separation between two galaxies) and in $\mu = \cos{\theta}$ (being $\theta$ the angle between the separation vector and a reference direction). The estimator above uses a data catalogue containing $N_{\rm D}$ objects for which we wants to estimate the 2PCF and a random catalogue of $N_{\rm R}$ objects following the geometry of the data catalogue. The term ${\rm DD}(r,\mu)$ represents the number of pairs of objects in the data catalogue inside the bin centred on $r$ and $\mu$, ${\rm RR}(r, \mu)$ represents the same but for a random catalogue following the geometry of the data catalogue, and ${\rm DR}(r,\mu)$ is the number of cross-pairs between the two catalogues. The usual rule is to use a random catalogue which contains at least $50$ times more objects than the data catalogue, to avoid being sensitive to shot noise. In our \covmos~catalogues, we have a number density of $0.02\,(\mathrm{Mpc}/h)^{-3}$ in a box of $1\, (\mathrm{Gpc}/h)^{3}$. This means that we would need $10^9$ random points. It has been shown that it is possible to resort to the splitting method proposed by~\citet{keihanen_19} to significantly reduce the computation time of both the cross-term ${\rm DR}(r,\mu)$ and the auto-term ${\rm RR}(r,\mu)$ with a minimal loss in accuracy. However, for our purpose, this would still induce a prohibitive computational cost. Indeed, we need to repeat this measurement on $10\, 000$ realisations for each of the four geometries at the two considered redshifts ($z=0$ and $z=0.5$) for a total of $80\, 000$ evaluations. 

To tackle this issue we rely on the fact that we have rather simple geometries (sphere or cone) and we designed an adapted estimator. In Appendix~\ref{appendix:2pcf}, we review the key steps that allowed us to show that the estimator we chose to use is equivalent to the Landy \& Szalay estimator; however, it also implies that we only need to compute the pairs in the data catalogue. Thus, we can compute the cross-pairs between data and random catalogue with a single sum over the data catalogue itself. The gain in time is tremendous because we do not need to resort to a random catalogue at all. 

Based on Eq.~\eqref{eq:pk_conv}, we can show that the 2PCF should, in principle, be affected by the presence of the window function via
\begin{equation}
\tilde \xi(\vec r) = \xi(\vec r) \, g(\vec r) \; , 
\end{equation}
where 
\begin{equation}\label{eq:win_func}
g(\vec r) := \frac{1}{V}\int |W_{\vec k}|^2 \, {\rm e}^{\mathrm{i} \vec r\cdot \vec k}\,\dif^3 k \;.
\end{equation}
The function $g(\vec r)$ is basically the 2PCF induced by the survey geometry. This is exactly what the $\rm{RR}$ term is correcting for in the Landy \& Szalay estimator (see Eq.~\ref{LS}). Further discussion on this aspect can be found in Appendix~\ref{appendix:2pcf}. As a result, contrary to the power spectrum estimator, in the case of the 2PCF the survey geometry does not affect directly the measurement.

\section{Two-point statistics with a survey window function}
\label{sect:measurements}

\subsection{Modelling}
\label{sect:model}

In this section, we describe how we modelled the measured two-point statistics, which is relevant for the fitting procedure presented in Sect.~\ref{sect:param}. This procedure allowed us to ensure that no bias originating from the modelling would affect the parameter inference.
Since we chose \halofit\ as our input power spectrum for the \covmos\ realisations, in this step, we do not aim to validate the model itself;  rather, we want to check that we are able to handle correctly the forward modelling of the effect of the window function.

\subsubsection{Fourier space}\label{sect:model_pk}

For the power spectrum, there are two main effects that need to be taken into account. First, there is the fact that in non-periodic geometries, the estimated power spectrum is not the true one, but it is the window-convolved power spectrum in Eq.~\eqref{eq:pk_conv}. Second, the high-$k$ filter inherent to the \covmos~method can be exactly predicted.

The high-$k$ filtering of the power spectrum is a step that is needed for \covmos~to match the variance of the density field computed directly from the one-point PDF and the one estimated through the integration of the power spectrum itself. This filter can be exactly predicted and applied to the model power spectrum. We do not provide any further details here and refer instead to~\citet{baratta_19, baratta_22} for more.

The most simple way to convolve the model $P(k)$ with the window function is to make use of the convolution theorem, stating that the FT of a product of two functions is the convolution of the FTs of these two same functions. We can then take the Hankel transform of the predicted power spectrum to get the 2PCF, then multiply this predicted monopole 2PCF by the window function monopole (the monopole of Eq.~\ref{eq:win_func}), $g^{(0)}(r)$, and transform it back to Fourier space. We perform the Hankel transform with the public python package \texttt{mcfit}.\footnote{\url{https://github.com/eelregit/mcfit}} After some optimisation tests to obtain a fast and precise Hankel transform, we decided to have the integral performed on $2^{12}$ values of $k$ logarithmically spaced in the range $[10^{-3.5}, \,10^{3}]\, \invMpc$. 

In general, the convolution of the power spectrum with the window function involves a sum over the even multipoles of the 2PCF and those of the window function $g(\vec r)$, as described in~\citet{wilson_17}. However, as we are working in real space, only the monopole of the 2PCF (and power spectrum) is non-zero. Formally, the model-convolved power spectrum becomes 
\be
\tilde{P}^{(0)}(k) = \mathcal{H}^{-1}\left\{  \mathcal{H}\left[P(k)\right]\; g^{(0)}(r)  \right\}\; ,
\ee
where $\mathcal{H}$ is a Hankel transform, and $P(k)$ is the \texttt{Halofit} power spectrum with the \covmos~filter applied on it.

We then need to know the window correlation function $g^{(0)}(r)$, which, for the sphere, can be derived analytically from Eq.~\eqref{eq:wsphere_fourier}  (see Appendix~\ref{appendix:geom}) and it is expressed as
\be
g^{(0)}(r) = 1 - \dfrac{3}{4}\dfrac{r}{R} + \dfrac{1}{2}\left (\dfrac{r}{2R}\right )^3 \; .
\ee
We note that for a spherical window, $g(\vec r) = g^{(0)}(r)$. Indeed it only depends on the radial distance, $r$, as we are dealing with an isotropic geometry.

For a conoidal window, it is not possible to get a fully analytical form of $g(\vec r)$ and we refer to Appendix~\ref{appendix:geom} for more details on our procedure. Overall, this approach allowed us to obtain the monopole of the window correlation function (along with the other multipoles for completeness).

Finally, once the theoretical prediction was convolved with the survey window function, we needed to subtract the integral constraint (IC) contribution. This contribution comes from the fact that we estimate the mean number density from the limited surveyed volume and not from the whole Universe where $\langle \delta(\vec{x}) \rangle = 0$. For a surveyed volume, this is not true since $\delta^\mathrm{o}$ is a sub-sample of the true density field. This gives rise to an additional term in the convolved theoretical power spectrum which dominates at low $k$. The correction can be approximated as~\citep{wilson_17, demattia_19} 
\be\label{eq:IC}
\tilde{P}^\mr{IC}_{\vec k} = \tilde{P}_{\vec k} - \tilde{P}_{\vec 0} \dfrac{|W_{\vec{k}}|^2}{|W_{\vec{0}}|^2} \; ,
\ee
where $\tilde{P}^\mr{IC}_{\vec k}$ is the model that we use hereafter. More precisely, we used the monopole $\tilde P^{\rm IC}(k)$ of Eq.~\eqref{eq:IC}. 

\begin{figure}
\includegraphics[width=\linewidth]{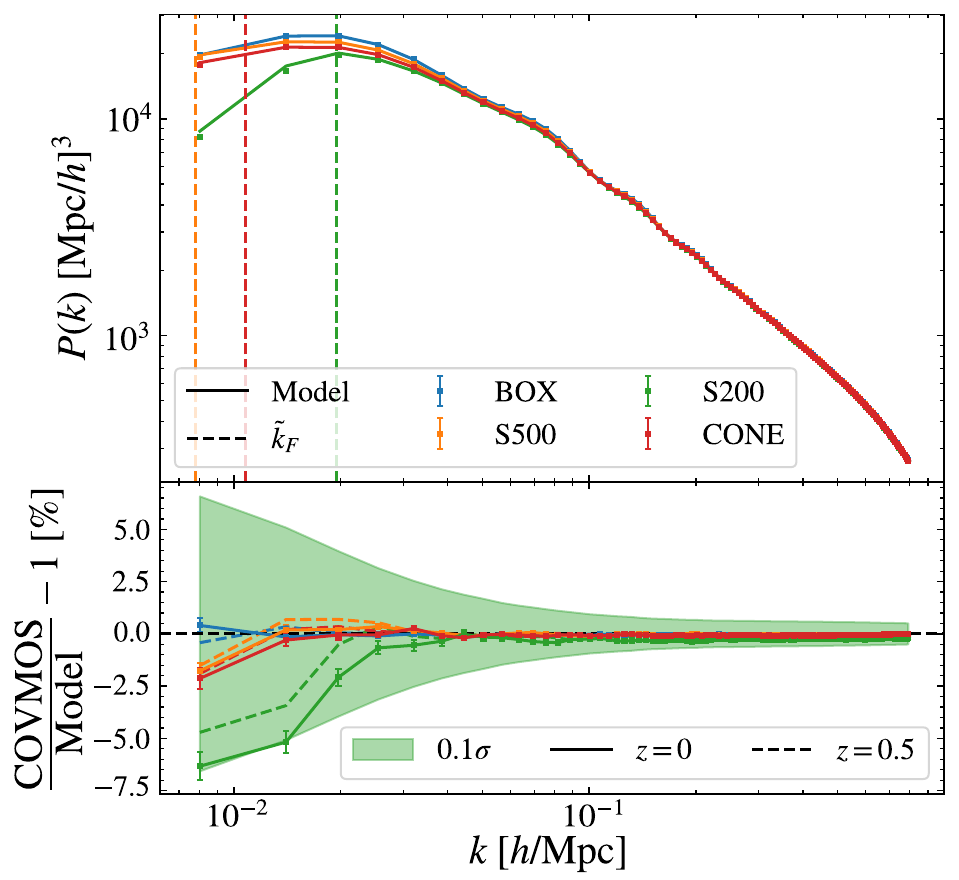}
\caption{Validation of the forward modelling of the window function on the power spectrum for all geometries. {\it Top}: Mean of the power spectra estimated from the $10\, 000$ \covmos~realisations for each geometry (squares with error bars) and the model convolved with the respective window function (lines). The vertical dashed lines represent the effective fundamental mode for each geometry. {\it Bottom}: Relative difference between the mean \covmos~power spectra and the convolved models for $z=0$ (solid lines) and $z=0.5$ (dashed lines). The error bars represent the error on the mean over the $10\, 000$ realisations (we only show it for $z=0$ to improve readability). The green area represents $10$\% of the error on a single realisation for S200 (we only show it for this case).}
\label{fig:geometries_pk}
\end{figure}

In Fig.~\ref{fig:geometries_pk}, we compare the mean \covmos~power spectrum for the different geometries compared to the model described above. For all non-periodic geometries and all redshifts, we see larger than $1$\% deviations for $k < \tilde k_{\rm F}$. This is not worrying, as this corresponds to scales larger than or similar to the size of the different survey window functions. We can also see a constant 0.25\% shift for the S200 geometry. Interestingly, this corresponds to the smallest volume. This might come from higher-order IC correction terms that are not taken into account in our modelling. Overall, we see that for both redshifts the relative difference in the mean over all realisations is within the $0.1\,\sigma$ region of a single realisation. We do not expect such small differences to affect parameter inference.

\subsubsection{Configuration space}\label{sect:model_xi}

In the configuration space, the treatment of the survey window function is easier as the density field is simply multiplied by the window function
(cf. Eq.~\ref{eq:rho_obs}). This contribution is directly divided out in the Landy \& Szalay estimator. The IC is a constant which is also directly subtracted from the estimation. Thus, we simply take the Hankel transform of the model power spectrum to get the model 2PCF. 

In Fig.~\ref{fig:geometries_xi}, we compare our model to the average 2PCF from the $N_{\rm s}=10\, 000$ \covmos~realisations for the different geometries. In particular, in the lower panel of Fig.~\ref{fig:geometries_xi}, we can see oscillations for scales smaller than $r = 20\, \Mpc$. This might be due to the integration in the Hankel transform, which we optimised to be fast so that it could carry out a larger number of evaluations for a Markov chain Monte Carlo (MCMC) run. As for the power spectrum described in the previous section, these are small effects that are visible when the average is done over such a large number of realisations. In Sect.~\ref{sect:param}, we  assume a minimum scale cut of $r_\mathrm{min} = 20\, \Mpc$ for the parameter inference; above this value, the mean is well within 10\% of the error on a single realisation for all geometries.

\begin{figure}
\includegraphics[width=\linewidth]{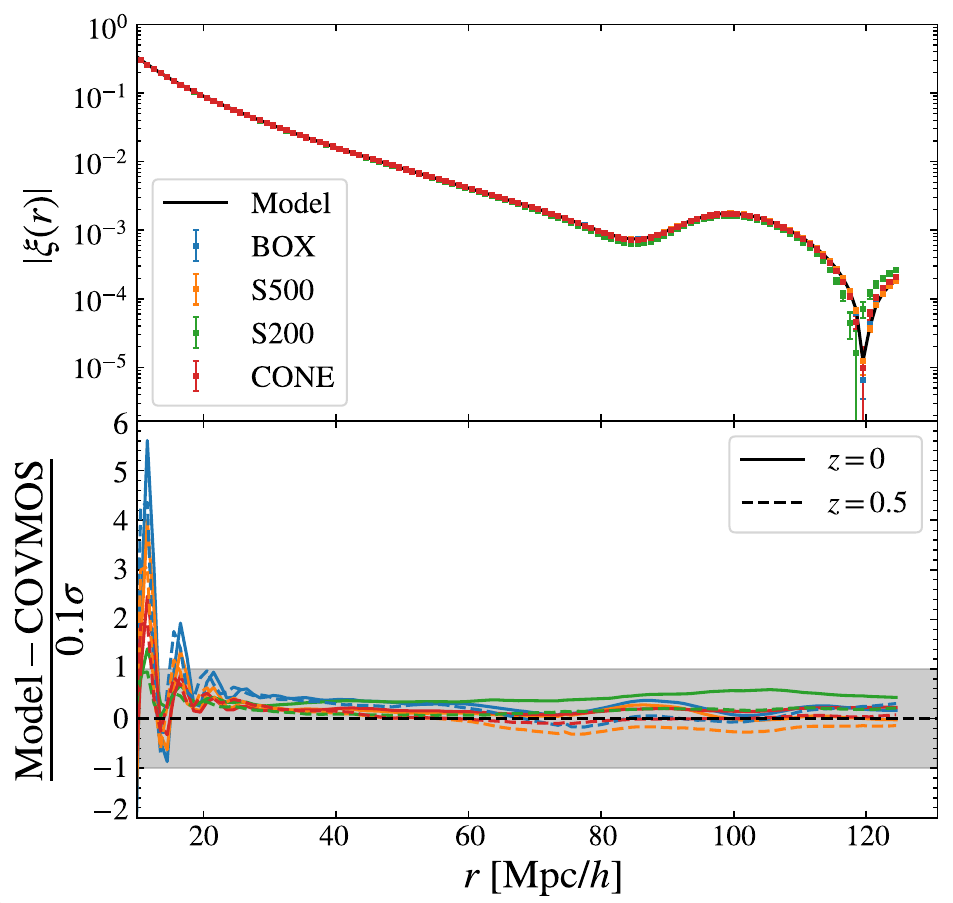}
\caption{Validation of the forward modelling of the window function on the 2PCF for all geometries. {\it Top}: Mean over the 2PCFs estimated from the $10\, 000$ \covmos~realisations for each geometry (squares with error bars) and the model (line), in absolute value. The error bars represent the error on the mean over the $10 000$ realisations.  {\it Bottom}: Residual in number of $\sigma$ considering $10$\% of the error on one realisation for each geometry.}
\label{fig:geometries_xi}
\end{figure}

\subsection{Skewness of the distribution of two-point statistics}\label{sect:2pt_skew}

In this section, we present the estimation of the skewness of both the power spectrum and 2PCF distributions from the $10\, 000$ \covmos\ realisations. We focus on the skewness here since, as we show in B-2025, this is the dominant contribution to the non-Gaussianity of the power spectrum distribution. Given the $n$-th realisation of the data vector $\vec X_n$, which can be either $P(k)$ or $\xi(r)$, and defining $X_n^{(i)}$ its $i$-th element, we estimate the reduced skewness $\hat S_3^i$ as 

\begin{equation}
    \hat S_3^i =   \frac{N_{\rm s}^{-1}\sum_{n=1}^{N_{\rm s}} \left[X_n^{(i)} - \bar X_i\right]^3}{\hat\sigma_i^3}\;,
\label{s3estimator}
\end{equation}
where $\bar X_i$ is the estimated average of $X^{(i)}$ over the $N_{\rm s}$ \covmos~realisations,

\begin{equation}
\bar X_i = \frac{1}{N_{\rm s}} \sum_{n=1}^{N_{\rm s}} X_n^{(i)}\;,
\label{mean}
\end{equation}
and its estimated standard deviation is

\begin{equation}
\hat\sigma_i = \sqrt{ \frac{1}{N_{\rm s}}\sum_{n=1}^{N_{\rm s}}\left [ X_n^{i} - \bar X_i \right]^2 }\; .
\label{std_dev}
\end{equation}

For the power spectrum, we want to compare this skewness estimation to its prediction in the case of a Gaussian field: $S_{3,\mr{G}} = 2/\sqrt{N_k}$ (see B-2025), where $N_k$ is the number of independent modes in the Fourier shell centred on the mode $k$. This number can be counted directly from the FFT grid or can be predicted as $N_k = V_k\,\tilde k_{\rm F}^{-3}/2$, where $V_k = 4\pi k^2 \tilde k_{\rm F}$ is the volume of the Fourier shells. 
In addition, in B-2025, we empirically found that $S_{3,\mr{G}} = 2.2/\sqrt{N_k}$ provides a better fit for non-periodic geometries and, therefore, we chose to consider the latter as the reference prediction in non-periodic geometries.

In Fig.~\ref{fig:skew_geom_pk}, we show the estimated skewness at $z=0$ of the power spectrum for the four geometries considered in this work compared to $S_{3,\mr{G}}$. On large scales, we can see that the skewness consistently rises both in the estimation and the prediction when decreasing the volume. This is consistent with a decrease of independent number of modes for a decreasing volume and is reflected in the $\tilde k_{\rm F}^{-2}=V^{-2/3}$ scaling of $N_k$. In addition, we also see an increase in the excess of skewness on small scales, shown in the subpanel of Fig.~\ref{fig:skew_geom_pk} by the difference between the estimated and predicted skewness. The latter effect was observed in other studies~\citep{upham_21} and this is due to the mixing of modes caused by the presence of a mask. In addition, in B-2025, we show that the IC is also partly responsible for this increase in skewness on small scales. Similar observations apply at $z=0.5$ with a lower excess of skewness on the same scales, as observed in B-2025.

\begin{figure}
\includegraphics[width=\linewidth]{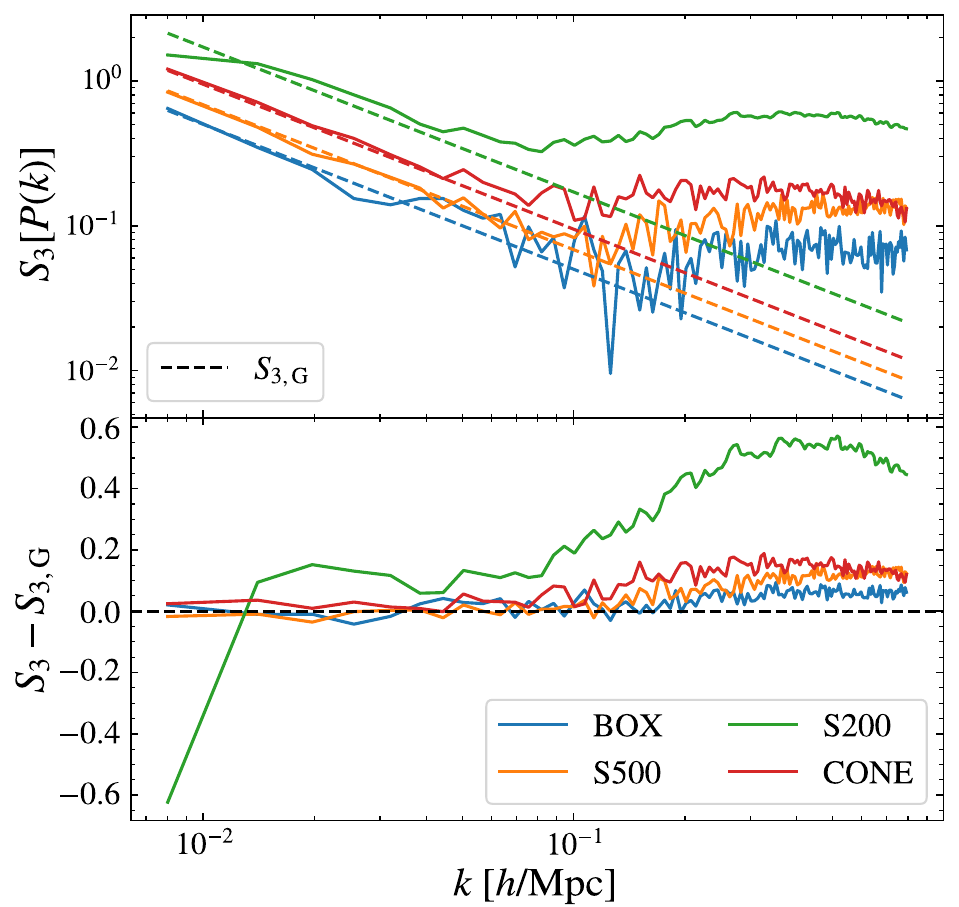}
\caption{Comparison of the estimated skewness and the expectation in the case of a Gaussian field. {\it Top}: Skewness of the power spectrum distribution estimated on the $10\, 000$ \covmos~realisations for each of the four geometries at $z=0$. The dashed lines show the skewness predicted in the case of a Gaussian field ; the BOX geometry uses the factor $2$ and the others the factor $2.2$ described in Sect.~\ref{sect:2pt_skew} {\it Bottom}: Difference between the estimated skewness and the prediction for a Gaussian field.}
\label{fig:skew_geom_pk}
\end{figure}

In Fig.~\ref{fig:skew_geom_xi}, we show the estimated skewness of the 2PCF for the the four geometries at $z=0$. However, in this case, we do not have a prediction to compare with. Although it is not straightforward to relate comoving scales $r$ to Fourier modes $k$, we can try to compare the different features we observe in configuration space to what we have learned in Fourier space to understand the behaviour of the 2PCF skewness with respect to scales and survey geometry. 

The largest pair separation for which we estimated the 2PCF ($r = 125\, \Mpc$) is much smaller than the scale associated to the smallest $k$ (corresponding to $k_{\rm min} = 2 \pi/L$) for which we estimated the power spectrum. As a consequence, we do not see the expected large-scale skewness. Indeed, it should arise from the low number of uncorrelated pairs at pair-separations of the order of the characteristic size of each geometry. However, we do see a distinctive peak of skewness on intermediate scales, around $r=5\, \Mpc$. This roughly corresponds to the excess skewness observed for the power spectrum, both in terms of scales (if we are considering the very approximate relation $k \simeq 2\pi/r$) and of amplitude of the skewness at the maximum, which is, for example, $S_3 \simeq 0.1 $ and 0.7 for the BOX and S200 geometries, respectively. On small scales, the skewness decreases as the number of pairs becomes so high that it dilutes the skewness.

Comparing the 2PCF skewness for the different geometries, we see a significant increase with decreasing volume, analogously to the power spectrum. We can see that apart from an overall increase in the amplitude, the peak at $r \simeq 5\, \Mpc$ is more pronounced for smaller volumes. This is similar to the power spectrum, for which the excess skewness with respect to the Gaussian field case is higher for smaller volumes. As for the power spectrum, similar observations are applicable to the case of $z=0.5$. 

\begin{figure}
\includegraphics[width=\linewidth]{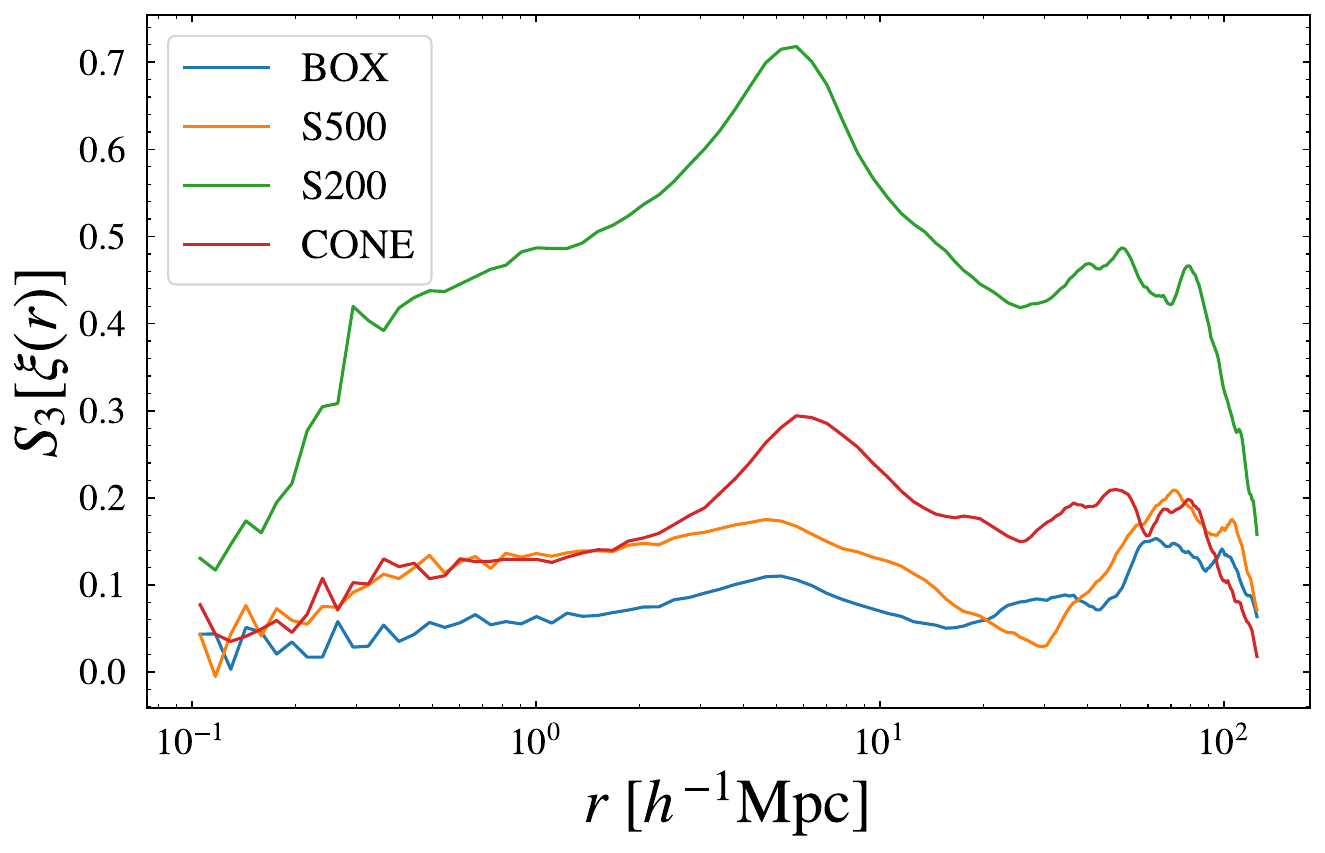}
\caption{Skewness of the 2PCF distribution estimated from the $10\, 000$ \covmos~realisations for each of the four geometries, at $z=0$.}
\label{fig:skew_geom_xi}
\end{figure}

\section{Likelihood Edgeworth expansion }\label{sect:nglik}

%In order to quantify the impact of using a more realistic, non-Gaussian likelihood on cosmological parameter inference, we turn to the Edgeworth expansion. 
In this section, we want to include the skewness of the distribution of two-point statistics measurements in the likelihood. It allows us to quantify the effect of wrongly assuming a Gaussian likelihood on parameter inference. 

The Edgeworth expansion is a technique used to approximate a non-Gaussian probability distribution by perturbing a Gaussian distribution based on its higher-order cumulants, such as the skewness and the kurtosis. This approach enables us to incorporate the non-Gaussian features observed in the two-point statistics while still retaining some of the computational advantages of the Gaussian distribution. For this work, we follow the prescription of~\citet{Amendola96} to build our non-Gaussian likelihood model.

We can start from the Gaussian likelihood, expressed as
\be\label{eq:gauss_like}
G(\Vec{x}, \tens{\lambda}) = \dfrac{|\tens{\lambda}|^{1/2}}{(2\pi)^{N/2}}\exp\left[-\dfrac{1}{2}\vec{x}^{T}\tens{\lambda}\vec{x}\right] \;,
\ee
where $\vec{x} = \vec{d} - \vec{t}$ is the difference between the data vector, $\vec{d}$, and the theory vector, $\vec{t}$, and $\tens{\lambda}$ is the inverse of the covariance matrix, also called the precision matrix. Applying the Edgeworth expansion up to first order (i.e. only including the skewness) to the Gaussian likelihood, we get the non-Gaussian likelihood
\be
   L(\vec{x}, \tens{\lambda}, \tens{k}_3) = G(\vec x, \tens \lambda)\left[ 1+H(\vec x, \tens \lambda, \tens{k}_3)\right] \; .
\ee
The multivariate Edgeworth expansion $H(\vec x, \tens \lambda, \tens{k}_3)$ is defined as 
\be\label{eq:edgew_full_exp}
   H(\vec x, \tens \lambda, \tens{k}_3) = \dfrac{1}{6} \sum_{i,\,j,\,k=1}^{N} k_{3,ijk}\,h_{ijk}(\vec x,\tens \lambda) + ... \; ,
\ee
where $N$ is the dimension of the vector, $\vec x$. The $h_{ij...n}$ refer to the Hermite tensors, defined as 
\be
h_{i_1...i_n}(\vec{x}, \tens{\lambda}) = (-1)^n\,G^{-1}(\vec x,\tens \lambda)\,\dfrac{\partial^n G(\vec x,\tens \lambda)}{\partial{x_{i_1}}...\,\partial{x_{i_n}}} \; ,
\ee
while $\tens k_n$ represents the cumulant tensors of order $n$. For $n=3$ and defining $x_i$ as the $i$-th element of $\vec x$, this corresponds to
\be
k_{3,ijk} = \langle x_{i} x_{j} x_{k} \rangle_\mathrm{c} \; ,
\ee
which leads to the skewness when $i=j=k$. Indeed, in our companion paper B-2025 we only considered the normalised diagonal of this tensor, $S_3 = \langle x^{3} \rangle_\mathrm{c} / \langle x^{2} \rangle_\mathrm{c}^{3/2}$. However, as shown by Eq.~\eqref{eq:edgew_full_exp}, off-diagonal elements are expected to contribute to the likelihood.

To simplify Eq.~\eqref{eq:edgew_full_exp} we can consider a change in the basis, such that the precision matrix becomes the identity. We call $\tens P$ the matrix that operates the change in the basis, such that 
\be\label{eq:lin_trans}
    \vec y = \tens{P}\, \vec x\;,
\ee
is a random variable with unit variance. We call this new basis the uncorrelated basis, as its covariance and precision matrix are the identity. This is very similar to what was done by~\citet{lin_20} in the context of the weak lensing likelihood in configuration space. 

In the uncorrelated basis, the first order of the Edgeworth expansion is expressed as
\be\label{eq:edge_final}
\begin{array}{rl}
&H(\vec x, \tens \lambda, \tens{k}_3) = \displaystyle \dfrac{1}{6} \sum_{i=1}^{N}\, \langle y_{i}^{3} \rangle_\mathrm{c}\, H_3(y_i)
\\&
\\& \displaystyle + \dfrac{1}{2} \sum_{j>i}\, \left[ \langle y_{i}^{2}y_{j} \rangle_\mathrm{c}\, H_2(y_i)H_1(y_j) + \langle y_{i}y_{j}^{2} \rangle_\mathrm{c}\, H_1(y_i)H_2(y_j) \right]
\\&
\\& \displaystyle + \sum_{k>j>i}\, \langle y_{i}y_{j}y_{k} \rangle_\mathrm{c}\, H_1(y_i)H_1(y_j)H_1(y_k) \; ,
\end{array}
\ee
where the $H_n$ are the Hermite polynomials of the order of $n$. We explicitly split Eq.~\eqref{eq:edge_final} in the three different sums involving the three different types of correlators: the one-point $\langle y_{i}^{3} \rangle_\mathrm{c}$, two-point $\langle y_{i}^{2}y_{j} \rangle_\mathrm{c}$ (and $\langle y_{i}y_{j}^{2} \rangle_\mathrm{c}$), and three-point $\langle y_{i}y_{j}y_{k} \rangle_\mathrm{c}$ correlators. We can notice that Eq.~\eqref{eq:edge_final} is not positive definite and its amplitude can potentially be higher than unity when the skweness becomes important. This might cause the likelihood expression to be negative; thus, we checked that this is not happening in the various cases that we report in the present work.

The third-order one-point correlator $\langle y_i^3\rangle_\mathrm{c}$ is the reduced skewness $S_3$ in the uncorrelated basis (i.e. after applying the linear transformation Eq.~\ref{eq:lin_trans}). Therefore, it is interesting to estimate the three-point correlators $S_3^{ijk}$ in both basis. We used the following estimator,
\begin{equation}
\hat S_3^{ijk} = \dfrac{1}{N_{\rm s}}\,\dfrac{\sum_{n=1}^{N_{\rm s}}\, (X_n^{(i)}-\bar X_i)\,(X_n^{(j)}-\bar X_j)\,(X_n^{(k)}-\bar X_k)}{\hat\sigma_i\,\hat\sigma_j\,\hat\sigma_k} \; .
\label{estimk}
\end{equation}
In Fig.~\ref{fig:2ptcorr}, each triangle in the upper left displays the two-point correlators, $S_3^{ijj}$ (for which two of the three vector elements are the same), in the original basis both for the power spectrum (left) and 2PCF (right). It shows that on top of having non-zero diagonal elements (also shown in Figs.~\ref{fig:skew_geom_pk} and~\ref{fig:skew_geom_xi}), there is a large fraction of two-point correlators exhibiting a non-negligible amplitude. We also estimated the three-point correlators and their amplitudes follow the same trend.

As we designed the non-Gaussian likelihood model to be used in the uncorrelated basis, we also wanted to estimate the skewness tensor in that basis. This result is displayed in the lower-right triangle of each panel in Fig.~\ref{fig:2ptcorr}. We can notice a surprisingly result, namely, when estimating the two-point correlators in the uncorrelated basis, the non-Gaussian contributions are reduced to a level compatible with the noise in the estimator. They fluctuate around zero with a standard deviation of $0.01$. The exact same drastic reduction happens when considering the three-point correlators. 

\begin{figure*}
\includegraphics[width=\linewidth]{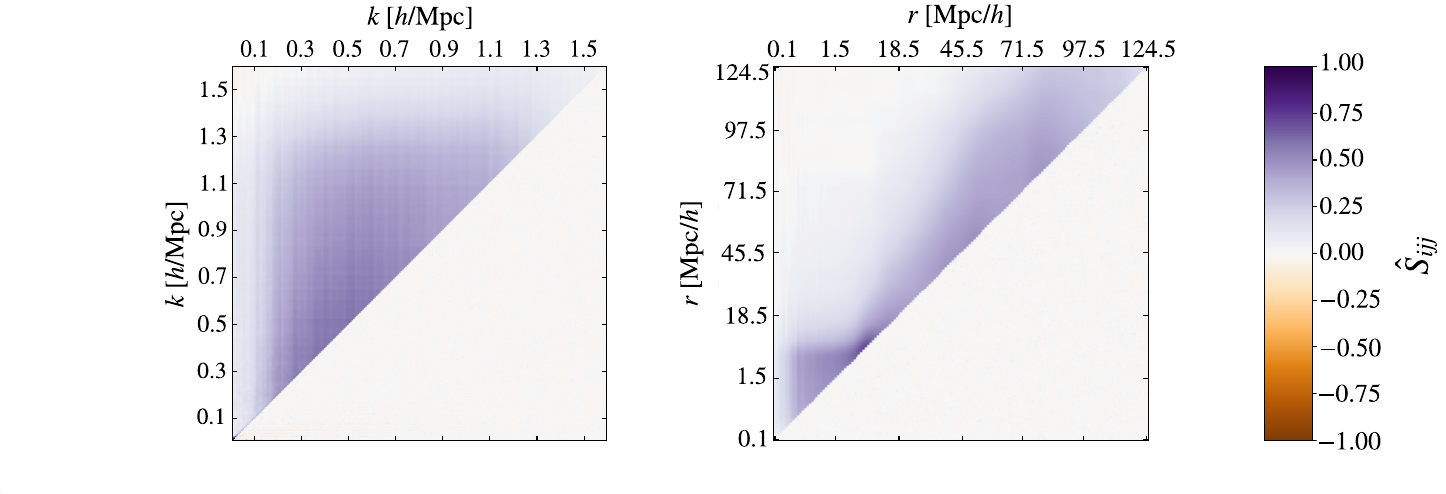}
\vspace{-0.5cm}
\caption{ Matrix of the reduced two-point correlators at an order of $3$ of the power spectrum (left panel) and 2PCF (right panel), estimated with Eq.~\eqref{estimk}, for the S200 geometry. In each panel,  elements on the the upper left display the correlators according to the standard basis ($S_3^{ijj}=\langle x_ix_j^2\rangle_\mathrm{c}\,\sigma_i^{-1}\,\sigma_j^{-2}$),
while the lower right shows the corresponding matrix when estimated after turning to the basis in which the covariance matrix is the identity ($S_3^{ijj}=\langle y_iy_j^2\rangle_\mathrm{c}\,\sigma_i^{-1}\,\sigma_j^{-2}$). }
\label{fig:2ptcorr}
\end{figure*}

However, there are data elements that continue to be characterised by a significant amount of one-point skewness, $S_3$. This is highlighted in Fig.~\ref{fig:uncorr_s3}, where we show the one-point skewness of the power spectrum (original basis) together with the one-point skewness of the data vector after the change in basis (Eq.~\ref{eq:lin_trans}) has been applied to the elements of the power spectrum (uncorrelated basis). We note that once the change in the basis is applied, it becomes difficult to make a clear correspondence between the scales involved. Indeed, each element of the data-vector in the uncorrelated basis is a linear combination of the elements of the data-vector in the original basis. However, it appears that each row of the transformation matrix $\tens P$ is highly peaked. Thus, we used the index of the element at the position of that peak to assign the corresponding scale in the uncorrelated basis. Figure~\ref{fig:uncorr_s3} shows that when the skewness is estimated in the uncorrelated basis, the small-scale excess of skewness (around $k = 0.4\, \invMpc$) present in the original basis is impressively suppressed. It drops by one order of magnitude and tends to the expected level of skewness corresponding to a Gaussian field. This means that only the large-scale modes continue carrying on a significant skewness contribution. For the sake of clarity, we only represent the S200 geometry, but we checked that all other geometries exhibit the same behaviour.

\begin{figure}
\includegraphics[width=0.9\linewidth]{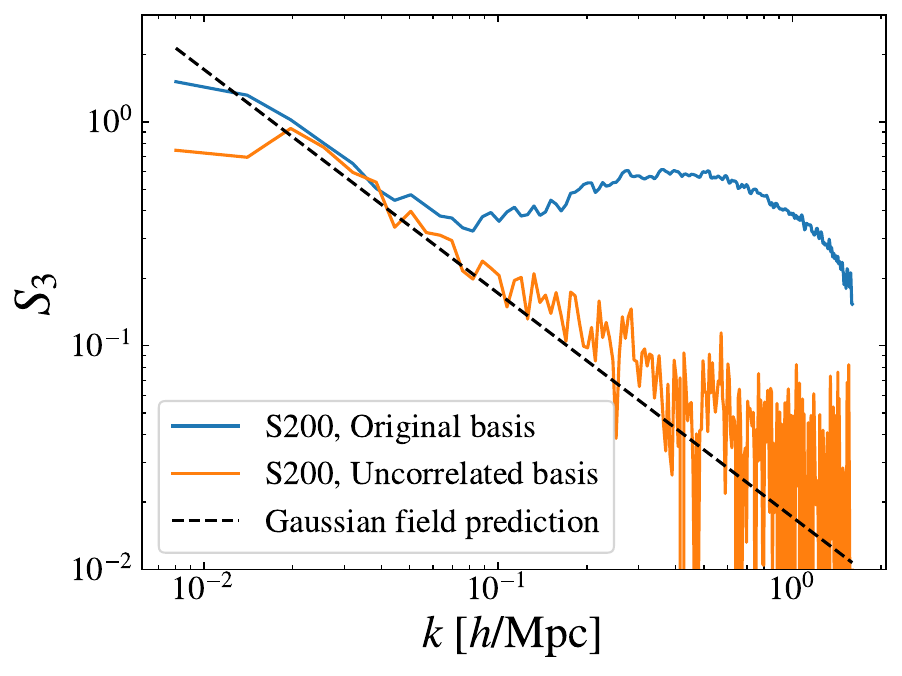}
\caption{Reduced skewness estimated for the S200 geometry in the original and uncorrelated basis. 
The black dashed line is the prediction in the case of a Gaussian field, $S_{3,\mathrm{G}} = 2.2/\sqrt{N_k}$ (see Sect.~\ref{sect:2pt_skew}).}
\label{fig:uncorr_s3}
\end{figure}

As a result, it is interesting to devise two different scenarios for modelling the non-Gaussian likelihood. In scenario-1, we take into account only the one-point correlators in Eq.~\eqref{eq:edge_final}; whereas in scenario-2, we keep all terms up to three-point correlators.

\section{Parameter inference with a non-Gaussian likelihood}\label{sect:param}

To gauge the impact of the Gaussian likelihood approximation at the level of parameter inference, we performed a series of MCMC runs with three different likelihood models: the Gaussian and the two scenarios of the Edgeworth expansion described in the previous section.

We used the \texttt{MontePython}~\citep[][version 3.3.0]{montepython_3} software to run the MCMCs. Our data vector can be the power spectrum or the 2PCF and is taken from one of the $10\, 000$ \texttt{COVMOS} realisations at redshifts $z=0$ and $0.5$. 

The covariance matrix, $\hat C_{ij}$ (which is used to evaluate the precision matrix $\lambda$), was obtained from the remaining $9999$ \texttt{COVMOS} realisations for each redshift. We always fit for the two redshifts simultaneously but we did not consider any correlation between the two redshifts as they emerge from independent realisations. We chose to fit these two redshifts simultaneously to get a reasonable constraining power on  cosmological parameters. In addition, these two redshifts have been chosen to maximise the excess of skewness in the likelihood (cf. B-2025). 

The modelling of the power spectrum and 2PCF has been described in Sect.~\ref{sect:model}. Since we are working in a co-moving space (and not in the observed redshift space), we do not include the Alcock--Paczynski distortion. Given the observations in Sect.~\ref{sect:model_xi}, the 2PCF scale range is fixed to $\rmin = 20\, \Mpc$ and $r_\mr{max} = 124\, \Mpc $, which corresponds to $105$ $r$-bins. Regarding the power spectrum, we study the dependence of the cosmological results with respect to the maximum wave mode, $\kmax$, that we take into account in the likelihood. The different $\kmax$ values we considered are $0.1$, $0.15$, $0.2$, $0.25$, $0.3$, and $0.35\, h/$Mpc, corresponding to $15$, $23$, $31$, $39$, $47$, and $55$ $k$-bins, respectively. When considering geometries that differ from the BOX case, we also vary the minimum wave mode $k_\mr{min}$ of the analysis. We took it to be the fundamental mode for the periodic box ($k_{\rm F} = 0.008\, \invMpc$) and to be the effective fundamental mode ($\tilde k_{\rm F}$) otherwise.

In practice, we started by setting up our scale cuts, which define the data vector. Then we estimated the covariance matrix from the \covmos~realisations and took its inverse to find the change in the basis $\tens P$ for which the precision matrix is the identity. Finally, we compute the corresponding third-order cumulant tensor in the uncorrelated basis. As a result, during the MCMC we apply the change in the basis to the vector $\vec x$ which is the difference between the measurement and the tested model, before computing the likelihood. The likelihood can then take three forms: $L_\mathrm{G}$ as the Gaussian likelihood, along with $L_\mathrm{E_1}$ as the Edgeworth likelihood in scenario-$1$ and $L_\mathrm{E_2}$ in scenario-$2$. By taking $-2\ln$ of these three likelihoods and discarding the normalisation factor of the Gaussian part (see Eq.~\ref{eq:gauss_like}) given the covariance is kept fixed during the MCMC, we get the following expressions:
\begin{align}
    &-2\ln{L_\mathrm{G}} = \chi^2 \; ,\\
    &-2\ln{L_\mathrm{E_1}} = \chi^2  -2\ln{(1 + H_\mathrm{1pt})} \; ,\\
    &-2\ln{L_\mathrm{E_2}} = \chi^2  -2\ln{(1 + H_\mathrm{1pt} + H_\mathrm{2pt} + H_\mathrm{3pt})} \; ,
\end{align}
where $\chi^2 = \sum_i y_i^2$ is just the sum of the square of the element of the difference vector between the data and the model in the uncorrelated basis. The terms denoted as $H_{n\mathrm{pt}}$ correspond to the three terms in Eq.~\eqref{eq:edge_final} that involve the one-, two-, and three-point correlators.

\renewcommand{\arraystretch}{1.5}
\begin{table}
    \centering
     \caption{Priors and fiducial values for the cosmological parameters. }
        \begin{tabular}{l | c c c }
        % \hline
        \multicolumn{1}{l |}{$\bs{\theta}$}  & Priors & Fiducial value\\
    	\hline
    	\hline
    	\multicolumn{1}{l |}{$\omegab$} & \makecell{$\mathcal{N}(0.0224,\ 0.0004)$\\ or $\mathcal{U}(0.0050,\ 0.0390)$} & 0.0224\\

    	\multicolumn{1}{l |}{$\omegac$} & $\mathcal{U}(0.01,\ 0.20)$ & 0.12\\

    	\multicolumn{1}{l |}{$h$} & $\mathcal{U}(0.40,\ 1.00)$ & 0.67\\
    	% \hline
        
        \end{tabular}
        \label{tab:priors}
        \tablefoot{The baseline choice for $\omegab$ is the Gaussian BBN prior.}
    \end{table}
\renewcommand{\arraystretch}{1.5}

The cosmological parameters we chose to vary are $\omegac$, $\omegab$, and $h$. We did not expect to be able to simultaneously constrain these three parameters, so we applied a tight Gaussian prior on $\omegab$ centred on its fiducial value with a width coming from Big Bang nucleosynthesis (BBN) results~\citep{cook_17},
\be
\omegab = 0.0224 \pm 0.0004 \; .
\ee
Nonetheless, we show in the following that for the two largest volumes (BOX and S500) we are actually able to constrain $\omegab$ along with the other parameters. Thus, we also included a case where $\omegab$ is varied in a wide uniform prior. This case allows us to gauge the effect of priors on the likelihood modelling. The fiducial values and priors are shown in Table~\ref{tab:priors}.

In Fig.~\ref{fig:tri_pk_prior}, we show the result, for the power spectrum case, of the MCMCs in the Gaussian likelihood approximation for the biggest and smallest volumes, the BOX, and the S200 geometries, respectively. We can see that for both geometries there is no significant bias with respect to the fiducial cosmology. The contours are slightly shifted in the $(\omegac,\ \omegab)$ plane for the BOX geometry, but this is most likely due to sample variance. In any case, this modest shift with respect to the fiducial cosmology is not important in view of the main goal of this study, namely, the difference between the constraints obtained in the Gaussian likelihood approximation or with a non-Gaussian likelihood. 

In the BOX geometry, we have enough constraining power to not have prior-dominated constraints on all three cosmological parameters, thanks to the large volume. However, there is a clear degeneracy between the three parameters and $\omegab$ is completely unconstrained for the S200 geometry. Thus, in the following, our baseline analysis always includes a BBN prior for all geometries.

\begin{figure}
\includegraphics[width=\linewidth]{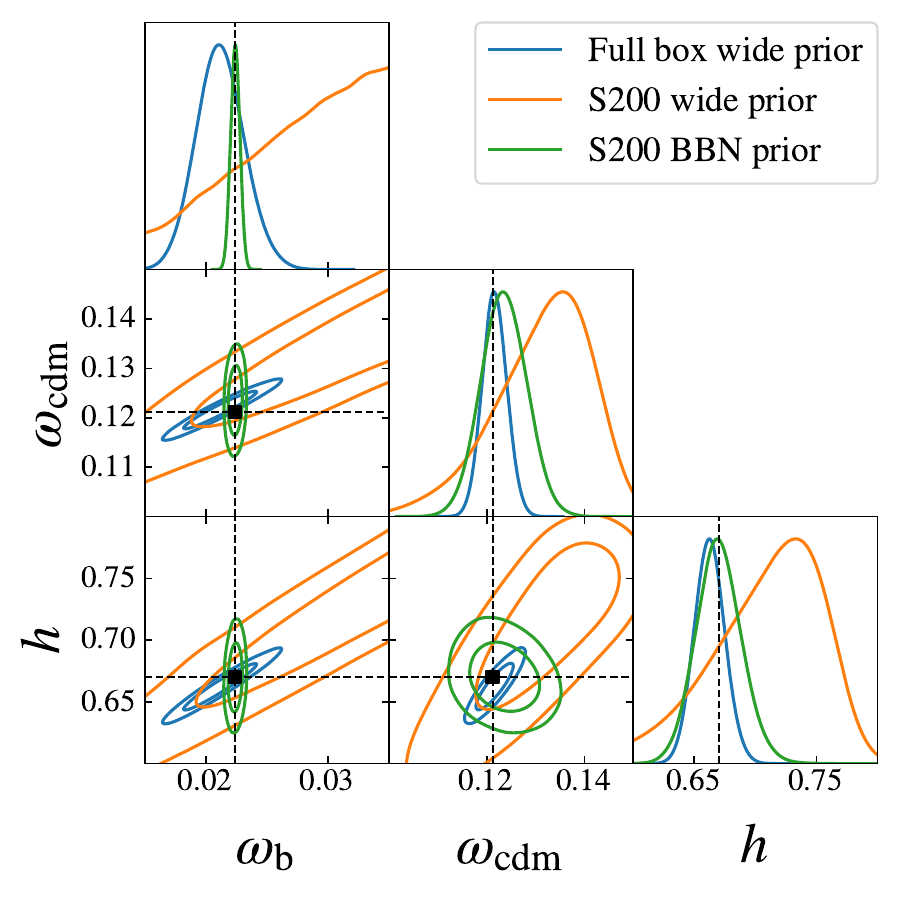}
\caption{1D and 2D marginalised posterior distribution of cosmological parameters estimated in the Gaussian likelihood approximation, with the power spectrum in  the S200 geometry for a wide or BBN prior on $\omegab$, and in the BOX geometry only for the former prior. This result is obtained for $\kmax = 0.2\, h/\mr{Mpc}$.}
\label{fig:tri_pk_prior}
\end{figure}

\begin{figure}
\centering
\hspace{-0.45cm}
\includegraphics[width=0.4\textwidth]{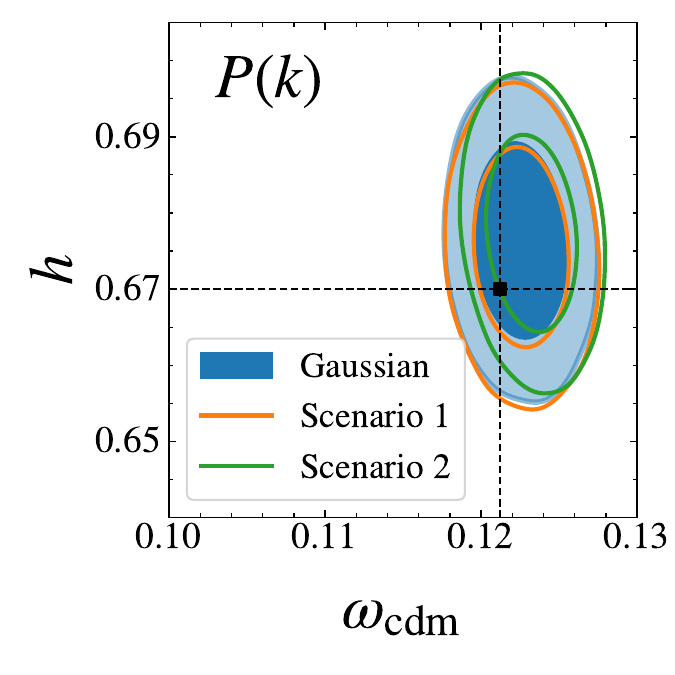}
% \hspace{26pt}
\includegraphics[width=0.4\textwidth]{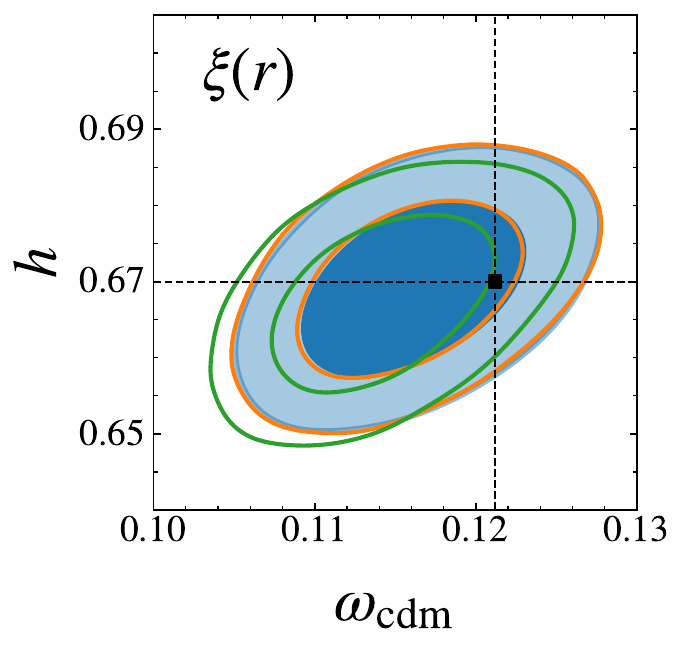}
\caption{2D marginalised posterior distribution of cosmological parameters estimated in the CONE geometry for the Gaussian likelihood approximation and the two scenarios of the Edgeworth expansion. The top and bottom panels show the results obtained in Fourier and the configuration space, respectively. The corresponding scale cuts are $\kmax = 0.2\, h/\mr{Mpc}$ in Fourier space and $\rmin = 20\, \mr{Mpc}/h$ in configuration space.}
\label{fig:tri_xi_pk_cone}
\end{figure}

In Fig.~\ref{fig:tri_xi_pk_cone}, we show the effect of modelling the likelihood with the Edgeworth expansion compared to approximating it as a Gaussian in Fourier and configuration space. We show the results obtained in the two scenarios of the Edgeworth expansion: only accounting for the one-point correlator (scenario-$1$) and also accounting for the two- and three-point correlators (scenario-$2$). We only show the case of the CONE geometry because it is the one exhibiting the most significant differences between the posterior distributions of cosmological parameters.

We first want to stress that we have found very similar results in Fourier and configuration space. The main difference is in the overall constraining power, which is lower for the 2PCF as we did not include scales as large as in Fourier space. In both cases, the Gaussian and scenario-$1$ constraints are almost identical and it is only when including the two- and three-point correlators of the Edgeworth expansion (scenario-$2$) that we start to see some differences with the two other cases. Indeed, the contours are slightly shifted away from the two other cases. In Fourier space, the ellipse is also compressed in the direction of $\omegac$.

To understand why scenario-$2$ differs from the other cases, we need to recall that the two- and three-point correlators we are introducing seem to be largely dominated by a Gaussian noise of mean $0$ and standard deviation $0.01$. We note that the dispersion of the Gaussian noise is independent from the considered geometries and is due to the limited number of realisations we used. Thus, it is probable that the shift and the reduction of the width of the contours we see in Fig.~\ref{fig:tri_xi_pk_cone} are random and due to the inclusion of noise in the likelihood model. This would be similar to what happens when using a noisy estimate of the covariance matrix~\citep{Taylor_13, Dodelson_13, gouyou_25}.

To verify this hypothesis, we generated three random realisations of the two- and three-point correlators from a Gaussian distribution with the same characteristics as the one we estimated (i.e. centred on $0$ and with a standard deviation of $0.01$) and we assigned to the diagonal elements $k_{3,iii}$ the skewness estimated from the \covmos~simulations. We repeated the parameter inference from scenario-$2$, but with these three random realisations instead of the correlators we estimated from the \covmos~simulations. We show the result of this test in Fig.~\ref{fig:tri_pk_random}. We can see that the three contours obtained in scenario-$2$ with a random realisation of the two- and three-point correlators are shifted with respect to each other in different directions. The width of the contours can also be compressed depending on the realisation. Actually, one of the contours appears to be very close to the original scenario-$2$, while the other one is closer to the Gaussian likelihood case. This indicates that the noise in the estimation of the third-order cumulant tensor (mostly in the non-diagonal part of it) is propagating at the level of the posterior distribution. It perturbs the width and the position of the contours, as happens with a noisy estimate of the covariance matrix. More precisely, this result demonstrates that any potential systematic shift is, at most, comparable in size to those caused by noise. A definitive test would require a significantly
larger number of realisations. Furthermore, while the FoM appears unaffected, a
shift in the recovered mean values could still be masked by the noise level.
As a result, in the following, we assume the ansatz that scenario-$1$ is enough to describe the non-Gaussian behaviour of the likelihood of two-point statistics.

\begin{figure}
\includegraphics[width=\linewidth]{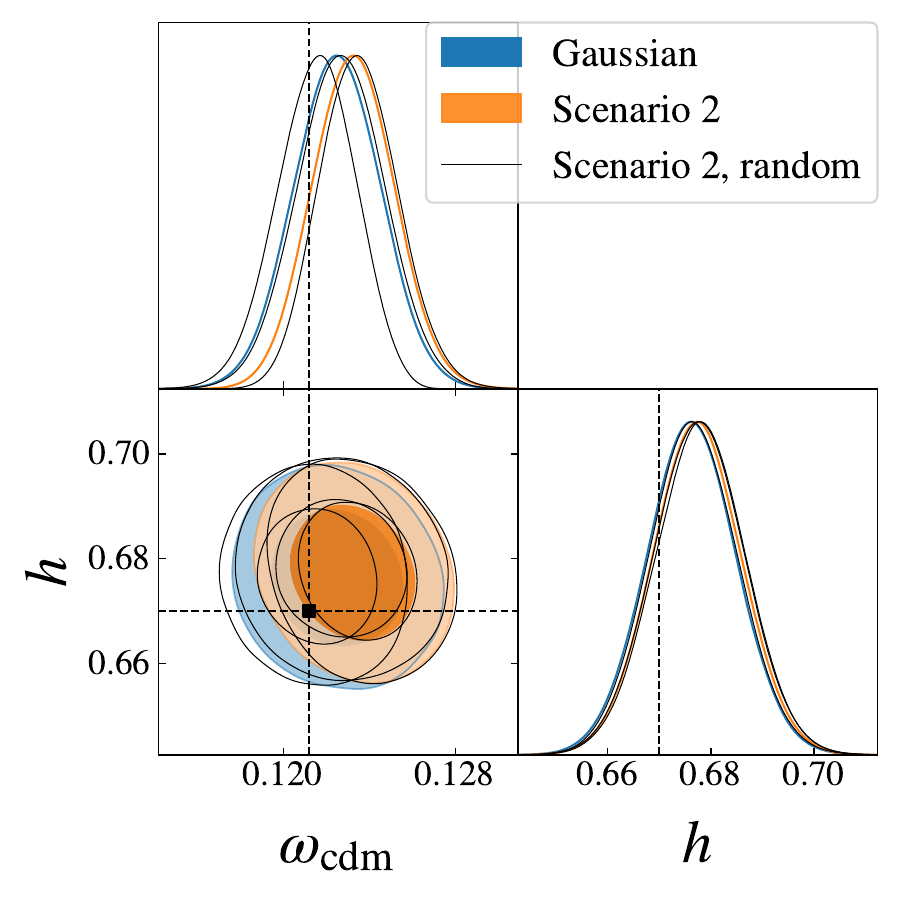}
\caption{$1$D and $2$D marginalised posterior distribution of cosmological parameters estimated in the CONE geometry at $\kmax = 0.2\, h/\mr{Mpc}$ for the Gaussian likelihood (filled blue contours) and scenario-$2$ using the estimated two- and three-point correlators (empty orange contours) of a random realisation drawn from a Gaussian distribution (empty black contours). }
\label{fig:tri_pk_random}
\end{figure}

In the remainder of this section, we focus on the figure of merit (FoM) to summarise the effect of considering an Edgeworth likelihood, rather than the Gaussian approximation in various configurations. The FoM is defined as\footnote{It is inversely proportional to the hyper-volume delimited by the $2\,\sigma$ contours in the full parameter space. The higher the FoM, the tighter the constraints.} \citep{Wang_08}
    \be
        \mathrm{FoM} := \dfrac{1}{\sqrt{\det(\hat{\tens{\Phi}})}} \; , 
    \ee
where $\hat{\tens{\Phi}}$ is the estimated parameter covariance matrix, including $\omegac$, $h$, and $\omegab$ when the latter is not constrained by the BBN prior. Here, the FoM is an acceptable metric of the overall constraining power as we have seen in Figs.~\ref{fig:tri_xi_pk_cone} and \ref{fig:tri_pk_random} that the posteriors are Gaussian (except when $\omegab$ is varied in a wide uniform prior and cannot be constrained, but we do not consider this case in the rest of the section). 

Figure~\ref{fig:fom_geom} shows the ratio of the FoM obtained with the Gaussian or Edgeworth likelihood (scenario-$1$) for all geometries in Fourier and configuration space. In addition, we also show the cases where we let $\omegab$ to vary in a wide uniform prior, for the geometries where it can be constrained (BOX and S500). For all the cases studied, we never observe a difference in the FoM larger than $5$\%. We do not see a specific trend with respect to the volume of the geometries. Furthermore, including the BBN prior does not seem to specifically push the FoM ratio in a particular direction.

\begin{figure}
\includegraphics[width=\linewidth]{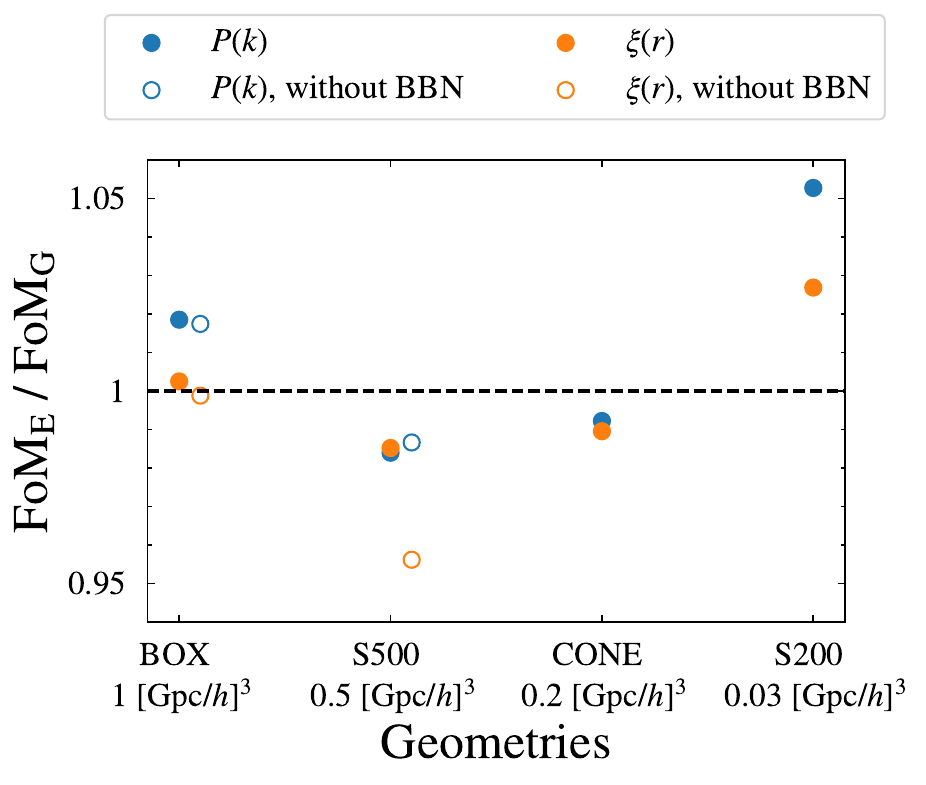}
\caption{Ratio of the FoM obtained with the Gaussian and Edgeworth expansion likelihood, for both the power spectrum and the 2PCF and the four different geometries (with their corresponding volumes below their names). For the power spectrum $\kmax=0.2\,  h/\mathrm{Mpc}$ and for the 2PCF $r_\mathrm{min} = 20\, \mathrm{Mpc}/h$.}
\label{fig:fom_geom}
\end{figure}

After varying the geometry, we went on to vary the scale cut used in the analysis. We varied $\kmax$ in the Fourier analysis between $0.1$ and $0.35\, \invMpc$ in the BOX geometry. Figure~\ref{fig:fom_kmax} shows the comparison between the Gaussian and the Edgeworth likelihood again in terms of the ratio of the FoM obtained in the two cases, for the different $\kmax$ values. We can see from this figure that there is still no impact on cosmological constraints of the non-Gaussian features of the two-point statistics likelihood even when including small scales ($\kmax = 0.35\, \invMpc$). Indeed the difference between FoM$_\mr{E}$ and FoM$_\mr{G}$ is never larger than 5\% and we cannot see any particular trend with $\kmax$. The question of whether the BBN prior is included or not does not change this conclusion.

\begin{figure}
\includegraphics[width=\linewidth]{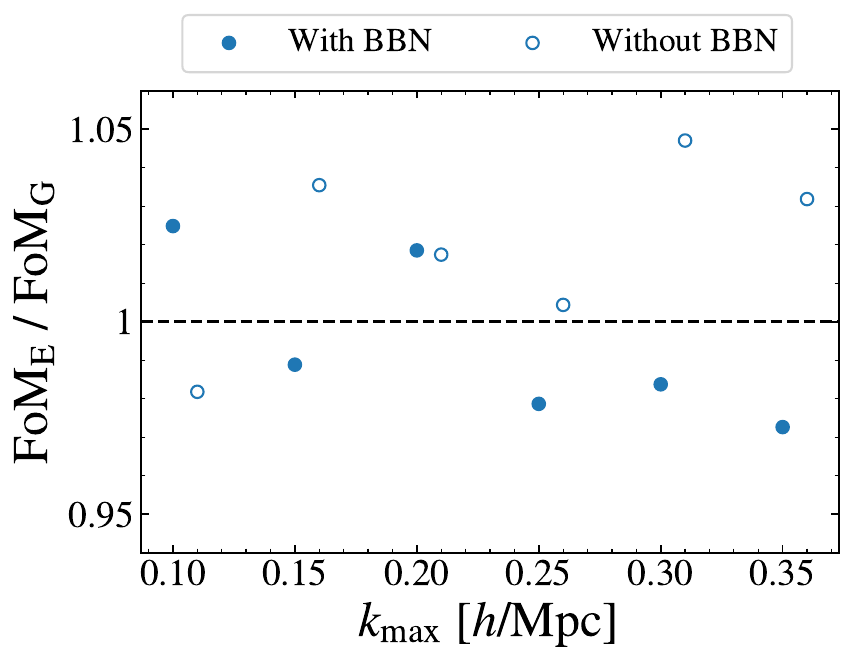}
\caption{Ratio of the FoM obtained with the Gaussian and Edgeworth expansion likelihood for the power spectrum in the periodic box. This ratio is shown for increasing $\kmax$ and including or not the BBN prior.}
\label{fig:fom_kmax}
\end{figure}

These results can be understood in light of the findings of Sect.~\ref{sect:nglik}, which show how the change in the basis affects the skewness tensor. When shifting to the uncorrelated basis, the skewness of two-point statistics is reduced to a level that is compatible with the skewness expected for two-point statistics of a Gaussian field. That is a skewness which is non-negligible on large scales, but that vanishes on intermediate scales, and a skewness tensor that has vanishing off-diagonal elements. Thus, when performing a parameter inference in the uncorrelated basis, the only modification to the Gaussian likelihood that the Edgeworth expansion is bringing is a level of skewness equivalent to the case of a Gaussian field. We find that this level of skewness has no impact on parameter inference, even for the highest level of skewness corresponding to the smallest volume. This result is in line with the findings of~\citet{lin_20} who performed a similar analysis for weak lensing two-point statistics. Of course, if we consider parameters that are especially constrained by the largest scales (e.g. $f_{\rm{NL}}$ related to primordial non-Gaussianity) the conclusion might changed slightly.

\section{\Euclid\/ context}\label{sect:pinoc}

The simulations we have explored up to now and in B-2025 have been very useful because they allow us to easily explore different settings and find the ones that were maximising the skewness. However these simulations are far from the real \Euclid data we will get in the future, because they are composed of dark matter particles and not of galaxies; in addition, they will be presented in terms of volume, shot noise, and redshift distribution. Thus, we wish to look at how the two-point statistics distribution would appear in a more realistic context.

For this purpose, we considered the ELM~\citep{EP-Monaco1}, a set of fast simulations generated with \texttt{PINOCCHIO}~\citep{monaco_13}, which are characterised by properties that resemble those of the spectroscopic galaxy sample we will get from \Euclid~\citep{EuclidSkyOverview}. These simulations have the geometry of a cone that has a half aperture angle of $30\degree$ and span a redshift range going from 0.9 to 1.8, divided into four redshift bins. The redshift distribution and the mean density of the simulated galaxies match the ones expected for \Euclid's ${\rm H}_\alpha$ emission line galaxies. We only considered the lowest redshift bin, with edges $0.9 < z < 1.1$, to maximise the skewness of the sample (cf. B-2025).

We have only $1000$ of these simulations, which is not enough to get a precise estimation of $S_3$. Thus, we cloned these 1000 ELM with \covmos\ to get a larger set of $10\,000$ realisations in both real and redshift space. All the details of the cloning procedure are described in Appendix~\ref{appendix:pinoc}.

From these $10\,000$ clones of the ELM, we can now estimate the skewness of the galaxy power spectrum distribution in both real and redshift space, as shown in Fig.~\ref{fig:pinoc_s3}. We can see that from these simulations that we did not detect any significant amount of skewness at high $k$, so that the estimated skewness of the monopole follows the expectation in the case of a Gaussian underlying field. Here, we do not show the prediction for the multipoles in redshift space, but they can be predicted (and we give them in B-2025).

This absence of the high-$k$ skewness is mainly due to the fact that the density in the ELM (close to what we expect for \Euclid~) is much smaller than the density we have in the \covmos~dark matter simulations, meaning that the shot noise is higher. We explain in B-2025 that the shot noise attenuates the skewness on small scales because at these scales the power spectrum distribution is affected by a Poissonian noise. In addition, the type of tracers of the matter field is expected to play a role. Since (as we show in B-2025) that the skewness depends on the ratio between the penta-spectrum (i.e. a $6$-point correlation function in Fourier space) and the power spectrum, it becomes clear that changing the multi-point distribution of the density field, would change the expected level of skewness. Still, it is not obvious to predict that a biased tracer of the matter would decrease the expected skewness because this is highly dependent on the mapping between matter and the considered tracer.

We can also notice that the angular mask considered in the current ELM is an over simplification of what will eventually be the \Euclid angular footprint on the sky. However, increasing the volume reduces the skewness of the distribution of two-point statistics. Thus, despite the fact that the expected \Euclid footprint will eventually be more complicated, we can guess that the skewness cannot exceed the level of the studied S200 geometry.

As a result, in a more realistic setting such as the one presented in the current section, only the large-scale skewness can be measured. This demonstrates that for the considered tracers of the matter field that  \Euclid is targeting, we do not detect any excess of skewness at high $k$. The level of skewness we find here is thus similar to the level of skewness that is introduced in the Edgeworth likelihood for the parameter inference we presented in the previous section. We find this level of skewness to have no impact on cosmological parameter constraints.

\begin{figure}
\includegraphics[width=\linewidth]{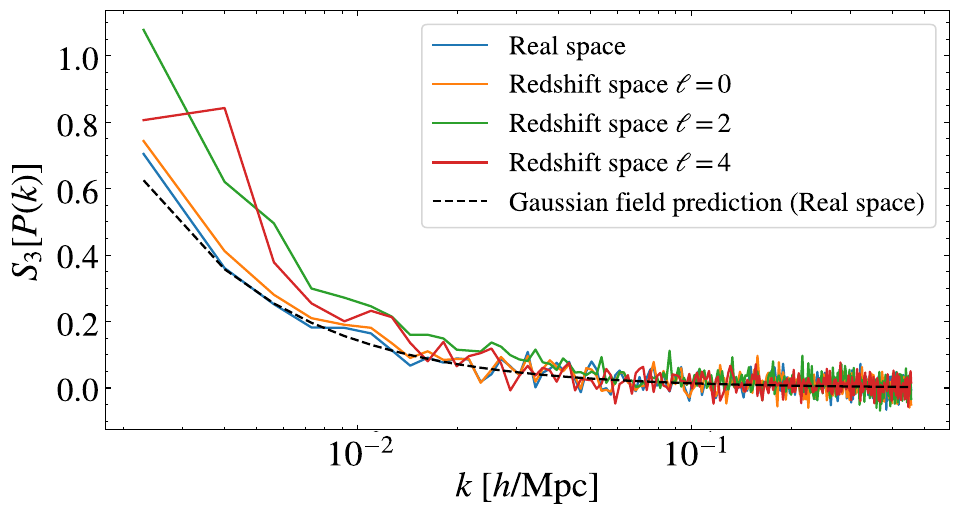}
\caption{Skewness of the power spectrum distribution in real space (blue) and redshift space for the first three even multipoles (orange, green and red) estimated from $10\,000$ ELM \covmos~clones in the first redshift bin with edges $0.9 < z < 1.1$. The black dashed line is the prediction of the skewness in the case of a Gaussian field in real space.} 
\label{fig:pinoc_s3}
\end{figure}

\section{Conclusions}
\label{sect:conclu}

In this second paper of the series, we investigate the impact of the non-Gaussianity of the likelihood on cosmological parameter inference. In the prior companion paper B-2025, we study the origin of the non-Gaussian features in the distribution of two-point statistics in detail and here, we aim to quantify their consequences for cosmological analyses, while relying on the assumption of a Gaussian likelihood.

To this end, we produced $10\, 000$ \covmos~realisations based on the \demcov~simulations, adopting a \halofit~power spectrum as input, and designed four different survey geometries (BOX, S200, S500, and CONE) to mimic the impact of the survey window function. We developed a fast estimator of the 2PCF that operates in both spherical and cone-like geometries without requiring random catalogues, and measured both the power spectrum and 2PCF from all realisations at two redshifts: $z=0$ and $z=0.5$ (thereby maximising the excess of skewness) .

We first quantified the scales over which unbiased cosmological constraints can be recovered from these two-point statistics, properly accounting for the effects of the survey mask and the IC. We then looked at the estimated skewness from all the measurements and found, as expected, an equivalent trend of the skewness with scales and survey volume in Fourier and configuration space. As already observed in B-2025 for the power spectrum, we confirmed that reducing the effective survey volume amplifies both the overall level of skewness and the excess skewness with respect to the case of a Gaussian field observed on small scales (around $k=0.5\, \invMpc$), for both the power spectrum and the 2PCF.

To model the likelihood of two-point statistics, we adopted an Edgeworth expansion including the skewness of the distribution, following \citet{Amendola96}. This expansion involves the complete skewness tensor where the diagonal is what we usually call the skewness. To simplify the calculation of the terms of this expansion, we performed a change in basis that transforms the precision matrix into the identity matrix. We then estimated the full skewness tensor in both the original and the new uncorrelated basis. In the original basis, we found significant non-diagonal elements in the skewness tensor, which would not be present if the underlying field was Gaussian. However, in the uncorrelated basis, these non-diagonal elements are reduced to a level compatible with Gaussian noise fluctuating around 0 with a standard deviation of 0.01. In this basis, only the large scale non-Gaussianity is still present in the distribution of 2pt statistics, as if the underlying field was Gaussian.

To perform the parameter inference using the Edgeworth expansion, we considered two scenarios: one including only the one-point correlator (the diagonal elements of the skewness tensor) and another accounting for the full skewness tensor. We found that including the full tensor introduces noise in the inference process due to the limited precision of its estimation, analogously to what happens when using a noisy covariance matrix. Consequently, we restricted our final analysis to the first scenario.

Within this framework, we show that even in cases with high measured skewness, the cosmological constraints are remarkably insensitive to the Gaussian assumption for the likelihood; namely, the FoM changes by less than $5$\%, regardless of the amount of non-Gaussianity. In addition, we have shown that the shift of the contours are negligible. These results stay unchanged when varying the scales included in the analysis or when restricting the parameter space with a BBN prior. This robustness can be understood in light of the level of skewness that was observed in the uncorrelated basis. These results are in line with \citet{lin_20} who performed a similar study in the context of the likelihood of weak lensing two-point statistics. However, in that work, the authors had only considered the diagonal of the skewness tensor in the Edgeworth expansion. Our results are also in good agreement with \citet{hahn_18}, who found that for the power spectrum analysis of BOSS the cosmological parameters are not affected by the Likelihood non-Gaussianity, except in the case of $f\sigma_8$, which exhibits a shift of $-0.44\sigma$. As we focussed on real space to maximise the skewness, we did not consider this parameter in our work.

Finally, to understand whether these conclusions would apply in a more realistic \Euclid context, we applied a cloning procedure to \Euclid realistic galaxy mock catalogues: ELM. From the $1000$ available mocks, we were able to estimate the PDF, the power spectrum, and the velocity dispersion of the galaxy field, which allowed us to use \covmos~to produce $10\, 000$ clones of the ELM. The estimated skewness from these more realistic mocks is compatible with the one expected for a Gaussian field. This is mainly due to the fact that the level of shot noise present in these mocks (close to what we will get with real \Euclid data) is high enough to dilute the non-Gaussian features on intermediate scales, such that only large scale non-Gaussianity remains This effect from the shot noise was observed in B-2025. The skewness measured in these \Euclid-like mocks is therefore comparable to the level included in our Edgeworth likelihood analysis, where we found no impact on the inferred cosmological parameters. Note that, since the skewness of the distribution of two-point statistics tends to be reduced when increasing the volume, the presented results are conservative.

Overall, we conclude that for \Euclid-like analyses based on two-point statistics, whether in Fourier or configuration space, the Gaussian likelihood assumption is amply justified. The small level of residual non-Gaussianity has a negligible effect on the recovered cosmological parameters and their uncertainties. This is in agreement with the recent results obtained by \citet{Oehl_2025} and \citet{Oehl_2026} in the context of weak-lensing. This validates the use of a Gaussian likelihood in the forthcoming \Euclid cosmological analyses, as implemented in the \Euclid likelihood code CLOE \citep[Cosmology Likelihood for Observables in \Euclid,][]{EP-ClOE3}.

\begin{acknowledgements}
  
\AckEC 
The project leading to this publication has received funding from Excellence Initiative of Aix-Marseille University -A*MIDEX, a French "Investissements d’Avenir" programme (AMX-19-IET-008 -IPhU). 
The DEMNUni-cov simulations were carried out in the framework of “The Dark Energy and Massive Neutrino Universe covariances" project, using the Tier-0 Intel OmniPath Cluster Marconi-A1 of the Centro Interuniversitario del Nord-Est per il Calcolo Elettronico (CINECA). We acknowledge a generous CPU and storage allocation by the Italian Super-Computing Resource Allocation (ISCRA) as well as from the coordination of the “Accordo Quadro MoU per lo svolgimento di attività congiunta di ricerca Nuove frontiere in Astrofisica: HPC e Data Exploration di nuova generazione”, together with storage from INFN-CNAF and INAF-IA2.
\end{acknowledgements}

%
% Here comes the reference list, generated via bibtex from
% your bibfile my.bib and Euclid.bib. Please make sure that
% the same paper is not referenced twice, one from your my.bib
% file, and once from Euclid.bib.
%

\bibliography{Euclid}

%
% Now you can add appendices.
% In this example, the appendices are in one column mode.
% If that is not requires, comment out \onecolumn
% Note that appendices in A\&A come {\it after\/} the references.

\begin{appendix}
  %\onecolumn %If you don't want single column for the Appendix, please
             %comment this out

\section{Two-point correlation estimator}
\label{appendix:2pcf}

In this appendix we review  the Landy \& Szalay estimator for the
2PCF. In practice, we show that for
specific survey geometries we can predict not only the Random-Random pair
counts but also the Data-Random. This, induces a considerable gain in the time
required to estimate the 2PCF from a galaxy catalogue. This is particularly relevant when estimating the covariance matrix of the Landy \& Szalay estimator from a large
sample ($\sim\!\! 10\,000$) of realisations.

In the following, we focus on estimating the multipoles, $\xi^{(\ell,m)}(r)$, of the
2PCF $\xi (\vec r)$. Indeed, without any lack in terms of generality, the 2PCF can be decomposed as a spherical harmonics expansion,

\begin{equation}
\xi (\vec r) = \sum_{\ell=0}^\infty\sum_{m=-l}^l \xi^{(\ell, m)}(r)\, Y_{\ell m}(\theta, \phi) \; , 
  \label{SHxi}
\end{equation}
where $\theta$, $\phi$, and $r$ define the separation vector $\vec r$
in spherical coordinates and the $Y_{\ell m}$ are the spherical harmonic
functions normalised such that their orthogonal property satisfies
\begin{equation}
\int_0^{2\pi}\int_{0}^{\pi} Y_{\ell m}(\theta, \phi)\,Y_{\ell'm'}^{*}(\theta,
\phi)\;\dif^2 \Omega = \frac{4\pi}{2\ell+1}\delta_{\ell \ell'}^K\delta_{mm'}^K \; ,
  \label{ortho}
\end{equation}
where $\dif^2\Omega$ is the differential element associated with the solid angle. With this choice of normalisation for the spherical harmonics, when $m=0$ they reduce to the Legendre polynomials $\lcal_\ell (\cos\theta)$ of order $\ell$ which is useful to map the spherical
harmonics decomposition into the more usual Legendre decomposition. 
Using the orthogonality property in Eq.~\eqref{ortho} we can show that the
coefficients $\xi^{(\ell, m)}$ of the harmonic decomposition can be
obtained from
\begin{equation}
  \xi^{(\ell, m)} = \frac{2\ell+1}{4\pi}\int \xi (\vec r)\, Y_{\ell m}^{*}(\theta,\phi)\;\dif^2 \Omega \; .
  \label{xilm}
\end{equation}

Starting with the most obvious estimator that we can derive from Eq.~\eqref{xilm}, which assumes ergodicity. 
Since the 2PCF is defined as $\xi (\vec r) := \langle \delta(\vec x)\delta (\vec x+\vec r)\rangle$, we could say that for a given configuration $\vec r$ we can estimate the 2PCF by simply taking the volume average of the product of two density contrast at separation $\vec r$. However, we must take care of performing this volume average within a volume containing always a pair of density contrasts. This means that this effective volume is smaller than the total volume, $V$, and it will depend on the configuration of the separation vector. Thus, let us define this volume as $V(\vec r)$. This effective volume is simply the volume intersection between the geometry and its replication but translated by vector $\vec r$. Indeed. we can define $V(\vec r) := \int W(\vec x)\, W(\vec x + \vec r)\;\dif^3\vec x$. 
Based on the definition of the observed number density $n^{\rm o}(\vec x)$ (see Eq.~\ref{eq:rho_obs}), the mean number density $\bar n_\mathrm{V}$ and the observed density contrast (see Eq.~\ref{eq:d_obs}) we can define the following estimator for the 2PCF as
\begin{equation}
  \hat \xi (\vec r) := \frac{1}{V(\vec r)} \int_{V(\vec r)} \delta^{\rm o}(\vec x)\,\delta^{\rm o}(\vec x+\vec r)\; \dif^3 \vec x \; .
  \label{hatxi}
\end{equation}
From that simple definition, we can immediately deduce an estimator for the harmonic coefficients,

\begin{equation}
\hat \xi^{(\ell, m)} := \frac{2\ell+1}{ v } \int_{v} \frac{Y_{\ell m}^*}{V(\vec r)} \int_{V(\vec r)} \delta^{\rm o}(\vec x)\, \delta^{\rm o}(\vec x+\vec r)\; \dif^3 x\,\dif^3r \; ,
  \label{hatxilm}
  \end{equation}
  where the second average is made over a spherical shell within $r_{\rm min}$ and $r_{\rm max}$, thus $v=4\pi/3\, (r_{\rm max}^3 - r_{\rm min}^3) $. From Eq.~\eqref{hatxilm}, we  can actually get to the Landy \& Szalay estimator and this is what we show in the following. Indeed, from the definition of the density contrast we can expand the integrand of Eq.~\eqref{hatxilm} as 
  \begin{equation}
  \bar n_{\rm V}^2\delta^{\rm o}(\vec x)\, \delta^{\rm o}(\vec x+\vec r)\! = \! n^{\rm o}(\vec x)\, n^{\rm o}(\vec x + \vec r)  - \bar n_\mathrm{V} \left [ n^{\rm o}(\vec x) + n^{\rm o}(\vec x+\vec r) \right ] + \bar n_\mathrm{V}^2 \; .
  \label{dxdr}
  \end{equation}
   It follows that we can split Eq.~\eqref{hatxilm} into four contributions as 
  \begin{equation}
   \hat \xi^{(\ell, m)}  = \hat I_1 - \hat I_2 - \hat I_3 + \hat I_4 \; .
  \label{nn}
  \end{equation}
   The last contribution $\hat I_4$  can be trivially computed indeed,
   
    \begin{equation}
      \hat I_4 := \frac{2\ell+1}{v}\int_v \frac{Y^*_{\ell m}(\theta, \phi)}{V(\vec r)}\;\int_{V(\vec r)}\dif^3 x\, \dif^3 r \; ,
      \label{i4}
    \end{equation}
    for which the integral over volume $V(\vec r)$ leads to a simple integral over the shell $v$, such that
    \begin{equation}
     \hat I_4 = \frac{2\ell+1}{v} \int_v\ r^2 Y^*_{\ell m}(\theta, \phi)\,Y_{00}(\theta, \phi)\;\dif^2\Omega\, \dif r \; .
    \label{i4b}
  \end{equation}
  Since the volume of the shell is $v=4\pi/3 \left ( r_{\rm max}^3 - r_{\rm min}^3\right )$ and using the orthogonality relation in Eq.~\eqref{ortho}, it turns out that $\hat I_4 = \delta_{\ell 0}^{\rm K}\delta_{m0}^{\rm K}$ which is $1$ only if $\ell$ and $m$ are zero (i.e only for the monopole), so that $\hat I_4$ is null for any other multipole.
  
  In order to make an explicit calculation for the three other terms $\hat I_1$, $\hat I_2$, and $\hat I_3$ we can further assume that we are dealing with a catalogue of $N$ objects thus the density field can be represented by a Dirac sum such that $n^{\rm o}(\vec x) = \sum_{i=1}^N\delta^{\rm D} (\vec x-\vec x_i)$. It can be substituted in Eq.~\eqref{nn} to explicitly compute the three missing terms. The first term is defined as
  \begin{equation}
    \hat I_1 := \frac{2\ell+1}{v \bar n_\mathrm{V}^2} \int_v \frac{Y_{\ell m}^*(\theta, \phi)}{V(\vec r)}\, \int_{V(\vec r)}  n^{\rm o}(\vec x)\, n^{\rm o}(\vec x+\vec r)\; \dif^3 x\, \dif^3  r \; ,
    \end{equation}
    which leads to
    \begin{equation}
       \hat I_1 = \dfrac{2\ell+1}{v \bar n_\mathrm{V}^2}\, \sum_{i=1}^N\sum_{j=1}^N\,\int_v \frac{Y_{\ell m}^*}{V(\vec r)}\, \int_{V(\vec r)} \!\!\!\! \dirac(\vec x-\vec x_i)\, \dirac(\vec x+\vec r - \vec x_j)\; \dif^3 x\, \dif^3  r \, .
     \end{equation}
     Henceforth, we can avoid explicitly writing out the arguments of the spherical harmonics when they refer to $\theta$ and $\phi$. In the above equation, we can first integrate over $V(\vec r)$ and then integrate over the shell which leads to
     \begin{equation}
     \hat I_1 = \frac{2\ell+1}{v \bar n_\mathrm{V}^2} \sum_{i=1}^N\sum_{j=1}^N \frac{Y_{\ell m}^*(\theta_{ij}, \phi_{ij})}{V(\vec r_{ij})}\Theta (r_{ij}) \; ,
   \end{equation}
   where we introduced the separation vector $\vec r_{ij} = \vec x_j - \vec x_i$. $\theta_{ij}$, $\phi_{ij}$,  and $r_{ij}$ are the spherical coordinates of the separation vector $\vec r_{ij}$ and the function $\Theta$ is null except when $r_{ij}$ is inside the considered shell in which case it is $1$. Finally, by multiplying by $V$ inside the sum and dividing by $V$ in front, the estimator can be expressed as a weighted sum over the pairs of objects that belong to the shell,
   
   \begin{equation}
     \hat I_1 = \frac{2\ell+1}{v N \bar n_\mathrm{V}} \sum_{ij} w_{ij}\, Y_{\ell m}^*(\theta_{ij}, \phi_{ij})\, \Theta (r_{ij}) \; .
     \label{sumi1}
     \end{equation}
     The weight, $w_{ij} := w(\vec r_{ij}) := V/V(\vec r_{ij})$, can be seen as a correction due to the fact that since there is a finite geometry not all the pairs have the same occurrence probability. Indeed, given a separation vector, $\vec r$, we define a region of exclusion, where we cannot find a counter part with orientation $\vec r$ because it will be systematically outside the survey volume. Yet, another way of seeing this is from a Monte Carlo integration point of view in which the weight to be applied is the inverse probability distribution of the sampled multi-dimensional function. Thus, we can finally express the result as
     \begin{equation}
      \hat I_1 = \frac{{\rm DD}^{(\ell, m)}}{{\rm RR}}  \; ,
       \end{equation}
       where ${\rm DD}^{(\ell, m)}:= (2\ell+1)\, \sum_{ij} w_{ij}\, Y_{\ell m}^*\,(\theta_{ij}, \phi_{ij})\, \Theta (r_{ij}) $ and ${\rm RR} := 4\pi N \,\bar n_\mathrm{V}\, ( \rmax^3-\rmin^3)$ are the spherical harmonic decomposition of the weighted pairs inside the shell and the corresponding expected number of pairs in a random catalogue, respectively, if there were no survey boundaries. We note that if we neglect the wide angle effects in the survey (distant observer approximation), which consists in assuming a single line-of-sight across the survey, then each pair will contribute twice to the multipoles ($i$ and $j$ can be reversed). Instead, we could use the above equation to estimate odd-order multipoles without assuming that each pairs are contributing in a symetric way, in that case the angle $\theta_{ij}$ and $\phi_{ij}$ should be considered as being the angular coordinates of the separation vector $\vec r_{ij}$ in a local spherical frame centred on position $\vec x_i$. In such configuration we should double count each pair because they won't contribute in the same way to the multipoles. 

       Regarding terms $\hat I_2$ and $\hat I_3$ defined, respectively, as
       \begin{equation}
        \hat I_2 := \frac{2\ell+1}{v \bar n_\mathrm{V}} \int_v \frac{Y_{\ell m}^*}{V(\vec r)} \int_{V(\vec r)} n^{\rm o}(\vec x)\, \dif^3 x\;\dif^3 r \; , 
         \label{I2}
         \end{equation}
         and
         \begin{equation}
        \hat I_3 := \frac{2\ell+1}{v \bar n_\mathrm{V}} \int_v \frac{Y_{\ell m}^*}{V(\vec r)}\, \int_{V(\vec r)} n^{\rm o}(\vec x + \vec r)\; \dif^3 x\,\dif^3 r \; .
         \label{I2}
         \end{equation}
         Here, we notice that they are related one to another. Indeed, if we define the number of objects, $N_{\rm in}$, inside volume $V(\vec r)$, we can then compute it as
         \begin{equation}
           N_{\rm in} (\vec r) := \int_{V(\vec r)} n^{\rm o}(\vec x)\; \dif^3 x \; . 
           \label{nin}
           \end{equation}
           Introducing the window function $W(\vec x)$  ($1$ inside the survey and $0$ outside) we can show that
           \begin{equation}
             N_{\rm in} ( \vec r) = \int n(\vec x)\, W(\vec x)\, W(\vec x +\vec r)\; \dif^3 x \; , 
             \end{equation}
             where the integral is made over an infinite domain. Thus, we can change integration variables to $\vec x' = \vec x + \vec r$, such that
             \begin{equation}
             N_{\rm in} ( \vec r) = \int n(\vec x'-\vec r)\, W(\vec x' - \vec r)\, W(\vec x')\; \dif^3 x' \; . 
             \end{equation}
             As a result, we can express
             \begin{equation}
              \hat I_2 = \frac{2\ell+1}{v \bar n_\mathrm{V}} \int_v \frac{Y_{\ell m}^*}{V(\vec r)}\, N_{\rm in}( \vec r )\; \dif^3 r \; ,
             \end{equation}
             and
              \begin{equation}
              \hat I_3 = \frac{2\ell+1}{v \bar n_\mathrm{V}} \int_v \frac{Y_{\ell m}^*}{V(\vec r)}\, N_{\rm in}(-\vec r)\; \dif^3 r \; .
             \end{equation}
             Note that in the symmetric case of the plane parallel approximation, we can show that $\hat I_2 = \hat I_3$ because the sum is made over all possible orientation within the shell. 
             
             Be $f (\vec r)$ the average density inside the volume $V(\vec r)$, thus $f(\vec r) := N_{\rm in}(\vec r)/V(\vec r)$.
             We can express $\hat I_2$ and $\hat I_3$ with respect to their spherical harmonic decompositions,
             
             \begin{equation}
              \hat I_2 = \frac{3}{(\rmax^3-\rmin^3) \bar n_\mathrm{V}} \int_{\rmin}^{\rmax}r^2 f^{(\ell, m)}(r)\; \dif r \; . 
               \end{equation}
               Again this can be further simplified when considering a catalogue of objects. Indeed, we get
               \begin{equation}
               \hat I_2 = \frac{2\ell+1}{v \bar n_\mathrm{V}} \sum_{i=1}^N \int_v  \frac{Y_{\ell m}^*}{V(\vec r)} \int_{V} \dirac (\vec x- \vec x_i)\, W(\vec x - \vec r)\; \dif^3 x\, \dif^3 r \; ,  
              \end{equation}
              where we can compute the integral over $\vec r$ and $\vec x$. Then, defining
              \begin{equation}
              f_{\ell m}(\vec x) := \dfrac{(2\ell+1)}{v}\int_v w(\vec r)\, Y_{\ell m}^*\, W(\vec x - \vec r)\;\dif^3 r\; ,
              \end{equation}
              we obtain
              \begin{equation}
                \hat I_2 =  \frac{1}{N} \sum_{i=1}^N f_{\ell m}(\vec x_i) \; ,
                \label{hati2}
               \end{equation}
               which means that each point in the catalogue is contributing a certain amount to the various scales. And this contribution depends on the multipole considered. Once the weight function $f_{\ell m}$ is known, then it is a matter of computing a simple sum over the catalogue of objects.
               That term can indeed be related to the Data-Random term in the Landy \& Szalay estimator. After some calculations we can show that
               \begin{equation}
                   \hat I_2 = \dfrac{N_{\rm R}^2}{N_{\rm R}N}\dfrac{{\rm DR }^{(\ell, m)}}{{\rm RR}}\; .
               \end{equation} 
               As a result, computing the weight $f_{\ell m}$ allows us to extract in a very efficient way the cross counts between Data and Randoms. Finally, it is convenient to write the estimator of the multipoles of the 2PCF in a more standard way,
               
               \begin{equation}
               \hat \xi^{(\ell, m)} = \frac{{\rm DD}^{(\ell,m)}}{{\rm RR}} - 2\frac{{\rm DR }^{(\ell,m)}}{{\rm RR}} + \delta_{\ell 0}^{\rm K} \delta_{m0}^{\rm K} \; , 
               \label{xihatlm}
               \end{equation}
               where
               \begin{equation}
                 {\rm DR }^{(\ell, m)} := \bar n_\mathrm{V}\, (2\ell+1) \int_v N_{\rm in}(\vec r)\, w(\vec r)\, Y_{\ell m}^*\; \dif^3 r \; .
                 \end{equation}
                 The above expression can be directly computed from a single weighted sum over the catalogue as
                 \begin{equation}
                   {\rm DR }^{(\ell, m)} = \hat I_2\, {\rm RR} = 4\pi/3\; \bar n_\mathrm{V} (\rmax^3- \rmin^3)\, \frac{1}{N}\sum_{i=1}^Nf_{\ell m}(\vec x_i) \; .
                   \label{drterm}
                   \end{equation}
                   In addition, we can further re-write the estimator as
                   \begin{equation}
                   \hat \xi^{(\ell, m)} = \hat \xi^{(\ell, m)}_{\rm N} - 2\,\hat \xi_{\rm X}^{(\ell, m)} \; , 
                     \end{equation}
                     where $\hat \xi^{(\ell, m)}_{\rm N}$ is the natural estimator for the auto-correlation and $\hat \xi_{\rm X}^{(\ell, m)}$ is the natural estimator for the cross-correlation between Data and Randoms. Indeed,
                     \begin{equation}
                       \hat \xi^{(\ell, m)}_{\rm N} := \frac{{\rm DD}^{(\ell, m)}}{{\rm RR}} - \delta_{\ell 0}^{\rm K}\delta_{m0}^{\rm K} \; ,
                       \label{natural}
                       \end{equation}
                       and
                       \begin{equation}
                         \hat \xi^{(\ell, m)}_{\rm X} := \frac{{\rm DR }^{(\ell, m)}}{{\rm RR}} - \delta_{\ell 0}^{\rm K}\delta_{m0}^{\rm K} \; .
                         \label{cross}
                       \end{equation}
                       
                       We can notice that despite the fact that the estimator in Eq.~\eqref{xihatlm} has a very similar form as the Landy \& Szalay estimator, there is an interesting fundamental difference in the way the multipoles are estimated. Indeed, usually, to compute the multipoles of the Legendre expansion of the 2PCF we can define a binning in separation $r$ (same as we do) and an additional binning in $\mu = \cos(\theta)$ needs to be set up. The general problem in setting that binning is that it has to be thin enough to evaluate the angular integration without introducing too much errors but it cannot be too thin because the number of random pairs needs to be large enough in order to get a meaningful value that will not propagate noise in the angular integration. With the estimator in Eq.~\eqref{xihatlm}, we do not need to introduce an angular binning. Instead, we use the distribution of the pairs themselves to estimate the angular average as we would do in a Monte Carlo integration. We note that when evaluating the odd multipoles by dropping the distant observer approximation, it is necessary to keep both $\hat I_2$ and $\hat I_3$ which will no longer be equal.

\section{Specific geometries}
\label{appendix:geom}

We show how to compute the relevant quantities $V(\vec r)=Vg(\vec r)$ for the Data-Data pairs and $f_{\ell m}(
\vec x)$ for the Data-Random pairs necessary to apply our estimator of the 2PCF in the case of the sphere and the cone in the plane-parallel approximation. 

\subsection{Sphere}

Let us start with the definition 
\begin{equation}
g(\vec r) := \frac{1}{V} \int W( \vec x )\, W(\vec x + \vec r)\; \dif^3 x \; .
\label{gfunc}
\end{equation}
This is the volume of the intersection between two spheres of radius $R$ separated by the vector $\vec r$ normalised with respect to the volume of the sphere. This means that it depends only on the modulus of $r=|\vec r|$, the separation vector. Thus, only a monopole $g^{(0)}(r)$ is expected. This function can be computed either by directly integrating in Eq.~\eqref{gfunc} or by going to Fourier space. It leads anyway to 
\begin{equation}
g^{(0)}(r) = 1 - \frac{3}{4} \frac{r}{R} + \frac{1}{2}\left (\frac{r}{2R} \right )^3
\end{equation}
if $r<2R$ and $g^{(0)}(r)=0$ else. This condition is trivially explained by the fact that if the distance between the two spheres of radius $R$ is greater than $2R$ then their intersection is empty.

This is enough to compute the natural estimator (i.e. Eq.~\ref{natural}). However, when computing the Landy \& Szalay estimator, the next elements that are required are the weights that need to be applied when computing the Data-Random (see Eq.~\ref{drterm}). That weighting function should depend on the considered position $\vec x_i$ of each individual data point, on the considered multipole and on the separation. When considering the monopole it is straightforward (from consideration about the spherical symmetry of the problem) to show that the weights to be used are depending only on the distance $x$ of the data point from the centre of the sphere. In general,
\begin{equation}
 f_R^{(\ell m)}(\vec x) = \frac{2\ell+1}{v}\int_v w(\vec r)\, W(\vec x - \vec r)\, Y_{\ell m}^*(\theta, \phi)\;\dif^3 r\; ,                 
 \label{flm}
\end{equation}
where $\theta$ and $\phi$ are the angular coordinates of the separation vector $\vec r$. In practice it is more convenient to split the angular summation from the radial one using the auxiliary function $h^{(\ell m)}(\vec x, r)$ defined such that
\begin{equation}
 f_R^{(\ell m)}(\vec x) = \frac{3}{\rmax^3-\rmin^3}\int_{\rmin}^{\rmax} r^2\, h^{(\ell m)}(\vec x, r)\; \dif r \; .           
 \label{flmb}
\end{equation}
This way Eq.~\eqref{flmb} can be evaluated with a great accuracy from a Gauss--Legendre quadrature of order $N_{\rm GL}$ as soon as the spherical expansion $h^{(\ell m)}$ is know for each data point at position $\vec x$. From the above definition we can express $h^{(\ell m)}$ as
\begin{equation}
 h^{(\ell m)}(\vec x, r) = \frac{2\ell+1}{4\pi}\int_\Omega w(\vec r)\, W(\vec x - \vec r)\,  Y_{\ell m}^*(\theta, \phi)\; \dif^2\Omega \; ,             
\label{fzero}
\end{equation}
which might not always (i.e. for any window function) be computed analytically. However, in the case of a sphere, since the weighting function $w$ depends only on the modulus $r$ of the separation, it reduces to compute the spherical harmonic expansion $W^{(\ell m)}$ of the window function $W$ but shifted by vector $\vec x$. Thus, $h^{(\ell m)} = w(r)\, W^{(\ell m)}$, for which the monopole will depend only on the distance, $x$, from the data point to the centre of the sphere. This can be  calculated in a straightforward manner via

\begin{equation}
W^{(0)}(x,r) = \left \{ 
 \begin{array}{cll}
   1 & {\rm if} & x +r \leq R\;, \\
    0 & {\rm if} & |x-r| \geq R\;, \\
   \dfrac{1}{4xr}\,\left [ R^2 - (r-x)^2 \right ] & {\rm otherwise}\;.                               
 \end{array}
 \right .  
 \end{equation}
 In the case of the spherical window function, we can also predict the needed weight for any spherical harmonics of the Data-Random. By defining $\nu:=(R^2-x^2-r^2)(2xr)^{-1}$, we can show that for $\ell \neq 0$,
 
 \begin{equation}
W^{(\ell m)}(\vec x, r) = \left \{ 
  \begin{array}{cll}
    Y_{\ell m}^*(\theta_x, \phi_x)\, \dfrac{1}{2}\,\left \{ \lcal_{\ell+1} (\nu) - \lcal_{\ell-1}(\nu) \right \}& {\rm if} \, |\nu| < 1\;, \\        
   0 & {\rm else}\;,                                                                                   
 \end{array}
 \right .  
 \end{equation}
 where $x$, $\theta_x$, and $\phi_x$ are the spherical coordinates of the point $\vec x$. 
In Fig.~\ref{fig:dr} we show how the expected ${\rm DR}^{(\ell)}(r)$ compare with estimated one from a random catalogue. This is done for two different realisations of the data catalogues and we obtain a good agreement between the expectation and the estimation.
\begin{figure*}
\vspace{-2.5cm}
\centering
\hspace*{-0.8cm}
\includegraphics[width=74mm]{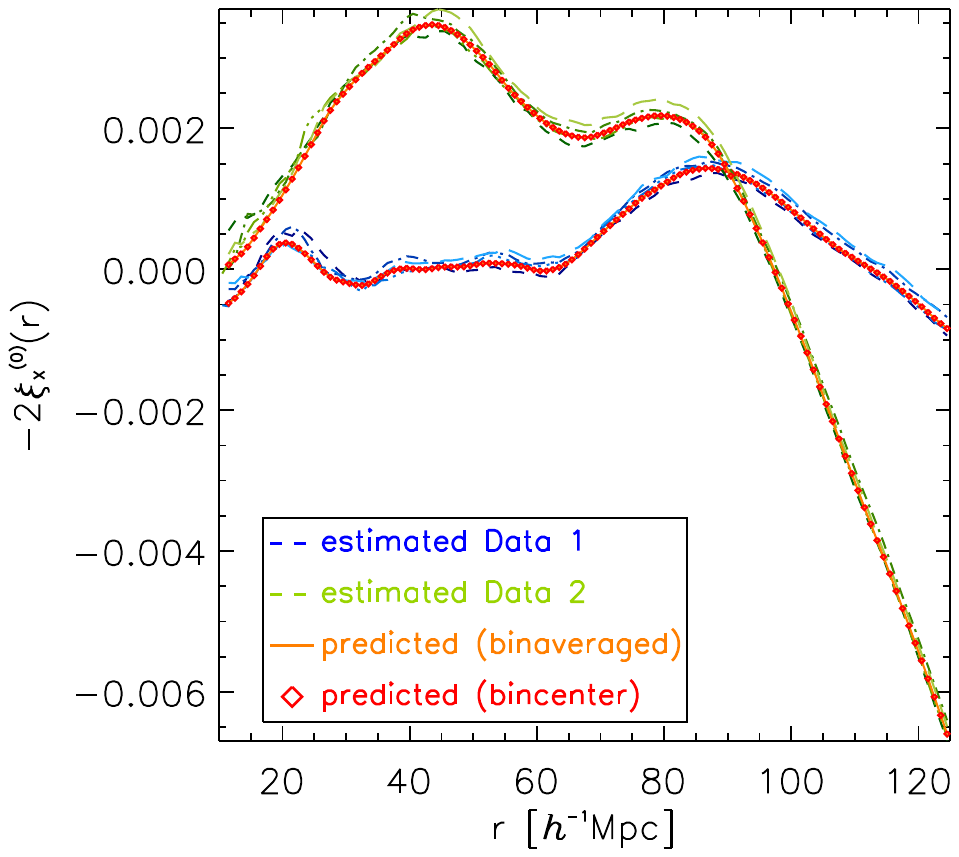}
\hspace*{-1.7cm}
\includegraphics[width=74mm]{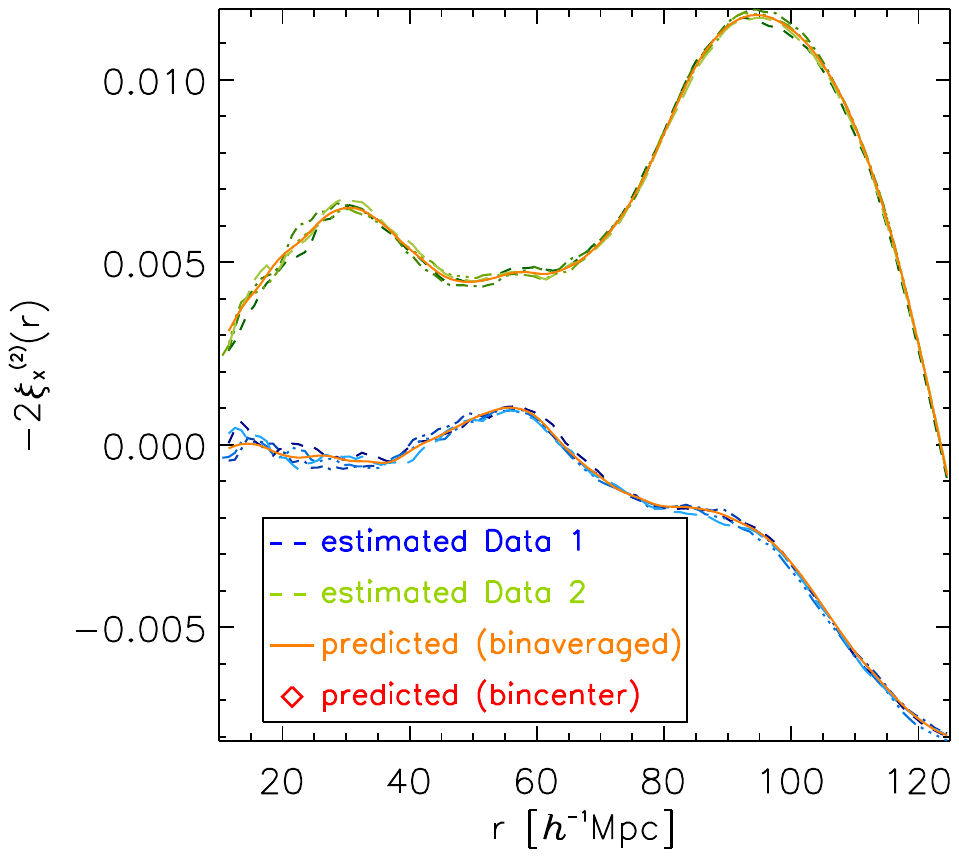}
\hspace*{-1.8cm}
\includegraphics[width=74mm]{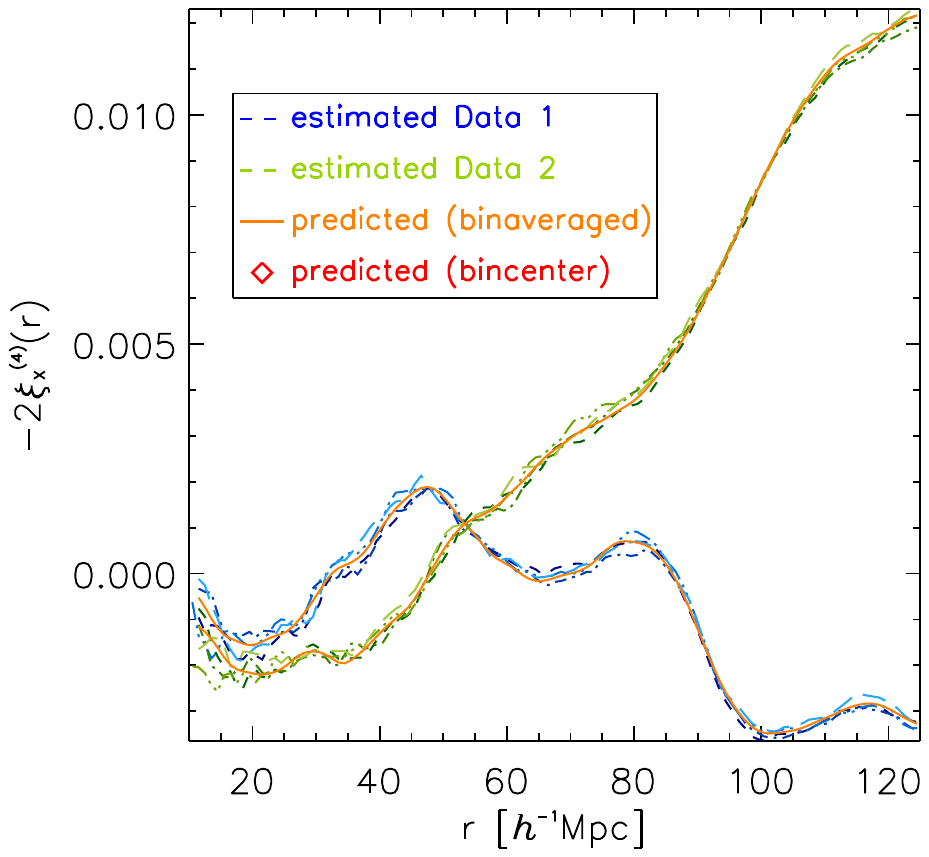}
\vspace*{-2.6cm}
\caption{\small Data-Random counts. {\it Left:} Monopole cross-correlation ($\xi_{\rm X}^{(0)}={\rm DR }^{(0)}/{\rm RR}-1$) between data and random catalogues.  The dashed lines correspond to four different random catalogues.  The shades of green and blue correspond to two different data catalogues. The red symbols show the prediction with the weighting method evaluated at the centre of each separation shell, the orange lines show the average (with a Gauss--Legendre quadrature) over each separation shell. {\it Middle:} Quadrupole cross-correlation ($\xi_{\rm X}^{(2)}={\rm DR }^{(2)}/{\rm RR })$ between Data and Random. {\it Right:} Hexadecapol cross-correlation ($\xi_{\rm X}^{(4)}={\rm DR }^{(4)}/{\rm RR })$ between Data and Random.} 
\label{fig:dr}
\end{figure*}

\subsection{Cone}

As in the case of the sphere it is necessary to evaluate the intersection volume between two identical cones of aperture $\alpha$ and truncated between $R_{\rm min}$ and $R_{\rm max}$ from the vertex of the cone. The separation vector $\vec r$ is applied between the vertices of the two cones and the axis $\vec e_{\rm z}$ of the cones remain parallel. Thus the axial symmetry is preserved, such that $g(\vec r) = g(r, \mu)$ where $\mu = \vec r\cdot \vec e_{\rm z}/r$.

Depending on $r$ and $\mu$ there are several cases that can appear but within all those cases the volume of the intersection can be computed by evaluating an integral,
\begin{equation}
V(r, \mu) = \int_{z_{\rm a}}^{z_{\rm b}} \scal(z)\;\dif z\, , 
\label{vol_cone}
\end{equation}
where the function $\scal(z)$ represents the surface of the intersection between two disks of radius $r_1$ and $r_2$ separated by distance $\lambda$. The distance between the two disks is fixed by the distance between the two axes of the cones, thus $\lambda = r\sqrt{1-\mu^2}$ and does not depend on the integration variable $z$. Instead, both $r_1$ and $r_2$ will be functions of $r$, $\mu$ and $z$. The disk intersection can be computed straightforwardly according to various cases
\begin{equation}
\scal_{12} = \left \{
\begin{array}{lcl}
\pi \; {\rm min}[r_1^2, r_2^2] & {\rm if } & \lambda < |r_1-r_2|\; , \\
\theta_1 r_1^2 + \theta_2 r_2^2 - \lambda r_2 \sin(\theta_2) & {\rm if} & |r_1-r_2| \leq \lambda \leq r_1+r_2\; ,\\
0 & {\rm if} & \lambda > r_1+r_2\; ,
\end{array}
\right .  
\label{disk}
\end{equation}
where $\cos\theta_1 := (\lambda^2 + r_2^2-r_1^2)(2\lambda r_2)^{-1}$ and $\cos\theta_2 := (\lambda^2 + r_1^2-r_2^2)(2\lambda r_1)^{-1}$. Notice that, despite the fact that Eq.~\eqref{disk} looks asymmetric in $r_1$ and $r_2$, since $r_1\sin\theta_1 = r_2\sin\theta_2$ it turns out that the symmetry is preserved. In practice, there are some cases for which we need to evaluate the intersection between two rings of external radii $r_1$ and $r_2$ and internal radii $r_1'$ and $r_2'$. In such cases the full intersection can be trivially obtained as $\scal = \scal_{12} - \scal_{1'2}-\scal_{12'}+\scal_{1'2'}$. 

The second important step is to be able to compute the cross-counts between the data catalogue and a random one. In order to maintain the numerical efficiency, this needs to be done without explicitly computing them using a random catalogue.
The idea is to use Eqs.~\eqref{hati2} and \eqref{drterm} in order to compute the ${\rm DR }^{(\ell m)}/{\rm RR }$ term by running a simple sum over the data catalogue. However that would require to compute the explicit expression of the weight $f_{\ell m}(\vec x_i)$ which needs to be applied at each object position $\vec x_i$ in the data catalogue.
Indeed,
\begin{equation}
f_{\ell m}(\vec x) := \frac{2\ell+1}{v}\int_v w(\vec r)\, Y_{\ell m}^*\, W(\vec x - \vec r)\;\dif^3 r \; , 
  \label{flm}
\end{equation}
which can be written in terms of the radial function $h^{(\ell)}(r, \vec x_i):= h^{(\ell\, 0)}(r, \vec x_i)$ as
\begin{equation}
  f_\ell(\vec x_i) = \frac{3}{\rmax^3-\rmin^3}\int_{\rmin}^{\rmax}r^2\, h^{(\ell)}(r, \vec x_i)\; \dif r \; , 
\end{equation}
where we focus on evaluating $m=0$ multipoles and even values of $\ell$. Thus,

\begin{equation}
  h^{(\ell)}(r, \vec x_i) = \frac{2\ell+1}{4\pi}(-1)^\ell\int_{\Omega}w(\vec r)\, \lcal_\ell(\mu)\, W(\vec x + \vec r)\; \dif^2\Omega \; ,
  \label{hl}
  \end{equation}
  where we exchange $\vec r$ into $-\vec r$. This means that the integrand of Eq.~\eqref{hl} can be null (due to the function $W$) if the vector $\vec y := \vec x +\vec r$ is outside the observed cone. This is also allowing us to compute the integral over the angle, $\phi$, which will take different values according to position, $\vec x_i$, on the considered data point.
  Let us define
  \begin{equation}
   G (r, \mu, x, \theta_x):=  \int_{0}^{2\pi}W(\vec y)\; \dif \phi \; ,
  \end{equation}
  where $r:= |\vec r|$, $\mu := \cos\theta$, $x:= |\vec x|$ and $\theta_x$ is the azimutal polar angle corresponding to vector $\vec x$. Given the simplicity of the window function $W$ which is unity if $\vec y$ is inside the cone-like geometry, we can say that there are three conditions to be satisfied,
  
  \begin{eqnarray}
    y &\le & R_{{\rm max}}  \label{c1}\;,\\
    y & \ge & R_{{\rm min}} \label{c2}\;,\\
    \cos\theta_y & \ge & \cos\alpha\;, \label{c3}
    \end{eqnarray}
    where $\theta_y$ is the azimuthal polar angle of vector $\vec y$ and $y:= |\vec y|$. In the following, we define $\nu := \cos\alpha$. Conditions~\eqref{c1} and \eqref{c2} can be expressed in terms of restrictions on the angle $\phi$.  Indeed, $y^2 = x^2 + y^2 + 2xr\cos\gamma$, where $\gamma$ is the angle between $\vec x$ and $\vec r$. Thus $\cos\gamma = \sin\theta_x\,\sin\theta\,\cos(\phi-\phi_x) + \cos\theta_x\,\cos\theta$.
    
    Introducing the auxiliary functions
        \begin{eqnarray}
          \lambda_{\max} & := & \frac{R_{\max}^2 - x^2 - r^2}{2xr}\; , \\
          \lambda_{\min} & := & \frac{R_{\min}^2 - x^2 - r^2}{2xr} \; ,
          \end{eqnarray}
          we can show that conditions \eqref{c1} and \eqref{c2} imply that
          \begin{equation}
          \frac{\lambda_{\min}-\mu_x\,\mu}{\sin\theta\,\sin\theta_x} \le \cos(\phi-\phi_x) \le \frac{\lambda_{\max}-\mu_x\,\mu}{\sin\theta\,\sin\theta_x} \; .
            \label{cophi}
          \end{equation}
          
          Introducing the functions
          \begin{eqnarray}
            l_{\min} & := & \frac{\lambda_{\min}-\mu_x\,\mu}{\sin\theta\,\sin\theta_x}\; , \\
            l_{\max} & := & \frac{\lambda_{\max}-\mu_x\,\mu}{\sin\theta\,\sin\theta_x} \; ,
          \end{eqnarray}
          we can see that depending on the values of $l_{\min}$ and $l_{\max}$, we will get different values for $G$, namely,
          %\begin{eqnarray}
          \begin{equation}
          \begin{array}{cccl}
            l_{\min}<-1 &  {\rm and} & l_{\max} > 1 & G=  2\pi \nonumber\;, \\
            l_{\min} > 1 &  {\rm or} &   l_{\max} < -1 & G =  0 \nonumber\;, \\
            l_{\min}\le -1 &  {\rm and} &  |l_{\max}| \le 1 & G =  2\left [ \pi - \acos(l_{\max})\right ]\nonumber\;,  \\
            |l_{\min}|\le 1 &  {\rm and}  &  l_{\max}\ge 1 & G = 2\,\acos(l_{\min}) \nonumber\;, \\
             |l_{\min}|\le 1 &  {\rm and} &  |l_{\max}| \le 1 & G =  2\left [ \acos(l_{\min}) - \acos(l_{\max}) \right ]. \nonumber
            \end{array}
            \end{equation}
            %\end{eqnarray}
            %
            However, we must also pay attention to condition \eqref{c3}, which can be written as
            \begin{equation}
              y \le \frac{x\cos\theta_x+r\cos\theta}{\cos\alpha} \; . 
              \end{equation}
              This allows us to introduce a new function,
              \begin{equation}
                  l_{\max}^*:= \frac{\lambda\left(\dfrac{x\mu_x+r\mu}{\nu}\right) - \mu\mu_x}{\sin\theta\,\sin\theta_x}\, ,
              \end{equation}
              where $\lambda(u) := (u^2-x^2-r^2)(2xr)^{-1}$. That means that in the above set of conditions, we need to substitute $l_{\max}$ by $l_{\max}^*$ if $l_{\max}^*< l_{\max}$.  In addition, if $l_{\min} < l_{\max}^*$ then $G=0$ because contrary to the fact that $l_{\min} < l_{\max}$ is guaranteed by the fact that $R_{\min}< R_{\max}$ there is no guaranty that the condition $l_{\min} > l_{\max}^*$ is satisfied.

              From those various cases, we can guess that despite the fact that $G$ will be continuous, it will be smooth only on certain segments of intervals. This prevents us from using a Gauss--Legendre quadrature or any other usual integration methods for computing the weights for each single object positions $\vec x_i$. However, we can guess that if the sum over objects is made before performing the integral in $\mu$, the resulting function will appear much smoother. This piece-wise continuous function will be added randomly over the survey geometry.  Thus, we can introduce the function,
              \begin{equation}
                  \gcal(r,\mu) := \frac{1}{N}\sum_{i=1}^N G(r, \mu, x_i,\theta_{x_i} )\; ,
              \end{equation}
              and formally express the expected normalised $\rm{DR}^{(\ell\, 0)}$ counts as
              \begin{equation}
                \hat I_2 = \frac{3}{\rmax^3-\rmin^3}\int_{\rmin}^{\rmax}\!\! r^2\, \underbrace{\frac{2\ell+1}{4\pi}(-1)^\ell \int_{-1}^1\!\!\lcal_\ell(\mu)\, \gcal(r, \mu)\; \dif \mu}_{:= S(r)}\,\dif r \; . 
                \end{equation}
                Then, we can evaluate separately the angular and radial integrals with Gauss--Legendre quadrature of order $N_{\rm GL}$  (with weights $w_j^{\rm GL}$) such that
                \begin{equation}
                  S(r) \simeq \frac{2\ell+1}{4\pi} (-1)^\ell\,\sum_{j=1}^{N_{\rm GL}} w_j^{\rm GL}\, w(r, \mu_j)\, \lcal_\ell(\mu_j)\,\gcal(r,\mu_j) \; , 
                  \end{equation}
                  and
                  \begin{equation}
                    \hat I_2 \simeq \frac{3/2}{(\rmax^2+\rmin)\,(\rmax+\rmin^2)} \sum_{k=1}^{N_{\rm GL}} w_k^{\rm GL}\,r_k^2\, S(r_k) \; ,
                    \end{equation}
                    where $r_k:= (\rmax-\rmin)\, \mu_k/2 + (\rmax+\rmin)\, \mu_k/2$ and $\mu_j$ (or $\mu_k$) are the zeros of the Legendre polynomials of order $N_{\rm GL}$. Note that the order $N_{\rm GL}$ does not need to be the same for the two integrations. Indeed, given that the radial integration is performed over thin bins, we can choose an order of $1$ or $3$ at maximum (noting that this had no effect in the spherical case). However, given the range in $\mu$ is large (between $-1$ and $1$) we use $N_{\rm GL} = 15$ which means that if the behaviour of the function to be integrated is captured by a polynomial of order $30$ then the integral is exact. Indeed, when increasing the number of objects to infinity we expect $w(\vec r)\, \gcal(r,\mu)\sim 1$ which means that we are integrating functions which are at most polynomials of order $4$, thus a Gauss--Legendre quadrature of order $15$ is more than enough.
                    
                    As in the spherical case, we made an explicit test to check our ability in predicting the cross-counts between data and randoms. To do so we computed the data-random count explicitly for two different data catalogues and each one with four different random catalogues. Each random catalogue contains ten times more object then in the data catalogue.
                    The result of that test is very similar to the spherical case previously described. We are able to evaluate efficiently the ${\rm DR}^{(\ell)}:={\rm DR}^{(\ell, 0)}$ term for the the monopole, quadrupole, and hexadecapole. 

\section{\covmos\ clones of the ELM}
\label{appendix:pinoc}

In this appendix, we review how we generate a series of $10\, 000$ \covmos\ realisations mimicking the power spectrum and one-point PDF of a set of $1000$ ELM realisations. As explained in \citet{baratta_19} and \citet{baratta_22} the required quantities to generate redshift-space realisations are the real-space power spectrum, the corresponding one-point PDF, the power spectrum of the divergence of the velocity field, the density-velocity dispersion relation, and the one-point velocity dispersion. In order to provide meaningful quantities we would need to evaluate them directly from the ELM catalogues themselves. However, both the power spectrum of the divergence of the velocity and the density-velocity dispersion relations are referring to the dark matter field which we do not have in the ELM realisations. Thus, those quantities ought to be evaluated following \citet{baratta_22} which uses the fitting procedure set by \citet{bel_19}. Instead, the power spectrum can be directly measured from the $1000$ ELM realisations. However, since the measurement are done in a cone-like geometry we can only measure the power spectrum convolved with the corresponding window function.

The galaxy monopole power spectrum can be estimated in the ELM mock in the first redshift bin with edges $0.9<z<1.1$ in real space (without applying RSD). The $30\degree$ cone consists of a cone with minimum comoving distance $R_{\rm min} = 2114\, \Mpc$ and a maximum comoving distance $R_{\rm max} = 2447\, \Mpc$ with a total aperture angle $2\alpha = \pi/3$. We keep only galaxies with ${\rm H}_\alpha > 2\times 10^{-16}$ erg$.$cm$^{-2}$ in our sample (where ${\rm H}_\alpha$ is the flux). We estimate the power spectrum in a bounding  box of size $L=3500\, \Mpc$ on a $512^3$ mesh. From each individual realisation, we apply on the catalogue a piece-wise continuous spline (PCS) mass assignment scheme in order to get a density on the mesh. We then Fourier transform the density grid and apply the interlacing technic to reduce the aliasing contribution and estimate the power spectrum in $256$ bin of $k$. We do not include IC because we estimate the mean number density in the catalogue over the $1000$ realisations. Thus, given the true power spectrum $P_{\vec k}$ we expect the measured power spectrum $\tilde P_{\vec k}$ with the cone-like geometry to be given by 
\begin{equation}
\tilde P_{\vec k} = \int P_{\vec k'}\, |W_{\vec k-\vec k'}|^2\; \dif^3 k' \; ,
\label{convolution}
\end{equation}
where $W_{\vec k}$ is the FT of the window function. There exist at least three possibilities to perform the convolution in Eq.~\eqref{convolution}. We can simply go to the configuration space, where the convolution becomes a simple product. This can be done, either in three dimensions or integrating the angular part and performing a Hankel transform in one dimension. Finally, the last method involves the computation of a mixing kernel, allowing us to compute in Fourier space the monopole power spectrum of the convolved power spectrum. After some tests with \covmos\ for which we know the true power spectrum and can apply a cone-like geometry we decided to use the least efficient in terms of execution speed but the most accurate method which is to apply the convolution as a product in the 3D configuration space \citep[see][]{sato_2011,rota_17}. 
Thus, the necessary steps are as follows: we put the true power spectrum $P_{\vec k}$ on a 3D Fourier grid and inverse Fourier transform it with an FFT to obtain the corresponding 3D 2PCF $\xi(\vec r)$. We can compute, for each grid point, the corresponding expected 2PCF affected by the window as $\tilde \xi(\vec r) = \xi(\vec r)\,g(\vec r)$. Then, we can apply an FFT in order to evaluate the convolved 3D power spectrum, $\tilde P_{\vec k}$, on which we could finally apply the same $k$-binning as in the measurements, leading to the monopole power spectrum, $\tilde P^{(0)}(k)$. 

Thus, with a given $\tilde P^{(0)}(k)$, we want to recover $P^{(0)}(k)$ in Eq.~\eqref{convolution}. To do so, we use the Rychardson--Lucy deconvolution method originally used to recover the shot noise corrected one-point PDF in count-in-cell analysis \citep{szapudi_pan_04, bel_16}. The idea is to start with an initial guess for the true power spectrum $P_0(k)$, then apply the convolution to it to obtain the corresponding $\tilde P_0^{(0)}(k)$, and compare it with the measured power spectrum $\tilde P^{(0)}(k)$ with a basic $\chi^2$ defined as 
\begin{equation}
\chi_0^2 := \dfrac{\sum_i \left[\tilde P_0^{(0)}(k_i) - \tilde P^{(0)}(k_i) \right]^2}{ \left[ \tilde P^{(0)} (k_i)\right]^2} \; .
\label{chi}
\end{equation}
This represents the starting point of this iterative method. Then, we can compute the ratio $R_0:= P_0(k)/\tilde P_0^{(0)}(k)$ and apply it to get a new guess for the step $n+1$ as $P_{n+1}(k) = R_{n}\, \tilde P^{(0)}(k)$. This process needs to be iterated enough times to obtain a good agreement between $\tilde P_n^{(0)}(k)$ and $\tilde P^{(0)}(k)$ quantified by a low $\chi^2$. The known caveat of this kind of process is that it is better to avoid too many iterations, which risks to cause non-physical behaviour in the deconvolved power spectrum. That is  why it is crucial to start with a good guess. To do so, we use the linear power spectrum and compute its convolved version to deduce the initial ratio $R$ such that $P_0(k) = R\, \tilde P^{(0)}(k)$. Indeed, since we expect the linear matter power spectrum to have the same behaviour as the galaxy power spectrum at low $k$ (where the galaxy bias is linear) and given that the convolution mostly affects low $k$-modes, we expect the effect of the convolution to be well captured by the initial ratio $R$. In practice with $n=4$ iterations we obtain a very good agreement between the measured $\tilde P^{(0)}(k)$ monopole and the predicted one $\tilde P_n^{(0)}(k)$ which indicates that $P_n^{(0)}(k)$ is close to being the deconvolved power spectrum. We basically stop the number of iterations when the $\vec k=\vec 0$ mode $P_{\vec 0}$ estimated and predicted agree within $1\,\sigma$. This corresponds to a decrease in $\chi^2$ of $0.2$\%. The final value of $\chi^2$ is $0.00041$ indicating a very good overall agreement. 
\begin{figure}
\vspace{-3.5cm}
\hspace{-1.cm}
\includegraphics[width=1.1\linewidth]{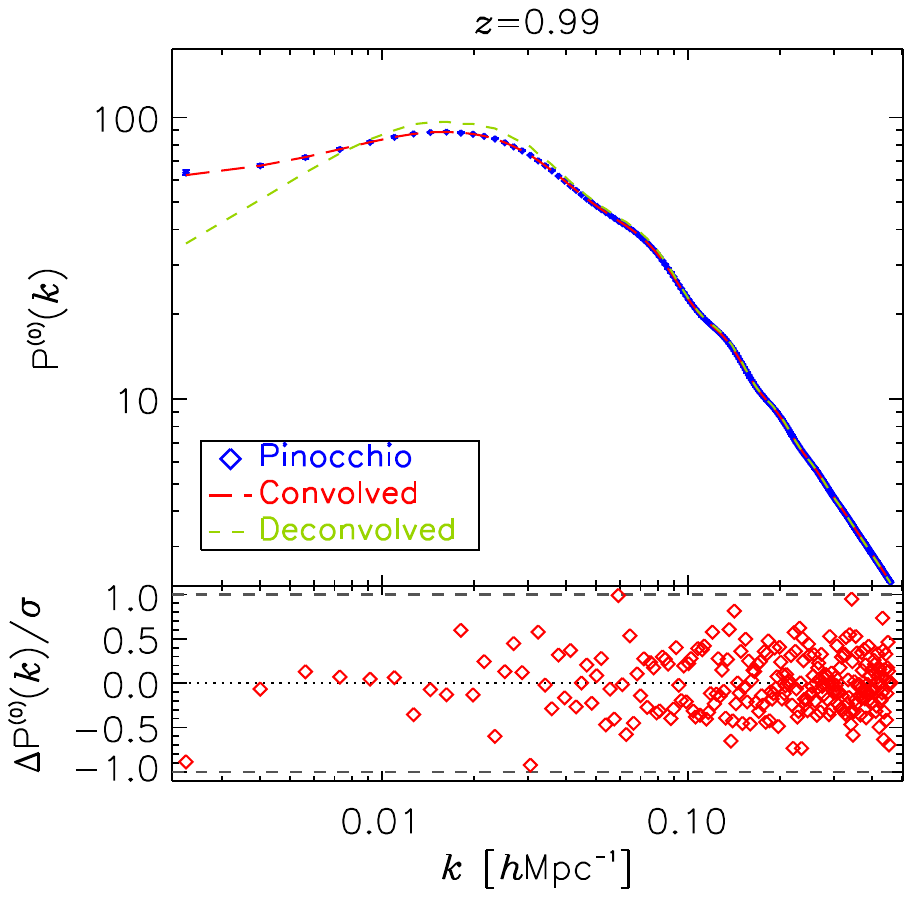}
\vspace{-3.8cm}
\caption{{\it Top:} Comparison between the monopole power spectrum in real space as measured in the $1000$ ELM (\texttt{Pinocchio}) and the convolved power spectrum obtained from our estimation of the deconvolved power spectrum. {\it Bottom:} Absolute difference between the measured power spectrum and the convolved one with respect to the error estimated from the mock catalogues.} 
\label{fig:deconvolution}
\end{figure}

In Fig.~\ref{fig:deconvolution} we show the original measured power spectrum monopole in the ELM catalogues at $z=0.99$ (i.e. the first redshift bin) together with our estimation of the deconvolved power spectrum. In the bottom panel, we show the agreement between the deconvolved power spectrum and the measured power spectrum once it is convolved.

In the following, we decribe how we evaluate the one-point PDF in the ELM catalogues. The most trivial way of estimating the one-point PDF would be to apply a count-in-cell algorithm on each catalogue and estimate the density contrast $\delta_N := N/\bar N - 1$ in each cell, where $N$ is the number of objects counted in the given cell and $\bar N$ is the mean number of objects per cell. We can get a very accurate estimate of the density PDF provided that the mean number of object is high, $\bar N > 20$. However, the mean number density in the catalogue is $\bar n = 0.134 \times 10^{-2}\, (\Mpc)^{-3}$.  Thus, on the scale $a=L/N \simeq 7\, \Mpc$ on which we need to reconstruct the PDF this would lead to an extremely low mean number of objects per cell $\bar N \simeq 0.4$. 
\begin{figure*}
\vspace{-2.5cm}
\centering
\hspace*{-1.1cm}
\includegraphics[width=76mm]{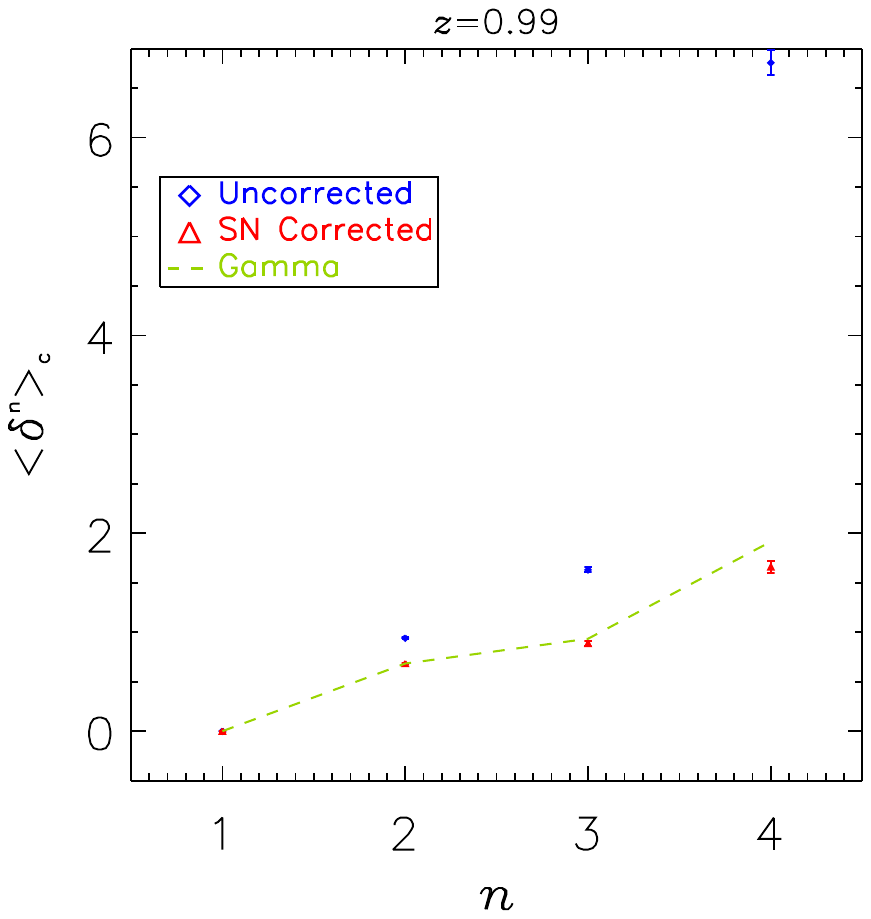}
\hspace*{-1.9cm}
\includegraphics[width=76mm]{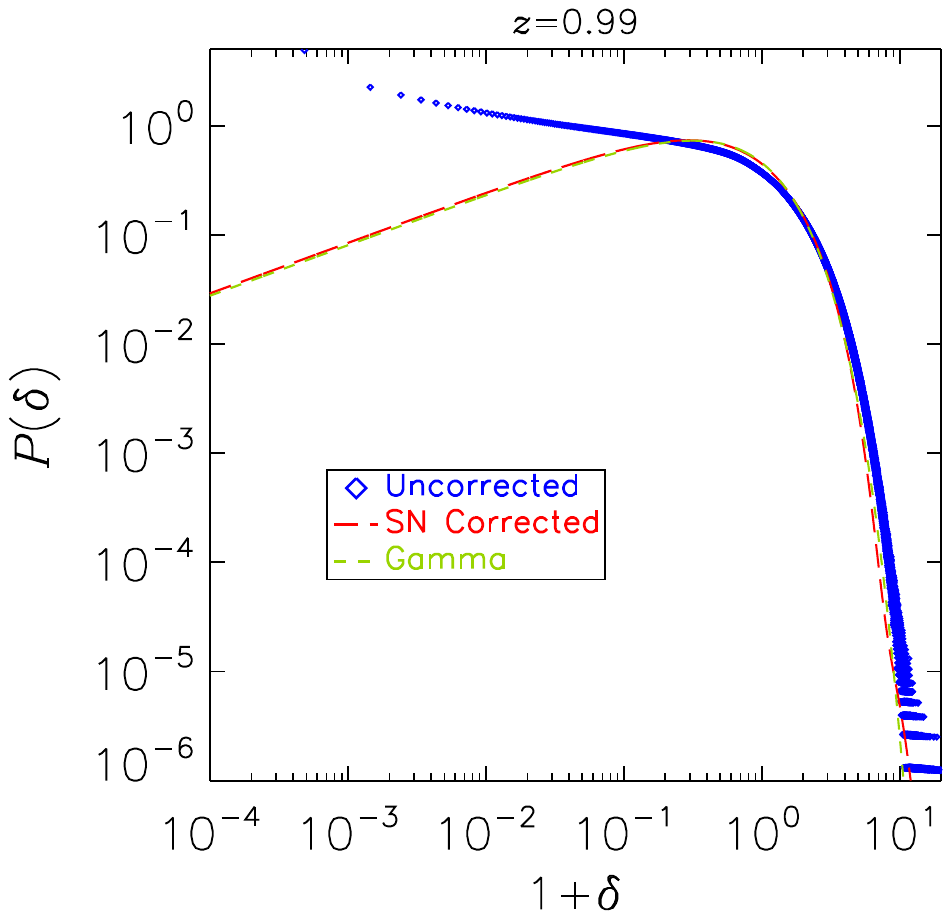}
\hspace*{-2.cm}
\includegraphics[width=76mm]{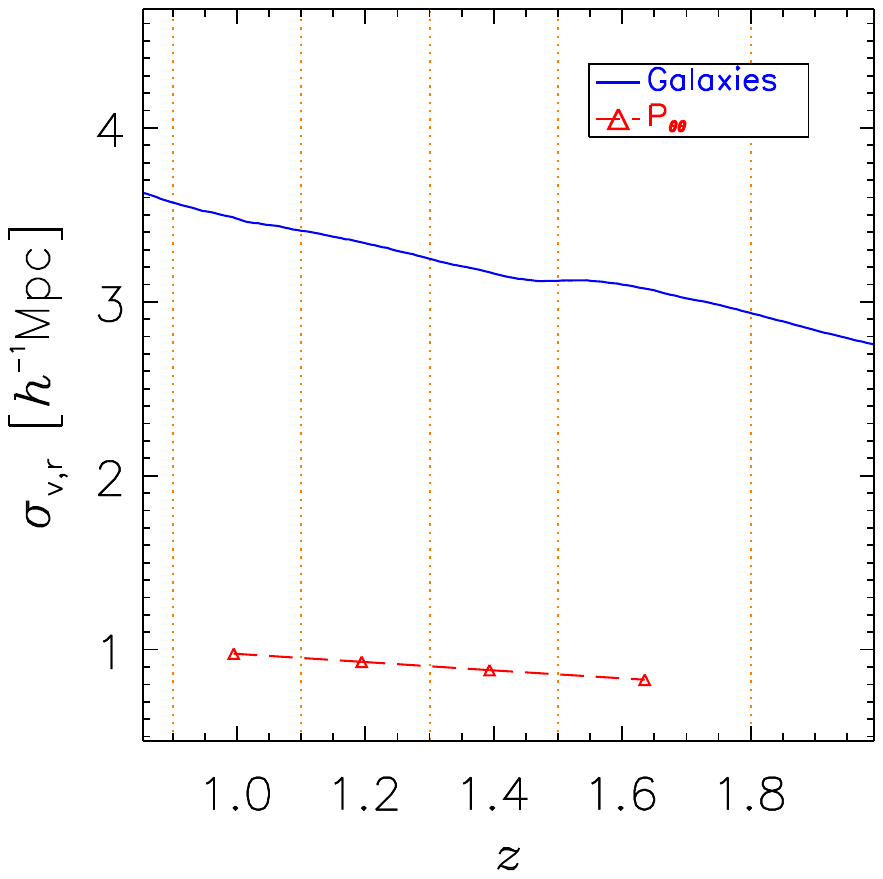}
\vspace*{-2.6cm}
\caption{ {\it Left:} Cumulant moments estimated in $200$ ELM realisations at $z=0.99$ without (blue diamonds) and with (red triangles) shot noise correction. The green dashed line shows the expected cumulant moments of a Gamma distribution which is matching the second-order ($n=2$) shot noise corrected cumulant moment. {\it Middle:} Probability density obtained after shot noise correction of the moments opposed to the measured PDF (blue diamonds).  {\it  Right:} Normalised velocity dispersion of galaxies in the ELM compared to the expected dispersion coming from the divergence of the large scale velocity field.} 
\label{fig:cumulent}
\end{figure*}

One possible way of reducing the effect of the shot noise is to use a higher-order interpolation scheme such as the PCS -- that we actually use to evaluate the power spectrum. It would reduce the shot noise contribution to the one we would obtain with ten times the mean number of objects per cell, but this is still too low. In addition, if we can directly map the true PDF into count-in-cell probability, it is not possible to link straightforwardly the true PDF to the PDF of the PCS density field. In turn, it is much easier to relate the moments of the PCS density field to the moments of the true underlying field, through standard shot noise correction. 

Let us describe how we proceed. Despite being fairly intuitive, those results have, to our knowledge, never been presented in literature yet. In the present derivation we focus only on one-point moments of a point process which have been split on a regular grid with a given mass assignment scheme $F$, and we define
\begin{equation}
\delta_F(\vec x) := \int \delta(\vec x')\, F(\vec x - \vec x')\; \dif^3 x' \; ,
\label{mas}
\end{equation}
where $F$ is normalised in such a way that $\int F(\vec x)\;\dif^3 x = 1$. As a result, we can also define $f := a^3 F$ as a dimensionless function. From Eq.~\eqref{mas} it is immediate to express the corresponding relation for the one-point cumulant moments of order $N$ of the field $\delta_F$ as
\begin{equation}
\langle \delta_F^N\rangle_\mathrm{c} = \int \dif^3  x_1\, ... \dif^3  x_N\; \xi_N(\vec x_1, ..., \vec x_N)\; F(\vec x_1)\,...\,F(\vec x_N) \; ,
\label{cumu}
\end{equation}
which has already been presented widely in literature \citep{bernardeau_02}, where $\xi_N$ is referring to the $N$-point correlation function. Since the $N$-point correlation function is affected by shot noise  we expect the cumulant moments $\langle\delta_F\rangle_\mathrm{c}$ to be modified too. Indeed, the shot noise contribution $\xi_N^{\,({\rm SN})}$ to the $N$-point correlation function is expected to arise each time there is a zero separation between any two-, three-, or $N$-point configurations; thus, we expect it to take the general form, 
\begin{equation}
\xi_N^{\,(\rm SN)} = \sum_{i=1}^N \frac{1}{\bar n^i}\sum_{\rm configurations}\xi_{N-i}\; \prod_{j=1}^i\delta^{\rm D}(\vec x_j-\vec x_q)  \; ,
\label{shotxi}
\end{equation}
where the sum over the configurations is running over all the possible configurations leading to a zero-lag at a given order $i$. As an example, we can get that at $N=2$ (i.e. for the 2PCF), there is only one correction term corresponding to two overlapping points. Thus
\begin{equation}
\xi^{(\rm SN)}(\vec r) = \frac{\delta^{\rm D}(\vec r)}{\bar n} \; ,
\label{xishot}
\end{equation}
where $\vec r:= \vec x_2-\vec x_1$ and $\xi:= \xi_2$. Then, at three points, there are two corrective terms ($i=1$ and $i=2$). The $i=1$ term leads to three possible zero-lag configurations $\vec x_1=\vec x_2$ or $\vec x_1=\vec x_3$ or $\vec x_2 = \vec x_3$ while the $i=2$ term leads to only one possible configuration, which is $\vec x_1 = \vec x_2 = \vec x_3$. Thus, we have
\begin{equation}
\zeta^{(\rm SN)}(\vec r, \vec s) =\frac{1}{\bar n} \left[ \xi(\vec s)\,\delta^{\rm D}(\vec r) + 2\; {\rm perm.}\right] + \frac{1}{\bar n^2}\,\delta^{\rm D}(\vec r)\,\delta^{\rm D}(\vec s) \; ,
\label{zetashot}
\end{equation}
where $\vec s := \vec x_3-\vec x_1$ and $\zeta := \xi_3$. Finally at order $4$ we get three contributions: $i=1$ shows all possible single collapse of two points among four, of which there are six possibilities, $i=2$ corresponds to the collapse of twice two points which are only three possibilities if the two pairs of collapsed point are separated, and then there are four possibilities of collapsing twice two points at the same position (i.e. collapsing three points at the same position) and finally one single possibility for collapsing the four points at the same position. Thus,

\begin{equation}
\begin{array}{rcl}
\omega^{(\rm SN)}(\vec r, \vec s, \vec t)\!\!\! & \!\!=\!\! & \!\!\! \displaystyle \frac{1}{\bar n}\left [ \delta^{\rm D}(\vec t)\,\zeta(\vec r,\vec s) + 5\; {\rm perm.} \right ] + \frac{1}{\bar n^3}\delta^{\rm D}(\vec r)\,\delta^{\rm D}(\vec s)\,\delta^{\rm D}(\vec t) \\
& & \\
& & \!\!\! \displaystyle + \frac{1}{\bar n^2} \left [ \delta^{\rm D}(\vec s)\,\delta^{\rm D}(\vec t)\,\xi(r) + 6\;{\rm perm.} \right ] \; , \label{omegashot}
\end{array}
\end{equation}
where $\vec t := \vec x_4 - \vec x_1$ and $\omega := \xi_4$. We stop at order $4$ since we want to reproduce correctly the global amplitude of the trispectrum in the ELM catalogues with the \covmos\ realisations. In general, in order to compute the shot noise contribution to the one-point cumulant moments, we need to compute the terms with collapsed configurations in Eqs.~\eqref{xishot} -- \eqref{omegashot}. This means that we need to compute 
\begin{equation}
g_{1,N}^{(\rm SN)} := \frac{1}{\bar n} \int \dif^3  x_1\,  ...\, \dif^3 x_N\; \xi_{N-1}\, \delta^{\rm D}_{N,1}\, F(\vec x_1)\,...\,F(\vec x_N) \; ,
\label{fn}
\end{equation}
where $\delta^{\rm D}_{N,1}:=\delta^{\rm D}(\vec x_N-\vec x_1)$, which leads to 
\begin{equation}
g_{1,N}^{(\rm SN)} := \frac{1}{\bar n} \int \dif^3 x_1\, ...\, \dif^3 x_{N-1}\; \xi_{N-1}\, F^2(\vec x_1)\,...\,F(\vec x_{N-1}) \; .
\label{fnb}
\end{equation}
This can be simplified by introducing the notation $F_n := F^n/\langle F^n \rangle_{\rm v}$ and $\langle F^n \rangle_{\rm v} := \int F^n(\vec x)\; \dif^3 \vec x$ which can be related to the normalised kernel $f$ as $\langle F^n\rangle_{\rm v} = a^{-3(n-1)}\int f^n(\vec s )\; \dif^3\vec s$, where $\vec s := \vec x/a$. Finally, we introduce the definition $\langle f^n\rangle_{\rm v} := \int f^n(\vec s )\; \dif^3\vec s$. In this way, we can see that any term similar to Eq.~\eqref{fnb} is a term, looking at the cross-correlation between two fields obtained with the two assignment schemes $F_2$ or $F$ at the same position, leads, respectively, to fields $\delta_{F_2}$ and $\delta_{F_1} = \delta_F$,

\begin{equation}
g_{1,N}^{(\rm SN)} = \frac{\langle F^2 \rangle_{\rm v}}{\bar n}\, \langle \delta_1^{N-1}\delta_2\rangle_\mathrm{c} \; .
\end{equation}
The other typical contribution that we need is either with three points collapsed at the same position $g_{2,N}^{(\rm SN)}$, or two pairs of points collapsed at two different  positions $h_{2,N}^{(\rm SN)}$ defined as 
\begin{equation}
g_{2,N}^{(\rm SN)} :=  \frac{1}{\bar n^2} \!\!\int \!\!\dif^3 x_1 \,...\, \dif^3 x_{N}\; \xi_{N-2}\,\delta^{\rm D}_{N,1}\, \delta^{\rm D}_{N-1,1}\; F(\vec x_1)\,...\,F(\vec x_{N}) \; ,
\label{gn}
\end{equation}
and
\begin{equation}
h_{2,N}^{(\rm SN)} := \frac{1}{\bar n^2}\!\! \int \!\!\dif^3  x_1 \,...\, \dif^3 x_{N}\; \xi_{N-2}\, \delta^{\rm D}_{N,1}\, \delta^{\rm D}_{N-1,2}\; F(\vec x_1)\,...\,F(\vec x_{N}) \; .
\label{hn}
\end{equation}
This leads to 
\begin{equation}
g_{2,N}^{(\rm SN)} = \frac{\langle F^3 \rangle_{\rm v}}{\bar n^2}\, \langle\delta_1^{N-2}\delta_3\rangle_\mathrm{c}\;,
\end{equation}
and
\begin{equation}
h_{2,N}^{(\rm SN)} = \frac{\langle F^2 \rangle_{\rm v}^2}{\bar n^2}\, \langle\delta_1^{N-3}\delta_2^2\rangle_\mathrm{c} \; .
\end{equation}

Defining $\hat \delta_n(\vec x) := \int \rho(\vec x')\, F^n(\vec x-\vec x')\;\dif^3 x' - \langle F^n \rangle_{\rm v}$ where $\rho$ is the discrete number density of an object, then we can show that the true cumulants are related to the measurable statistical quantities as
\begin{eqnarray}
  \langle\delta^2\rangle_\mathrm{c}\!\! &\!\! =\!\! & \!\!\langle \hat \delta_1^2\rangle_\mathrm{c} - \frac{\langle f^2\rangle_{\rm v}}{\bar N} \nonumber  \label{sn2 }\;,\\
 \langle\delta^3\rangle_\mathrm{c} \!\! &\!\! =\!\! & \!\!\langle\hat\delta_1^3\rangle_\mathrm{c} - \dfrac{3}{\bar N}\,\langle\hat\delta_1\hat\delta_2\rangle_\mathrm{c} + \dfrac{2}{\bar N^2}\,\langle f^3\rangle_{\rm v} \label{sn3}\;,\\
 \langle \delta^4\rangle_\mathrm{c}\!\! &\!\! = \!\!& \!\!\langle \hat\delta_1^4\rangle_\mathrm{c} - \dfrac{6}{\bar N}\, \langle \hat\delta_1^2\hat\delta_2\rangle_\mathrm{c} + \dfrac{1}{\bar N^2} \left ( 8\, \langle\hat \delta_1\hat \delta_3\rangle_\mathrm{c} + 3\,\langle \hat\delta_2^2\rangle_\mathrm{c} \right ) - \dfrac{6}{\bar N^3}\,\langle f^4\rangle_{\rm v} \nonumber\label{sn4} \; .
\end{eqnarray}
The above set of equations is the generalisation of the well established shot noise correction for count-in-cell statistics \citep{layzer_56, szapudi_92, angulo_08}. Note that the same kind of reasoning was used to correct shot noise in the context of marked correlation functions in \citet{karcher_25}.

In the particular case of the PCS mass assignment scheme, we can get $\langle f^2\rangle_{\rm v} = (151/315)^3$, $\langle f^3\rangle_{\rm v} = (1979/7560)^3$ and $\langle f^4\rangle_{\rm v} = (40853/270270)^3$. Since $\langle f^n\rangle_{\rm v}$ are of order $0.1$ it means that the PCS mass assignment scheme decreases the effect of shot noise by a factor of $10$ on all the cumulant moments.
In practice, we estimate in the ELM catalogues the fields $\hat f_1$, $\hat f_2$, and $\hat f_3$ by running over the objects in the three corresponding mass assignment schemes of $F$, $F_2$, and $F_3$. Finally, we can estimate all the statistical quantities relevant for computing the second-, third-, and fourth-order cumulant moments (see Fig.~\ref{fig:cumulent}).

In practice, we checked that the Gamma distribution offers a good description of the measured cumulant moments, thus we use a Gamma expansion \citep{gaztanaga_20, bel_16} in order to model the PDF in the ELM galaxy catalogues.

Finally, in the catalogues, we have access to the radial velocity $v_r$ of galaxies that can be normalised in $h^{-1}$Mpc with the expansion rate. We can thus estimate the one-point velocity dispersion by measuring the dispersion of the radial velocity as a function of the considered bin of redshift with
\begin{equation}
\sigma_v = \sqrt{ \frac{1}{N_{\rm g}}\sum_{i=1}^{N_g} v_{r,i}^2 } \; ,
\end{equation}
where $N_{\rm g}$ is the total number of galaxies that we have into the resdhift bin. In the right panel of Fig.~\ref{fig:cumulent}, the total velocity dispersion above is compared  to 
the expected variance of the coherent velocity flow $\sigma_{\theta\theta}^2 = 4\pi/3\int P_{\theta\theta}(k)\; \dif k$. It shows that the one-point velocity dispersion is roughly $3.5$ higher than the one given by the coherent velocities. It is therefore essential to include the one-point velocity dispersion in order to correctly reproduce the RSD.

\end{appendix}

\end{document}